\documentclass[a4paper,11pt,nohyper]{article}
\pdfoutput=1
\newtheorem{theorem}{Theorem}[section]

\newtheorem{rem}[theorem]{Remarks}

\usepackage{jheppub}

\usepackage[T1]{fontenc}
\usepackage[mathletters]{ucs} 
\usepackage[utf8x]{inputenc}
\usepackage{multirow,booktabs,array}
\usepackage{amssymb}
\usepackage{amsmath}
\usepackage{lmodern}
\usepackage{xcolor}
\usepackage{braket}
\usepackage{verbatim}
\usepackage{multirow}
\usepackage{graphics,graphicx}
\usepackage{psfrag}
\usepackage{empheq} 
\usepackage{mathtools}
\usepackage{cleveref}

\usepackage[makeroom]{cancel}

\usepackage{cases}
\usepackage{enumitem}

\def\a{\alpha}
\def\b{\beta}
\def\c{\gamma}

\def\g{\gamma}

\def\k{\kappa}                    
\def\l{\lambda}
\def\m{\mu}
\def\n{\nu}
  
\def\r{\rho}                                     

\def\u{\upsilon}
\def\x{\xi}

\def\G{\Gamma}

\let\a=\alpha \let\b=\beta \let\g=\gamma  
    \let\k=\kappa
\let\l=\lambda \let\m=\mu \let\n=\nu \let\x=\xi \let\r=\rho
    \let\c=\chi

    \let\G=\Gamma

 \def\bd{\begin{document}} \def\ed{\end{document}}
\def\ds{\documentstyle} \let\fr=\frac \let\bl=\bigl \let\br=\bigr
\let\Br=\Bigr \let\Bl=\Bigl
\let\bm=\bibitem
\let\na=\nabla
\definecolor{amethyst}{rgb}{0.6, 0.4, 0.8}
\definecolor{byzantine}{rgb}{0.74, 0.2, 0.64}

\newcommand{\boxedeq}[2]{\begin{empheq}[box={\fboxsep=6pt\fbox}]{align}\label{#1}#2\end{empheq}}
\newcommand{\be}{\begin{equation}}
\newcommand{\ee}{\end{equation}}

\newcommand{\bea}{\begin{eqnarray}}
\newcommand{\eea}{\end{eqnarray}}

\newcommand*\dif{\mathop{}\!\mathrm{d}}

\definecolor{ao(english)}{rgb}{0.0, 0.5, 0.0}

\usepackage{amsmath,latexsym,amssymb,slashed}
\usepackage{mathrsfs}

\title{General null asymptotics and superrotation-compatible configuration spaces in $d\ge4$}

 \author{F. Capone}
 \affiliation{Mathematical Sciences and STAG Research Centre, University of Southampton\\Highfield, Southampton, SO17 1BJ, UK.} 
\bigskip

\emailAdd{federico.capone@soton.ac.uk}

\abstract{We address the problem of consistent Campiglia-Laddha superrotations in $d>4$ by solving Bondi-Sachs gauge vacuum Einstein equations at the non-linear level with the most general boundary conditions preserving the null nature of infinity. We discuss how to generalise the boundary structure to make the configuration space compatible with supertanslation-like and superrotation-like transformations. One possibility requires that the time-independent boundary metric on the cuts of $\mathscr{I}$ is not fixed to be Einstein, while the other sticks to Einstein but time-dependent metrics. Both are novel features with respect to the four dimensional case, where time-dependence of the two-dimensional cross-sectional metric is not required and the Einstein condition is trivially satisfied. Other cases are also discussed. These conditions imply that the configuration spaces are not asymptotically flat in the standard sense. We discuss the implications on the construction of the phase space and the relationship with soft scattering theorems. We show that in even spacetime dimensions, the initial data compatible with such asymptotic symmetries produce maximally polyhomogeneous expansions of the metric and we advance a potential interpretation of this structure in terms of AdS/CFT and realizations of Ricci-flat holography.  
}

\begin{document}
	\flushbottom
	\maketitle	
\section{Introduction and motivational remarks}
Soft theorems characterise scattering processes in any theory of gravity with $d\ge4$ flat non-compact dimensions \cite{Sen2017,Sen2017b}. The Bondi-Metzner-Sachs (BMS) group of four-dimensional asymptotically flat spacetimes and its various extensions have been conjectured to be symmetries of semiclassical perturbative scattering because the action of the associated generating charges on the S-matrix  can be argued to imply gravitational leading (Weinberg \cite{Weinberg1965}) and subleading (Cachazo-Strominger \cite{Cachazo2014}) soft theorems \cite{Strominger2014,He2015,Kapec2014,Campiglia2014}. 
\\
\\
By definition, the BMS group \cite{Bondi1962,Sachs1962} preserves the universal structure of asymptotically flat spacetimes at either future $\mathscr{I}^+$ or past $\mathscr{I}^-$ null infinity, where $\mathscr{I}^{\pm}$ have $\mathbb{R}\times S^2$ topology and the universal structure is the pair formed by a null normal to $\mathscr{I}$ and the round sphere metric on the celestial sphere $S^2$. Such spacetimes can be called asymptotically Minkowski \cite{Geroch1977}. 
Under such conditions, the BMS group is the semidirect product of supertranslations (the Abelian factor), acting geometrically by arbitrarily shifting each point of $S^2$ along $\mathbb{R}$, and the proper orthocronous Lorentz group (the non-Abelian part), acting on $S^2$ as global conformal transformations. 

Extensions of BMS concern its non-Abelian part. The global conformal transformations of $S^2$ can be relaxed to either local conformal transformations (generated by two copies of Witt algebra) \cite{Barnich2010,Barnich2011a} or to arbitrary volume preserving smooth diffeomorphisms of $S^2$ \cite{Campiglia2014,Campiglia2015}. Local conformal transformations are usually called superrotations in this context. We will refer to them as BT-superrotations and use the name CL-superrotations\footnote{BT stands for Barnich-Troessaert, the authors that first studied the associated phase space \cite{Barnich2010,Barnich2011a}, while CL is for Campiglia-Laddha \cite{Campiglia2014,Campiglia2015}. Superrotations are more correctly called super-Lorentz transformations \cite{Compere2018}. Their divergence-free part is a generalisation of rotations (hence superrotations) while the rotation-free part is a generalisation of boosts (superboosts). BMS with BT-superrotations is sometimes called \emph{extended BMS}, BMS with CL-superrotations is called \emph{generalised BMS}. Recently also a Weyl-BMS group has been considered as a generalisation of Campiglia-Laddha proposal with Weyl rescalings \cite{Freidel2021}.} for the smooth volume preserving diffeomorphisms of $S^2$. Both extensions of BMS are obtained under appropriate relaxation of the asymptotic Minkowski conditions specified above. CL-superrotations need the universal structure to be defined as a pair involving a normal and a volume form \cite{Campiglia2014,Flanagan2019} over $S^2$ rather than a normal and a metric. BT-superrotations are long known to be naturally realised, for example, in boost-rotation symmetric spacetimes \cite{Bicak1989,Bicak1984} because 
these spacetimes possess topologically incomplete null infinity.%
\\
\\
The uncovered relations between the various form of BMS symmetries and soft theorems (as well as the related gravitational memory effects \cite{Zel1974,Chris_mem,Strominger2016,Strominger2017,Compere2018}, with which they form the triangular chains of equivalences called \emph{infrared triangles}) revived the interest in the asymptotically flat spacetime holography problem. 
Four-dimensional scattering amplitudes in Minkowski spacetime can be cast in terms of correlators of operators defined on the two-dimensional celestial sphere/the null boundary \cite{Cheung2016iub,Pasterski2017,Donnay2020,Bagchi2016,Banerjee2020}. This is supposed to hint to holographic realizations of gravity with vanishing cosmological constant \cite{deBoer2003} in the same way that quantum field theoy in AdS, rewritten in terms of boundary operators, relate to AdS/CFT in hindsight \cite{Penedones2017}. 
However, AdS/CFT is greatly more general than this rewriting \cite{Maldacena1997,MAGOO}. Analogously, the dynamical principles of flat spacetime holography are probably hidden beyond perturbative physics around Minkowski spacetime. The long-standing open question for holography with a vanishing cosmological constant is to find such a boundary \emph{structure X} (using Witten's terminology \cite{WittenStrings,Witten:2001kn}) that outputs the flat space S-matrix as well as the full richer physics of non-linear gravitational waves and black holes. 

We are thus led to approach the problem from more general points of view. A possible path to follow is considering more broadly the problem of Ricci flat holography \cite{deBoer2003} and its relations to holography in asymptotically locally AdS spacetimes \cite{SkenderisLec,Caldarelli2013aaa,Bagchi2010,Barnich2012,Bagchi2013,Ciambelli2018b,Costa2012fm,Costa2013vza}.
\\
\\
In this paper we adopt this more general point of view to explore higher dimensional realizations of BMS symmetries beyond perturbative analysis around Minkowskian backgrounds, with a particular focus on superrotations. As we explain in due course, this purpose soon imply that we have to abandon the strict notions of global asymptotic flatness at null infinity and resort to local Ricci flatness. In turn, the absence of a well defined boundary muddles the notion of asymptotic symmetries.
\\
\\
Our scope, results and the holographic perspective are better framed by recalling that 
a fundamental condition for both the infrared triangles and any holographic interpretations to hold consistently is the existence of a well-posed phase space structure. In a covariant phase space perspective \cite{Crnkovic1988}, where the phase space is built over a configuration space by endowing the latter with a symplectic structure, the problem is schematically divided as:
\begin{itemize}
	\item[1)] Definition of the configuration space of fields with given boundary conditions (in our case either at $\mathscr{I}^+$ or $\mathscr{I}^-$) with consistent asymptotic Killing fields,
	\item[2)] Definition of the phase space: charges associated to the asymptotic Killing fields must be finite and - in principle - integrable.
\end{itemize} 
In particular, since the ultimate goal is the scattering problem among data on disjoint boundaries, another condition must thus be met:
\begin{itemize}
	\item[3)] Matching conditions among the two disjoint boundaries must be defined for their asymptotic symmetries to form a unique algebra acting consistently and non-trivially (i.e. non vanishing charges) on the  S-matrix.
\end{itemize}
These requirements have been extensively discussed in $d=4$ and a monumental amount of literature has been produced, that cannot be all acknowledged here. For example, point 1) concerns the conditions satisfied by realistic radiative systems and is translated in the very definition of $\mathscr{I}$ as a smooth or polyhomogeneous surface \cite{Christodoulou1993,Chrusciel,Friedrich2017cjg}. Point 2) imposes very strict requirements and is usually to be discussed case by case. On the one hand, the lack of divergences is to be understood as the possibility to handle them with appropriate subtractions \cite{Gibbons1976} or - where available - (holographic) renormalization schemes \cite{Henningson1998}. On the other hand radiative phase spaces are in general incompatible with integrability. Supertranslations within smooth null infinity are special for both these aspects because they admit finite and integrable charges \cite{Wald2000} without any need of renormalization\footnote{Refer to \cite{Godazgar2020} for a corresponding analysis with (particular) polyhomogeneous fields.}. Superrotations instead require more care: the strategy of \cite{Wald2000} is not consistent when applied to BT-superrotations \cite{FlanaganBMS} and CL-superrotation charges need to be renormalized \cite{Compere2018,Flanagan2019}. Point 3) was the core intuition of the pivotal paper \cite{Strominger2014} and is further discussed in \cite{AshTalk,Prabhu2019}, all dealing only with supertranslations.
\\
\\
In higher dimensions, from the sole point of view of asymptotic symmetries of general relativity, the interest stems from the fact that standard definitions of asymptotic flatness - summarised in Section \ref{sec.standard} - do not experience the enhancement of the Poincaré symmetries to (any form of) BMS symmetries \cite{Hollands2003,Hollands2004,Hollands2005,Hollands2013cva,Tanabe2011,Tanabe2012}, while the existence of soft theorems in any number of dimensions suggest that there should exist a symmetry principle underlying them. The reason of the conundrum is the dependence of asymptotic symmetries on boundary conditions and asymptotic falloff behaviour of the fields\footnote{Here by boundary condition we mean the leading order of the asymptotic expansion, while the asymptotic falloff behaviour refers on how the field is developed in the bulk.}.

A satisfactory analysis of supertranslations in higher even dimensions has been given upon linearisation of the gravitational field equations and is based on the definition of a configuration space with Minkowskian asymptotics and the same falloff behaviour of the field as in four dimensions \cite{Kapec2015,Aggarwal2019}. Here we refer to this construction as KLPS configuration space,  KLPS standing for Kapec, Lysov, Pasterski, Strominger \cite{Kapec2015}. Despite these supertranslations are consistent asymptotic Killing vectors, \cite{Aggarwal2019} highlights severe issues in extending the phase space construction beyond the linear level. 

A natural expectation was that such a linear configuration space supports CL-super\-rotations by simply relaxing one of the asymptotic Killing conditions \cite{Avery2016,CaponeProcBMS}, as is the case in four dimensions\footnote{There is no natural algebraic definition of BT-superrotations in spacetime dimension greater than four, but see \cite{Capone2019} for a conjecture in terms of cosmic branes. We come back to this point later.}. However, the explicit analysis of \cite{Colferai2020} shows that CL-superrotations act inconsistently on the KLPS configuration space because they break its defining conditions at subleading orders, as we review in Section \ref{sec.KLPS}. As we see explicitly here, this inconsistency is not a consequence of the linearised analysis. Hence we need to reconsider point 1) from scratch in order to discuss higher dimensional superrotations.
\\
\\
The present paper accomplishes this goal. We discuss the most general, non-linear, Ricci flat configuration space at null infinity in any number of spacetime dimensions with appropriate boundary conditions supporting consistent actions of superrotation-like and super\-translation-like transformations. 
The postfix ``like'' is mandatory because we show (subsections \ref{sec.bc},\ref{sec.h1discussion}) that to accommodate smooth diffeomorphisms of the cross sections of null infinity (CL-superrotations) among the asymptotic Killing fields\footnote{Although the generators of CL-superrotations are not asymptotic Killing vectors in the standard sense (i.e. that they preserve asymptotically the metric), we will use this nomenclature.}, the boundary conditions need to be further extended beyond the conditions proposed by Campiglia and Laddha in four spacetime dimensions: the interpretation of these transformations as enhancements of translations and rotations/Lorentz boosts is in general lost.  These boundary conditions are discussed in \ref{sec.bc} and are shown to serve our purposes in subsection \ref{sec.h1discussion} via the claims \ref{clm.nonE}, \ref{clm.timeh}, \ref{clm.b0}, to which the experienced reader can immediately turn. They provide also potential solutions to the problems faced when extending KLPS-supertranslations beyond the linearised level in dimension greater than four. 

Both the linear KLPS configuration space, and the standard asymptotically flat configuration spaces, as well as configuration spaces with Robinson-Trautman solutions are subcases of the general ones considered here. However, even the milder extension of boundary conditions with respect to the standard asymptotically flat conditions produces new puzzles: as we discuss in subsection \ref{sec.limit}, we do not have any way - and we will not give any here - to assess the existence of stable global null infinity and well-posed limits to $i^0$. Both these observations are crucial aspects to consider for a definition of the scattering problem.

The minimal requirement for obtaining BMS-like symmetries in higher dimensions is to relax the falloff behaviour of fields with respect to what is assumed in customary treatments of higher-dimensional gravitational radiation. However, rather than engineering the falloff conditions for this purposes, we systematically derive them from the integration of Einstein equation. In this sense, the approach is similar to the one producing the most general asymptotic expansion of asymptotically locally AdS (AlAdS) spacetimes relevant for holography \cite{FG1985,SkenderisLec}. The extent to which this picture can be followed in our context is specified in section \ref{sec.aadis}.

We show that this produces non-trivial logarithmic terms in even dimensions greater than four, while the same can be trivially set to zero in odd dimensions (see summary in secion \ref{sec.res}). This structure suggests in subsection \ref{sec.leadinglog} a further comparison with the asymptotic expansion of AlAdS spacetimes and a possible holographic interpretation in view of various approaches to  Ricci flat holography. 

We may call the configuration spaces we discuss ``\emph{asymptotically locally flat}'' or - borrowing nomenclature from AdS/CFT literature - ``\emph{asymptotically locally Minkowskian}'' AlM\footnote{To further stress the nomenclature: we use the terminology ``null infinity'', ``$\mathscr{I}$'' with no mention that there is a global stable definition of these loci (as already known for the standard analsysis in odd dimensions).}. 
\\
\\
We stress that in this paper we limit ourselves to the discussion of the configuration space, hence point 1) of the previous list.  We hope to address the remaining points, as well as the holographic picture, elsewhere \cite{Capone2021}.
\\
\\
The paper is organised as follows. In section \ref{sec.recap}, which can be skipped by the expert reader, we recap the state of the art of the explorations on higher dimensional supertranslations and superrotations and explicate the issues mentioned in this introduction. Section \ref{sec.res0} is the core of the paper. After a discussion of the boundary conditions (subsection \ref{sec.bc}), we discuss in detail the results mentioned in this Introduction (subsection \ref{sec.res}) before moving to the proof of them in sections \ref{sec.aadis} and	 \ref{sec.ExpDetails}. The strategy of the solution of the main equations and an example (\ref{Sec.LeadingExp}) which captures several points of the main discussion is discussed in section \ref{sec.aadis}. Section \ref{sec.ExpDetails} delves into the details with a power-law ansatz for the integration of the equations, while section \ref{sec.maxpol} comments on the maximal polyhomogeneous expansions. Section \ref{sec.akvALM2} gives the details of the asymptotic Killing field computations and shows that generic diffeomorphisms of the cross sections of null infinty are impossible in higher dimensions without supertranslation-like transformations. We conclude in \ref{sec.concl} with some further comments and possible directions. In the first three appendices we give other computational details, while the latter is independent from the rest. It is a glossary of Geroch's definitions of asymptotic flatness \cite{Geroch1977} where we give a definition of Carroll structure with Campiglia-Laddha superrotations.

\paragraph{Note added in v4.} In this version the author corrects a mistake affecting paragraph \textbf{$\boldsymbol{d=6}$ and even} in section \ref{Sec.NR} of the previous versions and some other minor typos. The main conclusions of the paper are unaffected, but equation \eqref{eq.N6better0}/\eqref{eq.N6better} (a more explicit version of the already present \eqref{eq.N6}) is added and the discussion following them is corrected and enhanced. Two core statements already present in previous versions (in \ref{clm.nonE} and in the Conclusions) are clarified. These points are detailed in ``Erratum: General null asymptotics and superrotation-compatible configuration spaces in $d\ge4$'' to be published in JHEP.

\section{Bondi-Sachs problem, $d>4$ BMS current status}\label{sec.recap}
The Bondi-Sachs coordinate system $(u,r,x^A)$ is
defined so that the metric reads
\be
ds^2 = - {\cal U} e^{2 \beta} du^2 - 2 e^{2 \beta} du dr + r^2 h_{AB} (d x^A - W^A du)(d x^B - W^B du), \label{eq.bondi}
\ee
where $u=const$ picks a null surface, whose conormal is $l_\m=\partial_\m u$, and the other coordinates are defined by
\begin{equation}\label{gaugecond}
g_{rr}=g_{rA}=0\,,\quad \text{det}(h_{AB})=q(u,x),
\end{equation}
with capital latin indices running over the $(d-2)$ coordinates on the cross sections of the $u=const$ null surface and $q$ a fixed arbitrary function. Unless additional symmetry requirements are imposed, the metric functions $\mathcal{U}$, $\beta$, $W^A$ and $g_{AB}:=r^2h_{AB}$ are functions of all the coordinates. A configuration space is determined by the boundary conditions on these metric functions, taken as the asymptotic values as $r\rightarrow \infty$ and by the way in which these functions are expanded in $r$. 
\\
\\
Einstein equations take the form of a characteristic initial value problem where $h_{AB}$ is to be imposed on a initial null surface $u=const$ and the radial expansion of the functions $\b$, $W^A$ and $\mathcal{U}$ and $y_{AB}:=\partial_uh_{AB}$ is determined up to free functions ($\b_{(0)}$, $W_{(0)}^A$, $\mathcal{W}_{(d-1)}^A$, $\mathcal{U}_{(d-3)}$ and  $y_{(\frac{d-2}{2})AB}$), via the so-called main equations according to the scheme (in vacuum)
\begin{align}
&R_{rr}=0\Rightarrow &&\b(u,r,x)=\b_{(0)}(u,x)+b(u,r,x),\label{scheme0}\\
&R_{rA}=0\Rightarrow &&W^A(u,r,x)=W^A_{(0)}(u,x)+\frac{\mathcal{W}^A_{(d-1)}(u,x)}{r^{d-1}}+w(u,r,x),\label{scheme1}\\
&g^{BA}R_{AB}=0\Rightarrow &&\mathcal{U}(u,r,x)= \frac{\mathcal{U}_{(d-3)}(u,x)}{r^{d-3}}+\u(u,r,x),\label{eq.trace1}\\
&g^{DA}R_{AB}=0\Rightarrow && y_{AB}(u,r,x)=\frac{y_{(\frac{d-2}{2})AB}(u,x)}{r^{\frac{d-2}{2}}}+\tilde{y}_{AB}(u,r,x)\label{eq.traceless1}
\end{align}
The evolution in $u$ of $\mathcal{W}_{(d-1)}^A$ and $\mathcal{U}_{(d-3)}$ is determined algebrically  by the so-called supplementary equations
\begin{equation}\label{eq.suppl}
R_{uA}=\partial_r\left(r^{d-2}R_{uA}\right)=0,\quad R_{uu}=\partial_r\left(r^{d-2}R_{uu}\right)=0,
\end{equation}
respectively. The remaining equation
\begin{equation}\label{eq.triv}
R_{ur}\equiv 0
\end{equation} 
is an identity after \eqref{eq.trace1} by Bianchi identities (or vice versa) and is called the trivial equation.
\subsection{Radiation-compatible asymptotically Minkowski/Einstein boundary}\label{sec.standard}
Asymptotically Minkowski boundary conditions are those first imposed by Sachs \cite{Sachs1962a} in $d=4$ and adapted to the higher dimensional case by Tanabe and collaborators \cite{Tanabe2011}. We can summarise them as
\begin{itemize}
	\item[i)] Coordinate range: $u_0\le u\le u_1$, $r_0\le r\le\infty$, $x^A$ on the round-spheres (i.e in four spacetime dimensions $0\le \theta \le \pi $, $0\le \phi \le 2\pi$ with $\phi=\phi+2\pi$)\footnote{In the metric based approach, the topological restrictions, as observed by Sachs, follow from the assumed range of coordinates and the form of the asymptotic metric, but nothing is implied about the topology at $r< r_0$.} 
	\item[ii)] Over a characteristic, the metric functions behave asymptotically as
	\be\label{eq.bc}
	\lim\limits_{r\rightarrow\infty}rW^A=\lim\limits_{r\rightarrow\infty}\b=0\,\quad \lim\limits_{r\rightarrow\infty}h_{AB}=h_{(0)AB}(u,x)=\gamma_{AB}(x)
	\ee
	where we call $h_{(0)AB}$ the leading asymptotic order of $h_{AB}$, which according to these conditions is defined as the round metric $\gamma_{AB}$ on $S^{d-2}$. Einstein equations imply $\lim\limits_{r\rightarrow\infty}\mathcal{U}=1$ (it is proportional to the Ricci scalar of $\gamma$).
\end{itemize}
Along with these, Bondi-Sachs original conditions, define asymptotically flat fields to be non-polyhomogeneous in $r$, i.e.
\begin{itemize}
\item[iii)] The asymptotic $r$ expansion of each metric field does not contain logarithmic terms.
\end{itemize}

With the boundary conditions i) and ii), the configuration space is automatically consistent with linearised perturbations of the gravitational field if \cite{Hollands2004,Tanabe2011}
\begin{equation}\label{eq.rad}
h_{AB}- \gamma_{AB}=O(r^{\frac{2-d}{2}})
\end{equation}
because the free function $y_{(\frac{d-2}{2})AB}$ is traceless, symmetric and hence contains the right number of polarization modes and can be identified with the news tensor $N_{AB}$. At subleading orders $h_{AB}$ is imposed to be given as an expansion in inverse integrer powers of $r$ if $d$ even and integer as well as half-integers if $d$ is odd \cite{Tanabe2010,Tanabe2011}.

Both in four and higher dimensions, these conditions lead to well defined notions of mass and angular momenta from the asymptotic Killing fields that are defined so that both the gauge choice, the boundary conditions and the falloff conditions are preserved. In particular the condition 
\begin{equation}
\mathfrak{L}_\xi g_{AB}=O(r^{\frac{6-d}{2}})
\end{equation}
implies that Poincaré translations are enhanced to supertranslations only in $d=4$ (see section \ref{sec.akvALM2} paragraph ``BS \& $a=\frac{d-2}{2}$'').

When $d>4$ a more general definition of boundary conditions leading to a well defined Bondi mass involves  $\gamma_{AB}$ being any time-independent Einstein metric on the Euclidean cross sections $B^{d-2}$ of $\mathscr{I}$ \cite{Hollands2013cva}.
\subsection{KLPS, supertranslation-compatible configuration spaces and superrotations}\label{sec.KLPS}
The supertranslation-compatible configuration space of Kapec, Lysov, Pasterski and Strominger (KLPS) \cite{Kapec2015} is defined via the asymptotically Minkowski conditions i) and ii) and the imposition that $h_{AB}$ is expanded as in four dimensions without logarithmic terms
\begin{equation}\label{eq.nonrad}
h_{AB}=\gamma_{AB}+\frac{h_{(1)AB}}{r}+\frac{h_{(2)AB}}{r^2}+\dots
\end{equation}
and that all the other metric functions are expanded accordingly as in four dimensions with the assumption iii). With such assumptions, supertranslations are obtained automatically because of the condition (see section \ref{sec.akvALM2} paragraph ``BS in $d>4$ \& $a=1$'')
\begin{equation}\label{eq.st}
\mathfrak{L}_\xi g_{AB}=O(r).
\end{equation}
The analysis of KLPS is necessarily restricted to even-spacetime dimensions as no-half integer powers are included and is restricted to a linear study of the field equations.

The most important feature of the KLPS configuration space for our purposes is that Einstein equations fix $h_{(1)AB}$ to be time-independent
\begin{equation}\label{eq.KLPS}
\partial_u h_{(1)AB}=0,
\end{equation}
differently from the four-dimensional case where the time-dependence of $h_{(1)AB}$ is related to the news tensor (this is the radiative order in four dimension)\footnote{For $d=6$ $h_{(2)AB}$ is the radiative term related to the news tensor and is free. In higher dimensions equation \eqref{eq.KLPS} applies to all other orders before the radiative.}. As evident from sections \ref{sec.h1discussion} and \ref{Sec.NR}, this equation is not the effect of the linearised analysis, but only of the boundary conditions. 

Under these conditions, the covariant phase space analysis of points 2) and 3) of the Introduction was performed by Aggarwal in  \cite{Aggarwal2019}. The natural divergences of the charges due to the integrals over higher dimensinal spheres of the terms with slower falloff than the radiative order are cured by imposing a specific behaviour of the covariant derivatives of  $h_{(\frac{d-2}{2})AB}$ as $|u|\rightarrow\infty$ (the past and future boundaries of $\mathscr{I}$). Despite no first-principle derivation of these conditions was presented, they are satisfactory from the point of view of the scattering problem because they contribute to the correct soft theorem counting. As analysed in \cite{Aggarwal2019}, they are however not preserved beyond the linear level because a supertranslation transform $h_{(\frac{d-2}{2})AB}$ by additional pieces depending on $h_{(1)AB}$ which are $u$-independent by \eqref{eq.KLPS}.
\\
\\
A similar issue, rooted in \eqref{eq.KLPS} - but already at the configuration space level - affects the higher dimensional CL-superrotation analysis of \cite{Colferai2020}. Assuming that we can extend the definition of asymptotic Killings of the KLPS configuration space as done by Campiglia and Laddha in four dimensions,
\begin{equation}\label{eq.CL}
\mathfrak{L}_{\xi} g_{AB}=O(r^2).
\end{equation}
the resulting  $Diff(S^{d-2})$ smooth vector fields act on the configuration space breaking \eqref{eq.KLPS} because they induce $u$-dependent changes of $h_{(1)AB}$ \cite{Colferai2020} (see section \ref{sec.akvALM2} paragraph ``CL \& $a=1$, $d>4$''). Thus no consistent proposal of superrotation charge can be made on the KLPS configuration space. 
\\
\\
With a different take, a $u$-dependence for $h_{(1)AB}$ is obtained in the later linear analysis of \cite{Campoleoni2020} around Minkowski spacetime. Here, the time dependence of $h_{(1)}$ is due to time-independent deformations of the leading round sphere metric $\gamma_{AB}$ at the linear level.

In fact, this feature previously appeared in the five-dimensional non-linear analysis of \cite{Capone2019}, with the aim of exploring generalities of  the '4d cosmic-branes $\leftrightarrow$ superrotations' relationship. The arguments imply that the round sphere boundary metric $\gamma_{AB}(x)$ should be extended to a generic $h_{(0)AB}(x)$. The linearisation of the relevant equations of \cite{Capone2019} gives those of \cite{Campoleoni2020}, as can be checked in a covariant way in this paper (the linear version of \eqref{eq.core} with $\b_{(0)}=0$). In the next subsections we explicitly show that such metrics must not satisfy the Einstein condition, and hence in the linearised approach of \cite{Campoleoni2020}, the perturbation $\epsilon$ of $\gamma$ is such that $\gamma+\epsilon$ is not Einstein. 
\section{General Ricci flat asymptotics: summary and discussion of results}\label{sec.res0}
\subsection{General asymptotic conditions}\label{sec.bc}
A way out from \eqref{eq.KLPS} is obtained by appropriate generalisations of the boundary conditions, by which we mean the conditions imposed on the metric functions determining the leading form of the Bondi-Sachs metric, $h_{(0)AB}(u,x)$, $\b_{(0)}(u,x)$ and $W_{(0)}^A(u,x)$.  Here we discuss how much the assumptions made in ii) \eqref{eq.bc} can be generalised and what kind of physical situations may correspond to these boundary conditions. 

The generalisation of \eqref{eq.KLPS} is equation \eqref{eq.core0}. It is obtained as a consequence of the fourth main equation with the boundary conditions that we now describe.  We postpone the discussion of this result to subsection \ref{sec.h1discussion} in order to express it coherently to the order in which it appears in the asymptotic expansion (see section \ref{sec.res}) and avoid repetitions. There we see how various forms of these extended boundary conditions (claims \ref{clm.nonE},\ref{clm.timeh},\ref{clm.b0}) generalise \eqref{eq.KLPS}.

\subsubsection*{$\boldsymbol{W^A_{(0)}}$:} The condition $W^A_{(0)}=0$ is necessary to preserve the definition of $u$ ($g_{uu}<0$) because otherwise on-shell
\begin{equation}\label{eq.W0}
\lim_{r\rightarrow\infty}\frac{g_{uu}}{r^2}=h_{(0)AB}W^A_{(0)}W^B_{(0)}>0.
\end{equation}
\subsubsection*{$\boldsymbol{h_{(0)AB}}$:} 
The assumption is made in \eqref{eq.bc} that the cross-sections of $\mathscr{I}$ are spherical and that $h_{(0)AB}$ does not depend on $u$. Already Newman and Unti \cite{NU1962} lifted this topological restriction in their treatment of four dimensional asymptotically flat spacetimes and the relevant asymptotic symmetries have been studied in \cite{Foster1987}; an argument supporting the second assumption is instead the possibility of properly defining the stability of null infinity against perturbations\footnote{I am grateful to S. Hollands for having drawn this point to my attention.}, which is a fundamental requirement in the theory of asymptotics \cite{Geroch1977}.

\paragraph{Time-independent $\boldsymbol{h_{(0)}}$:} In four spacetime dimensions, with $\mathscr{I}$ homeomorphic to $\mathbb{R}\times S^2$, the assumption that $\hat{h}_{(0)AB}(x)$ (we use an hat to denote time-independent quantities) is the round-sphere metric is most natural because there exist a conformal transformation that maps any metric to the round sphere one \cite{Geroch1977} and the conformal factor can be absorbed by a redefinition of the radial and angular coordinates in the bulk. 
\\
\\
In higher dimensions, with $S^{d-2}$, a theorem by Kuiper states that any conformally flat metric can be mapped to the round sphere \cite{Kuiper1949}, then we can again absorb the conformal factor in a redefinition of the radial and the angular variables. Hence, under the conformally flat assumption, the asymptotically Minkowskian conditions imposed in \cite{Tanabe2011} are always possible. 

The higher-dimensional round sphere metric is an example of many inequivalent classess of Einstein metrics on spheres \cite{Bohm1998,Boyer2003}. The discriminating factor in the structure of the configuration space is whether the metric $h_{(0)AB}$ is Einstein or not. 
\begin{rem}\label{rem.Flg}
\emph{For the purposes of later discussions of $\b_{(0)}$, we remind the reader that in four dimensions the freedom to fix $\hat{h}_{(0)}$ to the round sphere metric supports the standard definition of asymptotic symmetries with $\mathfrak{L}_\xi g_{AB}=O(r)$ over the larger algebra generated by the Campiglia-Laddha condition $\mathfrak{L}_\xi g_{AB}=O(r^2)$. Following \cite{Flanagan2019}, we see a puzzle because the extension of BMS to CL-superrotations would be seen - from these geometrical considerations - as a fake enlargement of the asymptotic symmetries to include pure gauge transformations. This is in stark contrast with the relationship of CL-superrotations to subleading soft theorems. The resolution of the puzzle is obtained if a sound proof that the standard conditions fix degrees of freedom which are not truly gauge \cite{Flanagan2019} is obtained. See also \cite{Campiglia2020qvc}.}
\end{rem}
\paragraph{Time-dependent $\boldsymbol{h_{(0)}}$:}
Regardless of the considerations on stability of null infinity (which is anyway an evasive concept in higher - odd at least - dimensions), the most general metric $h_{(0)AB}$ on the cross sections of $\mathscr{I}$ is both time dependent and not necessarily Einstein.

Among the four-dimensional spacetimes with a time-dependent $h_{(0)AB}$, which are not asymptotically flat in the usual sense - but nonetheless locally possess null infinity -  Robinson-Trautman spacetimes \cite{RT1960} are particularly relevant in the theory of gravity waves. In the broader context of holography with negative cosmological constant and its relations with Ricci-flat holography, four dimensional AdS-Robinson-Trautman spacetimes have been discussed as dual of out of equilibrium phenomena \cite{Boonstra1999} and a flat limit was taken in \cite{Fareghbal2018}. Higher dimensional Robinson-Trautman spacetimes with any value of the cosmological constant have been defined in \cite{Podolsky:2006du} (see \eqref{eq.RT}).

The four-dimensional asymptotic analysis of \cite{HT1987,Hogan1985} accommodated Ricci flat Robinson-Trautman spacetimes in a Bondi-Sachs framework and more recently this has been partially reconsidered in \cite{Barnich2010}. Such spacetimes have been related to transitions sourced by superrotations \cite{Compere2018}\footnote{It should be noted however that Robinson-Trautman spacetimes do not really have a BMS asymptotic symmetry group because of the time-dependence of $h_{(0)}$. In this case the asymptotic coordinate transformations act on $u$ as $u\rightarrow U(u,x)$ and it is not possible to integrate the transformation so that $u\rightarrow\Omega(x)[u+\a(x)]$, where $\alpha$ parametrises the supertranslations and $\Omega$ is the conformal factor.}, in particular the impulsive limits of Robinson-Trautman spacetimes represents the creation/snapping of cosmic strings that have been interpreted as processes induced by BT-superrotations \cite{Strominger2017}. 

Despite these considerations, at the time of writing, no construction of asymptotic charges and the phase space has been given with the time dependent metric on the cuts of $\mathscr{I}$ (compare i.e. \cite{Barnich2011a,Compere2018,Freidel2021}). A brief account of the issues affecting a good definition of Bondi mass, Bondi mass loss, and news tensor in this case can be found in section 5 of \cite{HT1987}. In terms of the global existence of null infinity in such a case, the point is the one of stability mentioned before.

\subsubsection*{$\boldsymbol{\b_{(0)}}$:}
In discussing $\b_{(0)}$ we would like to recall the remark \ref{rem.Flg} on the gauge behaviour of the time-independent $\hat{h}_{(0)}$.

The scalar $\b$ is related to the expansion $\Theta= e^{2\b}/r$ of the null congruence generated by the rays in the null hyperusrfaces that foliate the spacetime and the standard boundary conditions \eqref{eq.bc} fix $\b_{(0)}$ to zero. There exist thus a diffeomorphism that gauges $\b_{(0)}$ away. Differently from the transformation that  reabsorbs the conformal factor in the mapping from $\hat{h}_{(0)AB}$ to the round sphere metric, the diffeomorphism required to gauge $\b_{(0)}$ away involves $u$ and thus deforms the initial null surface in the spacetime. This can be claimed to be unnatural from a characteristic initial value problem point of view \cite{Chrusciel}.

Such an argument is not enough to argue that $\b_{(0)}$ is more than a simple gauge freedom, but some further considerations support this claim. The situation is somewhat similar to the points remarked earlier on the gauge versus non-gauge character of the Campiglia-Laddha extension of the BMS group: does $\b_{(0)}$ encode some non-trivial physics? From the arguments below the answer seems in the affirmative, but the physics does not satisfy customary asymptotic flatness. This is an important difference with what Campiglia-Laddha conditions in four dimensions imply (our goal is to understand the higher dimensional case): they do modify the standard definitions of asymptotic flatness, but only in a mild way which do not manifestly alter the asymptotics. 

On the contrary, the effect of $\b_{(0)}$ is similar to the inclusion of a time-dependence of $h_{(0)}$. In fact, this correspondence was already stressed in \cite{Chrusciel} from the analysis of \cite{HT1987}. Furthermore, it is known that any metric in the class
\begin{equation}\label{eq.Ash}
ds^2 = e^{2 \beta}(-du^2 - 2 du dr) + r^2\g_{AB}dx^Adx^B
\end{equation}
with $\b=\b(u,r,x)$ can be conformally compactified in $d=4$ with a smooth $\Omega=r^{-1}$ and admits a smooth $\mathscr{I}$ \cite{Ashtekar1996}. This is another example of spacetime with a null infinity which is not asymptotically flat: its curvature is such that the stress-energy tensor does not satisfy the falloff requirements usually assumed for isolated systems in $d=4$ categorised as asymptotically flat \cite{Ashtekar1996}.  As can be checked from the explicit solutions in sections \ref{Sec.LeadingExp}, \ref{sec.solbWU}, the metric \eqref{eq.Ash} is the solution of the Bondi-Sachs problem with conditions $h_{(0)AB}(u,x)=\gamma_{AB}(x)$, $\mathcal{U}_{(d-3)}=W^A_{(d-1)}=0$ and $\partial_A\b_{(0)}=0$.

The metric \eqref{eq.Ash}, when considered in $d=3$ is the prototypical example of well-defined notions of asymptotic flatness as obtained from symmetry reductions of four dimensional cylindrical waves \cite{Ashtekar1996} and $\b$ is well understood to be related to the ``radiation'' content of the spacetimes \cite{Ashtekar1996}: different contents change the leading form of the metric, whereas in four dimensions the boundary metric is part of the universal structure, according to the standard (not CL) definition.
\\
\\
A latter motivation for keeping $\b_{(0)}$ explicit comes from holography.  This metric function was given a role in the dual CFT \cite{AMK,Compere2020} of AdS gravity in Bondi-Sachs gauge. From a flat limit perspective we would ideally like to keep track of the fate of the degrees of freedom included in $\b_{(0)}$ when taking the limit\footnote{A similar comment can be made for $W_{(0)}^A$, which should be kept when solving equations with $\Lambda\neq 0$, as \eqref{eq.W0} does not apply.}.

\subsection{Asymptotic expansion of $h$}\label{sec.res}
With a generic pair $(h_{(0)AB},\b_{(0)})$ in any $d>4$ we argue that the most general configuration space on a generic $u=const$ surface is built on $h_{AB}$ given by
\begin{equation}\label{eq.exph}
h_{AB}=h_{(0)AB}+\frac{h_{(1)AB}}{r}+\sum_{p\in \mathbb{N}}^{p<\frac{d-4}{2}} \frac{h_{(1+p)AB}}{r^{1+p}}+\frac{h_{(\frac{d-2}{2})AB}}{r^{\frac{d-2}{2}}}+\frac{\log r}{r^{\frac{d-2}{2}}}\texttt{h}_{(\frac{d-2}{2})AB}+\dots
\end{equation}
where the time dependence of each coefficient up to the order $\frac{d-2}{2}$ is determined by the fourth main equation in terms of the boundary data. The ellipsis hide further polyhomogeneous terms with integer and half-integer powers according to $d$. The proof that $p\in \mathbb{N}$ is in subsection \ref{Sec.NR}. It is important to notice that \eqref{eq.exph} is not an assumption, but it is the outcome of the fourth main equation (see also remarks at the end of this subsection). The procedure leading to \eqref{eq.exph} is exemplified in section \ref{Sec.LeadingExp} and all the details are in sections \ref{sec.L} and \ref{sec.partialu}. The behaviour of each of the terms in \eqref{eq.exph} is determined by such analysis and  can be here summarised in the following points 
\begin{itemize}
	\item The time-dependence of $h_{(0)AB}$ is fixed by the fourth main equation to be 
		\begin{equation}\label{eq.uh0}
		l:=\frac{\partial_u q}{2q}:=(d-2)\partial_u\varphi(u,x),\quad h_{(0)AB}(u,x)=e^{2\varphi(u,x)}\hat{h}_{(0)AB}(x),
		\end{equation} 
		Derivation at the beginning of section \ref{sec.partialu}. 
	\item The time dependence of $h_{(1)AB}$ is given by the first order differential equation
		\begin{equation}\label{eq.core0}
			\partial_u h_{(1)AB}-\frac{l}{d-2}h_{(1)AB}=\frac{2}{d-4}\left[e^{2\b_{(0)}}\mathcal{R}_{AB}+(d-4)\mathcal{B}_{AB}[\b_{(0)}]\right]
		\end{equation}
		where $\mathcal{R}_{AB}$ is the traceless part of the Ricci tensor of $h_{(0)AB}$ and $\mathcal{B}_{AB}$ is a traceless object built only on derivatives of $\b_{(0)}$, as specified in section \ref{sec.h1discussion}. There we further discuss the role of this equation in unlocking the issues discussed in the previous parts. The derivation of this equation can be found at the beginning of subsection \ref{Sec.NR}.
	
	\item  The term $h_{(\frac{d-2}{2})AB}$ is the radiative order and its time derivative is expressed in terms of the previous orders and a free function $N_{AB}$. This expression simplifies in four and in odd dimensions since $p$ in the sum of \eqref{eq.core} runs over positive integers. In these cases we have
	\begin{equation}\label{eq.radnews0}
	\partial_u h_{(\frac{d-2}{2})AB}-\frac{2l}{d-2}h_{(\frac{d-2}{2})AB}=N_{AB},\quad d=4 \text{ and } d=odd,
	\end{equation}
	which we take as the definition of $N_{AB}$.
	It is traceless and thus satisfies the minimal requirement for being a news tensor. In even dimensions the free function $N_{AB}$ is traceless and possess contributions from the radiative order and the overleading integer-power terms that couple with the generalised boundary conditions. For example, in $d=6$ 
	\begin{align}\label{eq.N6better0}
	N_{AB}&=\partial_u h_{(2)AB}^{(Tf)}-\frac{l}{2}h_{(2)AB}^{(Tf)}\nonumber\\
	&-h^E_{(1)(A}\mathcal{H}_{(2)B)E}+\frac{l}{4}h^E_{(1)(A}h_{(1)B)E}-\frac{h_{(0)AB}}{4}\left(-\mathcal{H}_{(2)C}^Dh_{(1)D}^C+\frac{l}{4}h_{(1)C}^Dh_{(1)D}^C\right),
	\end{align}
	where $h^{(Tf)}_{(2)AB}$ denotes the trace-free part of $h_{(2)AB}$ and $\mathcal{H}_{(2)AB}$ is minus the term in square brackets in \eqref{eq.core}. In particular, with a time-independent boundary metric and $\beta_{(0)}=0$ these additional terms are associated to the large gauge transformations (i.e. superrotations). See subsection \ref{Sec.NR} for the derivation of \eqref{eq.radnews0} and \eqref{eq.N6better0} and for more discussion on their status as proper news tensors.
	\item The logarithmic term $\texttt{h}_{(\frac{d-2}{2})AB}$ will be referred to as logarithmic term of the third kind (see Remarks below for the nomenclature). This term possess a non-trivial time dependence when $d=even$ both with the generic boundary conditions and the standard Minkowskian conditions (see equations \eqref{eq.nontriviallog0}, \eqref{eq.nontrivallog}); hence it is not natural to set this term to zero. The non-trivial time dependence is due to the presence of terms before the radiative order: with radiative falloff conditions this logarithmic term would be time independent (see subsection \ref{sec.rad}). The term $\texttt{h}_{(\frac{d-2}{2})AB}$ is also time-independent in four dimensions (because there is only one possibility: radiative) and in odd dimensions (because there are no half-integer powers before the radiative order). In these dimensions the leading log can be consistently set to zero by appropriate initial data, while in even dimensions differential constraints are imposed on the leading order terms by setting to zero such logarithmic behaviour. See subsection \ref{Sec.NR} for a derivation of these points and further discussion and subsection \ref{sec.leadinglog} for their potential interpretation in terms of AdS/CFT uplifts. 
	\item Once the leading logarithmic term is included, the time dependence of $h_{(0)AB}$ couples to a $r^{-\frac{d-2}{2}}\log^2 r$ term. By iteration, all powers of $\log r$ appear at the same order and at subleading. This is avoided if $h_{(0)AB}$ is time independent. See section \ref{sec.maxpol} for some details.
\end{itemize}
These results reduce to the four dimensional analysis with time dependent boundary metric \cite{Barnich2010} and the structure of the metric expansion with leading logarithmic term in four dimensions corresponds to \cite{Kroon2001}. After the next two remarks we discuss the structure summarised in this list. 
\begin{rem} \emph{\textbf{[On the assumptions about $h$].}}
\emph{In sections \ref{sec.standard} and \ref{sec.KLPS}, the equations \eqref{eq.rad} and \eqref{eq.nonrad} specify the asymptotic form of $h_{AB}$ for any $u$. The Bondi-Sachs problem is a characteristic initial value problem where $h_{AB}$ is specified on an intial $u=const$ surface, where also $\b$, $W^A$ and $\mathcal{U}$ are determined, and after which $h_{AB}$ is evolved in time by the fourth main equation to reiterate the procedure. Any a priori specification of $h_{AB}$ at \emph{any} $u$ may thus result in constraints for the leading order data (or boundary conditions). In pursuing the aim set in the introduction, one of the points of this paper is to explicate the constraints that a given ansatz for $h_{AB}$ induce, and thus find the most general form of $h_{AB}$ as generated upon time evolution from any specification of $h_{AB}$ at a previous time. Equation \eqref{eq.exph} is such a general form and the equations for each of its coefficients, for example \eqref{eq.core0}, are the outcomes of this general strategy.}

\emph{Particular initial values of $h_{AB}$ under which one may wish to discuss the time evolution for our current purposes are}
\begin{equation}\label{eq.initial}
h_{AB}=h_{(0)AB}+\sum_{p}\frac{h_{(a+p)AB}(u,x)}{r^{a+p}},
\end{equation}
\emph{where $p\in \mathbb{N}_0$ if $d$ is even and $p \in\mathbb{N}_0/2$ if $d$ is odd and $a$ parametrise the leading power: $a=\frac{d-2}{2}$ if we insist on an initial $h_{AB}$ starting from the radiative order, or $a=1$ if we stick to the most general solution of Einstein equations.}

\emph{We will thus see in section   \ref{sec.h1discussion}, as an example of these comments, that the Minkowskian boundary conditions where $h_{(0)AB}$ is the round sphere (or Einstein) and $\b_{(0)}=0$ can be understood as necessarily imposed by the dynamics if the radiation-compatible form of $h$ is assumed a priori at any $u=const$ slice.}
\end{rem}
\begin{rem}\emph{\textbf{[Nomenclature for logarithmic terms].}}
\emph{The logarithmic term appearing in  \eqref{eq.core} is one of three different kinds of logarithmic terms in the expansion of the metric coefficients. We choose to refer to it as the logarithmic term of the third kind.} 

\emph{The first kind stems from the term $r^{-n}h_{(n)AB}$ in the integration of \eqref{scheme1} and \eqref{eq.trace1} when $n$ is the appropriate order. These logarithmic terms affect the fourth main equation and induce logarithmic terms in $h_{AB}$ at order $m>n$. They are logarithmic terms of the second kind. Such logarithmic terms induce new further subleading logarithmic terms in the expansions of $\b$, $W^A$ and $\mathcal{U}$. When we include the logarithmic terms of the first and second kind in the asymptotic expansion we speak of \emph{minimally polyhomogeneous expansions} (see for example \cite{Chrusciel} in $d=4$). The logarithmic term showed in \eqref{eq.core} is of the third kind. This log is intrinsic to the fourth main equation, namely it can be generated upon time evolution even if we start with a non polyhomogeneous $h_{AB}$. The term shown constitute the most leading logarithmic contribution in $h_{AB}$, no further leading log terms are generated by Einstein equations unless they are induced by hands via initial conditions. An expansion with a log of the third kind is called \emph{maximally polyhomogeneous} (see \cite{Kroon2001} for the analogous usage of this name in $d=4$). Polyhomogeneous asymptotics in high dimensions have been analysed in \cite{Chrusciel2010}.}
\end{rem}

\subsubsection{Time-dependence of $h_{(1)AB}$}\label{sec.h1discussion}
As argued in the previous subsections, the analysis of BMS-like asymptotic symmetries to higher dimensions cannot overlook the interplay of boundary conditions on the behaviour of $h_{(1)AB}$. In this section we discuss this point starting from equation \eqref{eq.core0} which we now repeat for ease of presentation. The equation is
	\begin{equation}\label{eq.core}
		\partial_u h_{(1)AB}-\frac{l}{d-2}h_{(1)AB}=\frac{2}{4-d}\left[e^{2\b_{(0)}}\mathcal{R}_{AB}+(d-4)\mathcal{B}_{AB}[\b_{(0)}]\right],
	\end{equation}
	where $\mathcal{R}_{AB}$ is the traceless part of the Ricci tensor of $h_{(0)AB}$, such that
	\begin{equation}
	\mathcal{R}_{AB}=\hat{\mathcal{R}}_{AB}-(d-4)\varPhi_{AB}[\varphi]
	\end{equation}
	with 
	\begin{equation}\label{eq.TfGeroch}
	\varPhi_{AB}[\varphi]=(\hat{D}_A\hat{D}_B\varphi-\hat{D}_A\varphi\hat{D}_B\varphi)-\frac{\hat{h}_{AB}}{d-2}(\hat{D}^2\varphi-(\hat{D}\varphi)^2),
	\end{equation}
	when written in terms of $\hat{h}$ and $\varphi$.	$\mathcal{B}_{AB}$ is traceless and symmetric 
	\begin{equation}
	\mathcal{B}_{AB}=2e^{2\b_{(0)}}\left[\overset{\scriptscriptstyle (0)}{D_A}\partial_B \b_{(0)}+2\partial_A\b_{(0)}\partial_B\b_{(0)}-\frac{h_{(0)AB}}{d-2}\left(\overset{\scriptscriptstyle (0)}{D^2}\b_{(0)}+2\partial_C\b_{(0)}\partial^C\b_{(0)}\right)\right].
	\end{equation}
	In the rest of the paper we use also the definition 
	\begin{equation}\label{eq.defH}
	\mathcal{H}_{(2)AB}:=-e^{2\b_{(0)}}\mathcal{R}_{AB}+(4-d)\mathcal{B}_{AB}[\b_{(0)}].
	\end{equation}
We read equation \eqref{eq.core} in two ways: to analyse how much the imposition of radiative falloff behaviour (i.e. $h_{(1)AB}=0$ as well as all $h_{(1+p)AB}=0$ $\forall p<\frac{d-4}{2}$) constrains the boundary data, and to discuss which boundary conditions are compatible with a non-trivial time-dependence of $h_{(1)AB}$ and, in turn, with consistent actions of BMS-like asymptotic symmetries.
\begin{table}
	\begin{tabular}{ | p{7cm} | p{7cm} | } 
		\multicolumn{2}{|l|}{Radiative falloff $h_{(1)AB}\equiv0\qquad\qquad\qquad\Rightarrow\qquad\qquad\qquad \mathcal{H}_{(2)AB}=0$}\\[5pt]
		\midrule
		$l\sim\partial_u\varphi=0$ &\\[4pt]$a)\quad$ $h_{(0)}=\hat{h}$ Einstein\newline $\left.\qquad(\hat{\mathcal{R}}_{AB}\equiv0, \varPhi_{AB}\equiv0\right.$)  & $\mathcal{B}_{AB}=0\longrightarrow \partial_A\b_{(0)}=0$\\[4pt]
		$b)\quad$ $h_{(0)}$ Conformal to Einstein \newline $\left.\qquad(\hat{\mathcal{R}}_{AB}\equiv0,\,\partial_u \varPhi_{AB}=0\right.$)& $\mathcal{B}_{AB}=\varPhi_{AB}\longrightarrow \varphi(x)=2\b_{(0)}(x)$\\[4pt]
		$c)\quad$ $h_{(0)}$ Non-Einstein & $\mathcal{R}_{AB}\sim  (4-d)\mathcal{B}_{AB}$\\ 
		\midrule
		$l\neq0$  &\\[4pt]$d)\quad$ $h_{(0)}=\hat{h}$ Einstein $(\hat{\mathcal{R}}_{AB}\equiv0$)& $\mathcal{B}_{AB}=\varPhi_{AB}\longrightarrow \varphi(u,x)=2\b_{(0)}(u,x)$\\[4pt]
		$e)\quad$ $h_{(0)}$ Conformal to Einstein \newline $\left.\qquad(\hat{\mathcal{R}}_{AB}\equiv0\right.$)& $\mathcal{B}_{AB}=\varPhi_{AB}\longrightarrow \varphi(u,x)=2\b_{(0)}(u,x)$\\[4pt]
		$f) \quad$ $h_{(0)}$ Non-Einstein & $\mathcal{R}_{AB}\sim  (4-d)\mathcal{B}_{AB}$\\  
		\bottomrule
	\end{tabular}
	\caption{Synoptic view of various constraints imposed on the boundary data $(h_{(0)AB},\b_{(0)})$ by the request $h_{(1)AB}\equiv 0$ in $d>4$. The right column reports the conditions stemming from $\mathcal{H}_{(2)AB}$ under the assumptions in the left column and the simple arrow $\longrightarrow$ indicates a possible solution of the conditions.}
	\label{tab:h10}
\end{table}

It is important to recall that these equations are all related to a foliation in terms of $u=const$ surfaces, and hence equation \eqref{eq.core} is valid at one such surface. 

If we impose $h_{(1)AB}\equiv0$ on any $u=const$ surface, the right-hand-side of \eqref{eq.core} must vanish and hence this implies that the boundary data $(h_{(0)AB},\b_{(0)})$ must satisfy certain constraints which are summarised in table \ref{tab:h10}. Notice for example that imposing that $h_{(0)}$ is Einstein when $h_{(0)}$ depends on time is a much stronger condition than imposing it when $h_{(0)}$ is taken to be time-independent. Robinson-Trautman spacetimes defined in \cite{Podolsky:2006du} corresponds to case $d)$ with $\b_{(0)}=0$ and $\varphi(u,x)$ constrained so that $\Phi_{AB}=0$ at any $u$). Apart from the cases with non-Einstein $h_{(0)}$ and $\mathcal{B}_{AB}\neq 0$, the others reduce to case $a)$ or $d)$ when $\b_{(0)}$ is fixed by gauge choice to zero.
\\
\\
There is no a priori reason for imposing $h_{(1)AB}=0$ on all the $u=const$ surfaces. Equation \eqref{eq.KLPS} is obtained from the general equation \eqref{eq.core} if the right-hand side of this equation is constrained to vanish at any $u$: the standard asymptotically flat boundary conditions, i.e.  $\b_{(0)}=\varphi=0$ and $h_{(0)AB}$ Einstein are such a case. 
\begin{table}
	\begin{tabular}{c|c|c|l}
		\toprule
		\multirow{7}{*}{$\partial_u h_{(1)AB}\neq0$ }&\multirow{3}{*}{$l=0$}&\multirow{3}{*}{$\mathcal{H}_{(2)AB}\neq 0$}& $\hat{\mathcal{R}}_{AB}=0,\, \varphi(x)= 0\Rightarrow \mathcal{B}_{AB}\neq 0\qquad\qquad\,~\refstepcounter{equation}(\theequation)\label{Tuh1_1}$   \\[4pt]&&&$\hat{\mathcal{R}}_{AB}=0, \, \varphi(x)\neq 0\Rightarrow\varPhi_{AB}-\mathcal{B}_{AB}\neq 0\quad\,~\refstepcounter{equation}(\theequation)\label{Tuh1_2}$\\[4pt]&&&$\hat{\mathcal{R}}_{AB}\neq0\qquad\qquad\qquad\qquad\qquad\qquad\qquad~\refstepcounter{equation}(\theequation)\label{Tuh1_3}$\\[4pt]\cline{2-4}
		&\multirow{4}{*}{$l\neq0$}&\multirow{2}{*}{$\mathcal{H}_{(2)AB}= 0$}& $\hat{\mathcal{R}}_{AB}=0\Rightarrow \varPhi_{AB}=\mathcal{B}_{AB}\longrightarrow\varphi=2\b_{(0)}\quad~\refstepcounter{equation}(\theequation)\label{Tuh1_4}$\\[4pt]&&&$\hat{\mathcal{R}}_{AB}\neq0~\qquad\qquad\qquad\qquad\qquad\qquad\qquad\refstepcounter{equation}(\theequation)\label{Tuh1_5}$\\[4pt]\cline{3-4}&&\multirow{2}{*}{$\mathcal{H}_{(2)AB}\neq 0$}& $\hat{\mathcal{R}}_{AB}=0\Rightarrow\varPhi_{AB}-\mathcal{B}_{AB}\neq 0\hspace{62pt}~\refstepcounter{equation}(\theequation)\label{Tuh1_6}$\\&&&$\hat{\mathcal{R}}_{AB}\neq0\qquad\qquad\qquad\qquad\qquad\qquad\qquad\,~\refstepcounter{equation}(\theequation)\label{Tuh1_7}$\\
		\bottomrule
	\end{tabular}
	\caption{Various boundary data compatible with $\partial_u h_{(1)AB}\neq0$. The cases such that $\partial_u h_{(1)AB}=0$ can also be obtained from this under appropriate changes.}
	\label{tab:uh1}
\end{table} 
\\
\\
On the other hand,  $\partial_u h_{(1)AB}\neq 0$ is obtained for example when $(h_{(0)AB},\b_{(0)})$ satisfy the conditions in table \ref{tab:uh1}. From the table we can single out in particular the following cases corresponding to the smallest modifications of boundary conditions compatible with a time-dependent $h_{(1)AB}$:
\\
\\
\emph{In dimensions higher than four, the constraint \eqref{eq.KLPS} can be circumvented with}
	\begin{enumerate}[label=\textbf{C.\arabic*},ref=C.\arabic*]
	\item \emph{\label{clm.nonE} $\b_{(0)}=0$ and time-independent $h_{(0)AB}$ provided that it is not an Einstein metric;}
	\item \emph{\label{clm.timeh} $\b_{(0)}=0$ and Einstein $h_{(0)AB}$ provided that it depends on time;}
	\item \emph{\label{clm.b0} time-independent Einstein $h_{(0)AB}$ provided that $\b_{(0)}\neq 0$ ($\partial_A \b_{(0)}\neq 0$ actually);}
	\end{enumerate}
Mixed cases are in principle allowed.
Notice that when $h_{(0)AB}$ is Einstein and $\b_{(0)}=0$, $h_{(1)AB}$ depends on time through a conformal factor depending on $l$
\begin{equation}\label{eq.acase}
h_{(1)AB}(u,x)=\Omega(l)\hat{h}_{(1)AB}(x).
\end{equation}
A clarification about the statements above is in order. They constitute the ways in which \eqref{eq.KLPS} can be relaxed to the more general forms encoded in \eqref{eq.core}, which is a necessary ingredient to allow for extensions of BMS. They must also be understood under this perspective. Thus, \ref{clm.nonE} does not mean that $h_{(0)AB}$ is a fixed non-Einstein, rather that it cannot be fixed to be Einstein\footnote{Also \ref{clm.timeh}, \ref{clm.b0} are to be understood in the same way.}. In fact, CL-superrotations act as a diffeomorphism and a generic Weyl rescaling of $h_{(0)AB}$, according to \eqref{eq.h0killingcond}. As we further argue below,  \ref{clm.nonE} is the most conservative extension of the four-dimensional phase space to higher dimensions.
\subsubsection{On the limits to `spatial infinity' and the scattering problem}\label{sec.limit}
As highlighted, a time dependent $h_{(1)}$ in higher dimensions is incompatible with standard definitions of the asymptotically flat boundary conditions at null infinity. This necessarily affects any subsequent analysis of the scattering problem, which requires global notions of past and future null infinity and of spatial inifnity $i^0$, and well defined limits from $\mathscr{I}^+$, $\mathscr{I}^{-}$ to $i^0$. 

Even if such limits exist - which we would like to in order to define a scattering problem - we cannot expect that more general notions of null asymptotics, to which we land in seeking for generalisation of BMS, are compatible with the standard definitions of spatial infinity. These assumptions are bound to constrain the configuration space at null infinity, as we can easily argue. 

In all cases where the correspondence between BMS symmetries and scattering processes has been discussed with more details ($d=4$), a pivotal role is played by the Bondi mass aspect/angular momentum aspect evolution equation, which dictates how the asymptotic charges behave in the limits toward $i^0$. In our case, we can write the Bondi mass aspect equation, for example, as
\begin{align}\label{eq.mass}
[\partial_u+(d-1)\partial_u\varphi]m(u,x)=&-\frac{(d-2)\Omega_{(d-2)}}{\k^2}\left(\frac{1}{2(d-2)}(N_{AB})^2+\frac{1}{2}D_AD_BN^{AB}+\frac{1}{d-2}\overset{\scriptscriptstyle (0)}{D^2}\overset{\scriptscriptstyle (0)}{R}\right)\nonumber\\&+(d-1)\partial_u\varphi \mathfrak{M_1}+\mathfrak{M}_2+\mathfrak{b}
\end{align} 
where $\k^2=16\pi$, $\Omega_{d-2}=2\pi^{\frac{d-1}{2}}/\G(\frac{d-1}{2})$ and for notational purposes we split in the second line the contributions that automatically cancel for time-independent $h_{(0)}$ ($\mathfrak{M}_{1}$) from those that depends on terms above/below the radiative order ($\mathfrak{M}_{2}$) and those that vanish when $\b_{(0)}=0$ ($\mathfrak{b}$). The first line is the contribution coming from the radiative falloff conditions\footnote{We have further eliminated the global factors of $e^{2\b_{(0)}}$ as they are irrelevant now.}.  Needless to say, this equation is extremely hard to deal with, even when we restrict to one of the cases \ref{clm.nonE}, \ref{clm.timeh}, \ref{clm.b0}. 
Issues arise not only because of the presence of terms before the radiative orders, nor  because of the $\mathfrak{b}$ terms, but even just because of the time dependence of $h_{(0)}$ that prevents the interpretation of \eqref{eq.mass} as a mass loss equation\footnote{Notice that for Robinson-Trautman spacetimes in higher dimensions (which are defined with $h_{AB}=h_{(0)AB}$ Einstein), equation \eqref{eq.mass} collapse to $LHS=0$ \cite{Podolsky:2006du}.}, as said before. When integrating to get charges, the right-hand side of \eqref{eq.mass} would not in general be split in hard and soft parts as in \cite{Strominger2014} and we would not be able to adapt standard results of asymptotic quantization \cite{Ash2018,He2015}. The analysis is to be performed from scratch.

These difficulties can be traced back to the problem of defining a phase space that admits proper limits to $i^0$ over this configuration space. This can be seen in other ways, without referring to the Bondi mass aspect equation.

In the definition of asymptotic flatness at null infinity in higher even dimensions given in \cite{Hollands2013cva}, in addition to the assumption that $h_{(0)}$ is Einstein and time independent (and $\b_{(0)}=0$), a good limit to spatial infinity is assumed to be given by the condition that for some $u<u_0$ the metric  decays faster than the radiative order (in particular $O(r^{3-d})$) and approach a stationary solution. If we impose that on the initial $u=const$ surface $h_{AB}$ behaves accordingly, we obtain \eqref{eq.core} and in fact $h_{(1)AB}=0$ for any $u$ by the initial condition $h_{(1)AB}|_{u<u_0}=0$. Relaxing  this condition to a $u$-dependent Einstein\footnote{In the blown-up representation of $i^0$ \cite{Ashtekar1978b,Ashtekar1992,Beig1982}, the three-dimensional metric on the hyperboloid is constrained to be Einstein by the field equations \cite{Beig1982}, but its time dependence is not constrained.} $h_{(0)AB}$ with $h_{AB}$ falloff at rates $O(r^{>\frac{2-d}{2}})$ on the initial surface, equation \eqref{eq.acase} will be supplemented by the initial condition that $h_{(1)AB}|_{u=u_0}=0$ and hence again we get $h_{(1)AB}=0$ at any $u$.
\\
\\
We conclude that the assumption that $h_{(0)AB}$ is Einstein and that there is a region corresponding (in some sense) to $i^0$ is incompatible with having $\partial_u h_{(1)AB} \neq 0$ and consequently a consistent action of CL-superrotations on the configuration space.
\\
\\
These speculative conclusions are to be taken \emph{cum grano salis}. The point is similar to the debate around the meaning of BMS-like symmetries at spatial infinity in four spacetime dimensions \cite{Troessaert2017}, which is relevant to formalise the derivation of the antipodal matching conditions between $\mathscr{I}^+$ and $\mathscr{I}^-$ used in the definition of the scattering problem \cite{AshTalk,Prabhu2019,Nguyen2021}. Similarly we would need to reconsider the Being-Schmidt asymptotic analysis of spatial inifinity \cite{Beig1982} in higher dimensions to approach the problems here addressed. 
\subsubsection{Leading logs and holography}\label{sec.leadinglog}
The expansion \eqref{eq.exph} leads to a maximally polyhomogeneous expansion of the metric. As in four spacetime dimension \cite{Kroon2001}, this is a possibility implied by Einstein equations but whose physical interpretation is not clear. On the other hand, minimal polyhomogeneous expansions in four dimensions - i.e. those with the first and second kind of logarithmic terms - are for example understood in the Bondi-Sachs problem as a consequence of  the elimination of what Bondi and collaborators mistakenly assumed to be the Sommerfeld radiation condition \cite{Bondi1962,Chrusciel,Kroon}. Here we suggest a possible interpretation of the logarithmic term in \eqref{eq.exph} in terms of the well-understood logarithmic terms in the Fefferman-Graham expansion of asymptotically locally AdS (AlAdS) spacetimes \cite{SkenderisLec}.
\\
\\
The first logarithmic coefficient in the expansion \eqref{eq.exph} is time-independent in four and odd spacetime dimensions, while its time-dependence is determined by the previous orders of the asymptotic expansion in even dimensions. Suppose that at an initial $u=u_0$ $h_{AB}$ is given by \eqref{eq.initial} with $a=1$ (in particular $\texttt{h}_{(\frac{d-2}{2})AB}|_{u=u_0}=0$) and without the half-integer powers before the radiative order, then
\begin{equation}\label{eq.leadlog}
\texttt{h}_{(\frac{d-2}{2})AB}|_{u>u_0}\begin{cases}
= 0 \quad \text{in } d=4,\; d=odd\\
\neq 0\quad \text{in } d=even
\end{cases}
\end{equation}
With this, the structure of \eqref{eq.exph} is reminiscent of the asymptotic expansion of AlAdS$_{d}$ spacetimes in Fefferman-Graham gauge \cite{Henningson1998,deHaro2001}, which is known to play a central role in relating the quantum properties of conformal field theories and the geometric properties of spacetimes with negative cosmological constant.

The Fefferman-Grham metric is presented in Gaussian normal coordinates that foliate the spacetime with a family of timelike hypersurfaces including the conformal boundary and the metric reads
\begin{equation}
ds^2=\frac{\l^2}{z^2}(dz^2+g_{ij}(z,\c))d\c^id\c^j)
\end{equation}
where $\l$ is the AdS length scale, $\c=(t,x)$ and $g_{ij}$ satisfy an asymptotic expansion in terms of $z$ that contain logarithmic terms when the spacetime dimension is\footnote{We only describe the asymptotic expansion without writing down the explicit expression because it depends on the definition of the radial coordinate.} $d=odd$ \cite{Henningson1998,deHaro2001}. Two free data must be specified to uniquely determine the solution: the leading term, $g_{(0)}$, and (the traceless and  divergenceless part of) the coefficient multiplying the power of $z$ at the same order of the leading log. The metric $g_{(0)}$ represents the background metric of the dual CFT and the other datum is related to the one-point function of the CFT stress-energy tensor. The coefficient of the leading log, $g_{(log)}$, is equal to the metric variation of the holographic conformal anomaly. 
\\
\\
Given that the leading logarithmic term appears in $h_{AB}$ at the same order of the free datum of $h_{AB}$, there is a somewhat clear similarity between the expansion of $h_{AB}$ and that of $g_{ij}$, under a change of dimensions by one unit
\begin{equation}\label{eq.conjecture}
AlAdS_{d-1} \longleftrightarrow AlM_{d}
\end{equation}
This picture is coherent with the original approach of de Boer and Solodukhin based on the slicing of the interior and exterior of Minkowski spacetime in terms of Euclidean AdS and dS hypersurfaces \cite{deBoer2003,Costa2012fm}. This would justify why $h_{AB}$ in four-dimensional asymptotically flat spacetimes does not (necessarily) contain the leading logarithmic term: there are no logarithmic terms in three-dimensional AlAdS \cite{deHaro2001}.

A non-trivial aspect that is to be kept in mind is that, differently from the Fefferman-Graham metric, the Bondi-Sachs metric is not only given by $h_{AB}$, although it is built on it. Thus, despite suggestive, we cannot conclude much from the picture unless a deeper analysis is performed. This could go by pushing back the Bondi metric to a foliation in terms of the hyperboloidal surfaces, which are described asymptotically by the Fefferman-Grham expansion, and then produce a mapping between this and the Bondi coordinates.
\\
\\
Differently, a mapping between the Bondi-Sachs and Fefferman-Graham expansions has been obtained in \cite{AMK} in four-dimensional spacetimes with a negative cosmological constant and the flat limit of the phase space has been discussed in \cite{Compere2020}. This is a mapping of the form
\begin{equation}
\text{Fefferman-Graham } AlAdS_4\; \leftrightarrow\text{Bondi-Sachs } AlAdS_4\xrightarrow{\lambda\rightarrow\infty} \text{Asymptotically Flat}
\end{equation}
Due to the spacetime dimension considered, these papers could not identify the role of logarithmic terms in the Bondi-Sachs expansion as possibly related to anomalies of the dual holographic theory. In a $d$-to-$d$ mapping, the structure of \eqref{eq.exph}, \eqref{eq.leadlog} is at first sight puzzling. The question can be again answered by performing the integration of Einstein equations with the inclusion of the cosmological constant and analysing the fate under the flat limit of the logarithmic term we have been discussing\footnote{With a cosmological constant term only the third and fourth main equations are modified. The coupling to the cosmological constant produces further overleading terms with respect to what we obtain, but the structure of the integrals is not changed.}. 

As a final comment we remind that the precise structure of the asymptotic expansions is coordinate dependent. We should be cautious in giving a physical meaning to the various terms, unless they can be written in terms of gauge invariant quantities and recall that the holographic results are obtained after the procedure of holographic renormalization, where the anomaly term is expressed in terms of conformal invariants \cite{SkenderisLec}. Unless we specify precisely the transformation rules among the different coordinate systems/renormalization schemes we cannot push the analogy discussed here further. This is to be analysed in follow up works \cite{Capone2021}.

\section{Asymptotic analysis: discussion}\label{sec.aadis}
For ease of comparison with existing literature, we adopt conventions similar to \cite{Barnich2010}. We define the quantities (see Appendix \ref{app.Details})
\begin{equation}
l_{AB}=\frac{1}{2}\partial_u g_{AB}\,,\quad k_{AB}=\frac{1}{2}\partial_rg_{AB}\,,\quad n_A=\frac{1}{2}e^{-2\b}g_{AB}\partial_rW^B,
\end{equation}
and\footnote{Notice that \cite{Barnich2010} uses $K^A_B:=r^2\tilde{K}^A_B$.}
\begin{equation}
\tilde{n}_A=\frac{n_{A}}{r^2},\quad \tilde{K}^C_D:=\frac{1}{2}h^{AC}\partial_rh_{AD}, \quad \text{so that}\quad k^A_B=\frac{\delta^A_B}{r}+\tilde{K}^A_B\,.
\end{equation}
Einstein main equations take the form
\begin{equation}\label{eq.Rrr}
R_{rr}=0 \Rightarrow \quad \partial_r \b=\frac{r}{2(d-2)} \tilde{K}^A_B\tilde{K}^B_A\,,
\end{equation}
\begin{align}\label{eq.RrA}
R_{Ar}=0\Rightarrow\quad&\partial_r(r^d\tilde{n}_A)=\mathcal{G}_A(\b,\tilde{K})\, \\
&\mathcal{G}_A(h_{AB},\b)=r^{d-2}\left[\left(\partial_r-\frac{d-2}{r}\right)\partial_A\b-{}^{\scriptscriptstyle(d-2)}D_B\tilde{K}^B_A\right],\nonumber
\end{align}
\begin{align}\label{eq.trace}
g^{AB}R_{AB}=0\Rightarrow \quad& \frac{d-2}{r^2}\left[(d-3)+r\partial_r\right]\mathcal{U}=\mathcal{F}(h_{AB},\b,W^A)\,,\\
&\mathcal{F}(h_{AB},\b,W^A)= e^{2\b}\left[{}^{\scriptscriptstyle(d-2)}R-2({}^{\scriptscriptstyle(d-2)}D_AD^A\b+\partial^A\b\partial_A\b+n^An_A)\right]\nonumber\\&\qquad\qquad\qquad\qquad+\left(\partial_r+2\frac{d-2}{r}\right){}^{\scriptscriptstyle(d-2)}D_AW^A+2\frac{d-2}{r}l\,,\nonumber
\end{align}
\begin{align}\label{eq.traceless} 
g^{DA}R_{AB}=0 \Rightarrow  & \left(\partial_r +\frac{d-2}{r}\right)l^D_B+\left(\partial_u+l\right)k^D_B=\mathcal{H}^D_B\,,\\
\mathcal{H}^D_B&=-e^{2\b}\left[{}^{\scriptscriptstyle(d-2)}R^D_B-2\left({}^{\scriptscriptstyle(d-2)}D^D\partial_B\b+\partial^D\b\partial_B\b+n^Dn_B\right)\right]\nonumber\\
&- \left(\partial_r +\frac{d-2}{r}\right)\left(\frac{1}{2}{}^{\scriptscriptstyle(d-2)}D^DW_B+\frac{1}{2}{}^{\scriptscriptstyle(d-2)}D_BW^D-k^D_B\mathcal{U}\right)\nonumber\\
&-\left[{}^{\scriptscriptstyle(d-2)}D_C(W^Ck^D_B)+k^D_A{}^{\scriptscriptstyle(d-2)}D_BW^A-k^A_B{}^{\scriptscriptstyle(d-2)}D_AW^D\right].\nonumber
\end{align}
This involved set of equations is the main reason why the standard techniques of non-vanishing cosmological constant holography, such as holographic renormalization, does not extend naively to the case of null asymptotics. In AdS/CFT for example, we can organise the bulk Einstein equations in terms of Gauss-Coazzi equations for the bulk evolution of the timelike boundary surface. Einstein equations do not explictly contain terms involving the time derivative because they are hidden in the covariant derivatives along the timelike slices that foliate the spacetime, so that Einstein equations only involve the extrinsic curvature of such surfaces analogous to $k^A_B$ (as well as the induced metric).In the present case, instead, $k_{AB}$ is the (non-normalised) extrinsic curvature of $r=const$ timelike surfaces, but the null time direction is clearly distinguished.  The cuts of $r=const$ surfaces by $u=const$ null surfaces are spacelike and their extrinsic curvature is
\begin{equation}
Q_{AB}=l_{AB}+{}^{\scriptscriptstyle(d-2)}D_{(A}W_{B)}.
\end{equation}
We may write Einstein equations in terms of this derived quantity, but not much insight is gained\footnote{For example we get a shorter form of \eqref{eq.traceless}, but upon performing the manipulations \eqref{eq.kl} to remove $\partial_u k_{AB}$, the equation almost takes the same form as the one we discuss.}. The double null gauge is better suited to highlight the geometric structure of the equations. 
\\
\\
The solution of all the main equations can be given in a closed integral form depending on $\tilde{K}^A_B$ and encoding the asymptotic behaviour of $h_{AB}$.

\subsection{Integral solution of the main equations}\label{Sec.IntSol}
The integral form of the solutions of the main equations can be given following \cite{Barnich2010}. Here we somewhat imprecisely single out the relevant integration functions and leave integrals as indefinite when no chance of confusion arise. The solution of equation \eqref{eq.Rrr} is
\begin{equation} \label{eq.intb}
\b(u,r,x)=\b_{(0)}(u,x)+\frac{1}{2(d-2)}\int   r \tilde{K}^A_B\tilde{K}^B_A. 
\end{equation} 
A first integration of \eqref{eq.RrA} gives
\begin{equation}\label{eq.tilden}
\tilde{n}_{A}(u,r,x)=\frac{N_A(u,x)}{r^d}+\frac{1}{r^d}\int^r \mathcal{G}_A \dif s,
\end{equation}
and, from the definition of $\tilde{n}_A$,
\begin{equation}\label{eq.intW}
W^A(u,r,x)=W^A_{(0)}(u,x)+2\int^r\dif t\, e^{2\b(u,t,x)}h^{AB}(u,t,x)\tilde{n}_{B}(u,t,x).
\end{equation}
The integration of \eqref{eq.trace}
gives
\begin{equation}\label{eq.intU}
\mathcal{U}(u,r,x)=\frac{\mathcal{U}_{(d-3)}(u,x)}{r^{d-3}}+\frac{1}{r^{d-3}}\int^r\frac{\mathcal{F}}{d-2}s^{d-2}\dif s.
\end{equation}
Using
\begin{equation}\label{eq.kl}
\partial_uk^D_B=\partial_rl^D_B-2(l^D_Ak^A_B-k^D_Al^A_B)\,,
\end{equation}
the latter equation \eqref{eq.traceless} can be conveniently rewritten as
\begin{equation}\label{eq.dl}
	\partial_r l^D_B+\mathfrak{o}^{DA}_{CB}l^C_A=j^D_B
\end{equation}
where
\begin{equation}
\mathfrak{o}^{DA}_{CB}=\frac{d-2}{2r}\delta^D_C\delta^A_B-(\delta^D_Ck^A_B-k^D_C\delta^A_B)\,,\quad j^D_B=\frac{1}{2}\left(\mathcal{H}^D_B-lk^D_B\right)\,
\end{equation}
and the solution is given by Lagrange method\footnote{Given
	$
	\dot{y}(x)+f(x)y(x)=g(x),
	$
	the solution is 
	$
	y(x)=e^{-F(x)}(c+\bar{y}(x))
	$,
	where $F(x)$ is an antiderivative of $f(x)$ and $\bar{y}(x)$ is an antiderivative of $g(x)e^{F(x)}$ and $c$ is a constant.} as
\begin{equation}\label{eq.Lagrangesol}
l^D_B=e^{-\theta^{DA}_{CB}}\left(\frac{1}{2}N^C_A+\bar{l}^D_B\right)
\end{equation}
where the factor $1/2$ is chosen to cancel later factors of $2$ in the definition of $N_{AB}$,
with
\begin{equation}\label{eq.Lagrangesolpieces}
\theta^{DA}_{CB}=\int \mathfrak{o}^{DA}_{CB}\,,\quad \bar{l}^D_B=\int j^C_Ae^{\theta^{DA}_{CB}}\,.
\end{equation}
Notice that $\mathfrak{o}^{DA}_{CB}$ contains a term explicitly of order $r^{-1}$, which contributes with logarithmic terms when integrated. The integral defining $\theta^{DA}_{CB}$ is to be considered between a generic $r$ in the bulk (where the coordinate system breaks down) and a large $R$ to be sent to infinity. The potential logarithmic divergence is absorbed in a power of $r$.

Equivalently, to ease comparison with \cite{Barnich2010}, we can split
\begin{equation}\label{eq.def_l_l0_L}
l^A_B=l^A_{(0)B}+\tilde{L}^A_B\,,\quad l^A_{(0)B}=\frac{1}{2}h^{AC}_{(0)}\partial_u h_{(0)CB}\,,
\end{equation} 
and, to remove the explicit $r^{-1}$ piece from the operator acting on $\tilde{L}^A_B$, we can further define
\begin{equation}\label{eq.deftildeL}
\tilde{L}^D_B:=r^\frac{2-d}{2}L^D_B
\end{equation}
so that \eqref{eq.dl} becomes
\begin{equation}\label{eq.tildeL}
	\partial_r {L}^D_B+{\mathcal{O}}^{DA}_{CB}{L}^C_A={J^D_B}
\end{equation}
with
\begin{equation}\label{eq.finalO}
{\mathcal{O}}^{DA}_{CB}=-(\delta^D_C\tilde{K}^A_B-\tilde{K}^D_C\delta^A_B),
\end{equation}
and
\begin{equation}\label{eq.Jdef}
{J^D_B}:=r^{\frac{d-2}{2}}\tilde{J}^D_B:=r^{\frac{d-2}{2}}\left[j^D_B+\left(\delta^D_C\tilde{K}^A_B-\tilde{K}^D_C \delta^A_B\right)l_{(0)A}^C-\frac{(d-2)}{2r}\delta^D_C\delta^A_Bl^C_{(0)A}\right].
\end{equation}
The solution thus reads
\begin{equation}\label{eq.Lagrangesol}
	L^D_B=e^{-\Theta^{DA}_{CB}}\left(\frac{1}{2}N^C_A+\bar{L}^D_B\right)
\end{equation}
with
\begin{equation}\label{eq.Lagrangesolpieces}
\Theta^{DA}_{CB}=\int \mathcal{O}^{DA}_{CB}\,,\quad \bar{L}^D_B=\int J^C_Ae^{\Theta^{DA}_{CB}}\,.
\end{equation}
To reconstruct $l^D_B$ from $L^D_B$ one just uses
\begin{equation}\label{eq.deftildeL2}
l^D_B=r^{\frac{2-d}{2}}L^D_B+l^D_{(0)B}\,.
\end{equation}
\\
\\
Having solved the equations, we can complete the scheme \eqref{scheme0}-\eqref{eq.traceless1} as
\begin{align}
\b(u,r,x)&=\b_{(0)}(u,x)+b(u,r,x),\qquad && b=b[\tilde{K}]\nonumber\\
W^A(u,r,x)&=\frac{\mathcal{W}^A_{(d-1)}(u,x)}{r^{d-1}}+w^A(u,r,x),\qquad &&w^A=w^A[\b_0, \tilde{K}]\nonumber\\
\mathcal{U}(u,r,x)&= \frac{\mathcal{U}_{(d-3)}(u,x)}{r^{d-3}}+\u(u,r,x),\qquad &&\u=\u[\b_0,\tilde{K}, W_{(d-1)},\mathcal{U}_{(d-3)}]
\end{align}
and 
\begin{equation}
l^A_B(u,r,x)=l^A_B[\b_0, \tilde{K}, W_{d-1}, \mathcal{U}_{(d-3)}]
\end{equation}
The notation here indicates that the metric functions on the left depend of the quantities in square brackets on the right hand side, $\tilde{K}$ standing for any combination of $\tilde{K}^A_B$. All the solutions are given in terms of the auxiliary quantity $\tilde{K}^A_B$. Thus, in principle, the boundary conditions should not involve $h_{AB}$ directly but rather $\tilde{K}^A_B$ and $l^A_B$. Clearly the boundary condition
\begin{equation}\label{eq.bch0}
\lim_{r\rightarrow\infty} h_{AB}(u,r,x)=h_{(0)AB}(u,x)
\end{equation}
does correspond to the boundary condition
\begin{equation}
\lim_{r\rightarrow\infty}l^A_B=l^A_{(0)B}
\end{equation}
which allowed us to perform the splitting \eqref{eq.def_l_l0_L}, in terms of which, then, the boundary condition on $L^A_B$ is 
\begin{equation}
\lim_{r\rightarrow \infty}L^A_B=0.
\end{equation} 

These boundary conditions on $h_{AB}$ or equivalently $l^A_B$ imply a boundary condition on $\tilde{K}^A_B$
\begin{equation}
\lim_{r\rightarrow\infty}\tilde{K}^A_B=0
\end{equation}
However, as opposed to $l^A_B$ (or equivalently $L^A_B$), the radial behaviour of $\tilde{K}^A_B$ is not determined by any of the equations. 

We can pick virtually any form of $\tilde{K}^A_B$ as initial data and we are able to consistently solve the equations. Here we can spell the strategy leading to our discussion in section \ref{sec.h1discussion}: given some initial $\tilde{K}^A_B$ each order of the fourth main equation may impsose restrictions on the leading orders, through the conditions on $l^A_B$.
On the other hand, rather than imposing $\tilde{K}^A_B$ from the start, we can partially infer its behaviour from that of $l^A_B$ determined by the fourth main equation so that no restrictions on the leading data are obtained on later null surfaces. The procedure is cyclic. The next subsection exemplifies this point using the simplest initial condition and can be seen as a . 


\subsection{Example: leading behaviour and logarithmic terms}\label{Sec.LeadingExp}
To exemplify we take $\tilde{K}^A_B=0$. Hence $k^A_B$ reduces to $k^A_B=\delta^A_B/r$. We get
\begin{equation}
b=0\,,\quad w^A=\frac{W^A_{(1)}}{r}\,,		\quad\u=r\mathcal{U}_{(-1)}+\mathcal{U}_{(0)},
\end{equation}
where the explicit expression of the coefficients are in \eqref{eq.Wa1} and subsequent. In turn,  $\mathfrak{o}^{DA}_{CB}=\frac{d-2}{2r}\delta^D_C\delta^A_B$ and equation \eqref{eq.dl} reduces to
\begin{equation}
\partial_r l^D_B+\frac{d-2}{2r}\delta^D_C\delta^A_B l^C_A=\frac{j^D_{(1)B}}{r}+\frac{j^D_{(2)B}}{r^2}\,,
\end{equation}
where, stating from section \ref{sec.L} (see also section \ref{sec.h1discussion}),
\begin{eqnarray}\label{eq.first}
j_{(1)B}^D&=&\frac{1}{2}\left((d-2)\delta^D_B\mathcal{U}_{(-1)}-l\delta^D_B\right),\nonumber\\
j_{(2)B}^D&=&\frac{\mathcal{H}^D_{(2)B}}{2},\quad
\mathcal{H}^D_{(2)B}=-e^{2\b_{(0)}}\mathcal{R}_{AB}+(4-d)\mathcal{B}_{AB}[\b_{(0)}].
\end{eqnarray}
with $\mathcal{R}_{AB}$ and $\mathcal{B}_{AB}[\b_{(0)}]$ encountered in section \ref{sec.h1discussion} (see also \eqref{eq.Jsubleading2}). The solution of this equation is
\begin{eqnarray}\label{eq.example}
d > 4: \qquad l^D_B &=&\frac{2}{d-2}j^D_{(1)B}+\frac{2}{d-4}\frac{j^D_{(2)B}}{r}+\frac{N^D_B}{r^\frac{d-2}{2}},\nonumber\\
d=4: \qquad l^D_B &=&j^D_{(1)B}+\frac{1}{r}\left(j_{(2)B}^D\log r+N^D_B\right).
\end{eqnarray}
When we impose the boundary conditions on these solutions we get $l^D_B\equiv l^{D}_{(0)B}\Rightarrow$ \eqref{eq.uh0} and $\tilde{K}^A_B\equiv0\Rightarrow\partial_r l^A_B=0\Rightarrow(N^D_B=0,j_{(2)B}^D=0)$. The latter implies
\begin{equation}\label{eq.constraint}
\mathcal{H}^D_{(2)B}=0,
\end{equation}
which is trivial in $d=4$ and in any $d>4$ results in the constraint among $h_{(0)AB}$ and $\b_{(0)}$ we have discussed in table \ref{tab:h10} of section \ref{sec.h1discussion}. 

The conclusion of the toy solution presented here is the following. The metric
\begin{align}
ds^2=&-\left[e^{2\b_{(0)}}\left(r\mathcal{U}_{(-1)}+\mathcal{U}_{(0)}+\frac{\mathcal{U}_{(d-3)}}{r^{d-3}}\right)+h_{(0)AB}W^A_{(1)}W^B_{(1)}\right]du^2\\\nonumber&-2e^{2\b_{(0)}} dudr+ r^2h_{(0)AB}\left(dx^A+\frac{W_{(1)}^A}{r}du\right)\left(dx^B+\frac{W_{(1)}^B}{r}du\right)
\end{align}
is a solution of the field equations if \eqref{eq.constraint} holds (in particular with $\mathcal{W}^A_{(d-1)}$). Gauge fixing $\b_{(0)}=0$ (which implies $W^A_{(1)}=0$) we get
\begin{equation}\label{eq.RT}
ds^2=-\left(r\mathcal{U}_{(-1)}+\mathcal{U}_{(0)}+\frac{\mathcal{U}_{(d-3)}}{r^{d-3}}\right)du^2 -2 dudr+ r^2h_{(0)AB}dx^Adx^B,
\end{equation}
which is a Robinson-Trautman spacetime. Notice again that in $d>4$, \eqref{eq.constraint} must be satisfied, which implies in this gauge that $h_{(0)}$ is Einstein \cite{Podolsky:2006du}. 
\\
\\
With this example we see how we can use \eqref{eq.example} when relaxing the boundary condition on $\tilde{K}^A_B$ to infer the asymptotic behaviour of $\tilde{K}^A_B\neq0$ on generic null surfaces: if we prescribe that $\tilde{K}^A_B$ behaves asymptotically as the right-hand-side of \eqref{eq.example},  \eqref{eq.example} will be modified by the addition of further orders and repeating this procedure cyclically we get the most general expansion of $\tilde{K}^A_B$.  By virtue of the iterative procedure and the form of the equations, the new terms appearing with subleading powers do not alter the form of the previous terms. Only terms of the form $Ar^{-i}\log{r}$ modify the metric functions at order $r^{-i}$ by the addition of $A$-depending terms and possibly higher powers of $\log{r}$. 

In particular, this exemplifies two features of the asymptotic expansions: the first subleading order in $\tilde{K}^A_B$ is of order $r^{-1}$ both in four and higher dimensions, but in four dimensions we have a logarithmic term (of the third kind, following the nomenclature established in the Remarks of section \ref{sec.res}) at the same order.

We thus confirm what we have already discussed, that the request that the expansion starts from $r^{-\frac{d-2}{2}}$ in $d>4$ places constraints on the leading data, which are given by \eqref{eq.constraint}. 

\section{Asymptotic solutions: power-law seed}\label{sec.ExpDetails}
In this section we give the details of the solution constructed with a non-polyhomogeneous expansion of $h_{AB}$ of the form \eqref{eq.initial}, which we restate here
\begin{equation}\label{eq.exp}
h_{AB}=h_{(0)AB}+\sum_{p}\frac{h_{(a+p)AB}(u,x)}{r^{a+p}},
\end{equation}
where $p\in \mathbb{N}_0$ if $d$ is even and $p \in\mathbb{N}_0/2$ if $d$ is odd and $a$ parametrise the leading power: $a=\frac{d-2}{2}$ if we insist on an initial $h_{AB}$ starting from the radiative order, or $a=1$ if we stick to the most general solution of Einstein equations. Here we will discuss both these cases.

\paragraph{Determinant condition constraints.} With the given ansatz \eqref{eq.exp}, the determinant condition $h^{AB}\partial_r h_{AB}=0$, implies
\begin{equation}
h^{AB}_{(0)}h_{(a+p)AB}=0\quad \forall\, p<a
\end{equation} 
while the trace of the others $h_{(a+p)AB}$ ($p\ge0$)is determined in terms of the previous orders. To exemplify consider
\begin{equation}\label{eq.expex}
h_{AB}(u,r,x)=h_{(0)AB}(u,x)+\frac{h_{(a)AB}}{r^a}+\frac{h_{(a+p_o)AB}(u,x)}{r^{a+p_o}},\quad p_o=\begin{cases}
\frac{1}{2}\quad d\text{ odd}\\
1\quad d\text{ even}
\end{cases}.
\end{equation}
Its inverse is 
\begin{equation}
h^{AB}=h_{(0)}^{AB}-\frac{h_{(a)}^{AB}}{r^a}-\frac{h_{(a+p_o)}^{AB}}{r^{a+p_o}}+\frac{h_{(a)C}^{A}h_{(a)}^{CB}}{r^{2a}},
\end{equation}
where we understand that we have to discard all terms of order greater than $a+p_o$. We retain the order $2a$ because it may be equal to $a+p_o$ according to the cases: 
\begin{itemize}
	\item[R)] $a=\frac{d-2}{2}$, $d\ge 4$: $2a=a+p_o$ iff $d=4$, otherwise $a+p_o<2a$,
	\item[NR)] $a=1$, $d>4$: $a+p_o=2a$ for any $d$ even, $a+p_o<2a$ for any $d$ odd.
\end{itemize}
where $R$ and $NR$ stand for radiative and non-radiative.

The determinant condition thus imply
\begin{equation}
-\frac{a}{r^{a+1}}h_{(0)}^{AB}h_{(a)AB}-\frac{a+p_o}{r^{a+p_o+1}}h_{(0)}^{AB}h_{(a+p_o)AB}+\frac{a}{r^{2a+1}}h_{(a)}^{AB}h_{(a)AB}+\dots =0.
\end{equation}
So we get
\begin{equation}
h^{AB}_{(0)}h_{(a)AB}=0 \quad \forall a,\,d
\end{equation}
and
\begin{align}
d>4& \text{ odd }\;(p_o=\frac{1}{2}),\;\;a=\frac{d-2}{2}\quad \text{ or}&& a=1: \qquad&& h_{(0)}^{AB}h_{(a+p_o)AB}=0	\nonumber\\
d>4& \text{ even }\;(p_o=1),\;\;  a=\frac{d-2}{2}&&\quad\quad\,\;: && h_{(0)}^{AB}h_{(a+p_o)AB}=0	\\
d\ge4& \text{ even }\;(p_o=1),\;\; && a=1:  && h_{(0)}^{AB}h_{(2)AB}=\frac{1}{2}h_{(1)}^{AB}h_{(1)AB}.\nonumber
\end{align}
As we can see from this example, the consequences of the gauge choice in even $d>4$ with $a=1$ is really like $d=4$ and this is why many results can be transferred to such generic dimensions quickly. However, the situation changes with radiative falloff conditions.

Furthermore, when $d=4$ we can also use the relationship valid for any $2\times2$ traceless matrices\footnote{The symmetry of the matrix is not necessary for this condition to hold, but we clearly deal with symmetric matrices.} \cite{Barnich2010}
\begin{equation}\label{eq.2by2property}
M^A_CM^C_B=\frac{1}{2}\delta^A_B M^C_D M^D_C.
\end{equation}
It is easy to see that in higher dimensions this condition does not hold, in particular $M^A_CM^C_B$ is generically diagonal only for diagonal $M^A_C$ but is not proportional to the identity. 
\subsection{Asymptotic expansion of $\b$, $W^A$ and $\mathcal{U}$}\label{sec.solbWU}
With the ansatz \eqref{eq.exp} given $a$ and the boundary condition $W^A_{(0)}=0$ we get,
\begin{align}
\b&=\b_{(0)}+\sum_{p\ge0}\frac{\b_{(2a+p)}}{r^{2a+p}}\nonumber\\
W^A&=\frac{W^A_{(1)}}{r}+\sum_{p=0}^{d-2-a}\frac{W^A_{(a+1+p)}}{r^{a+1+p}}+\frac{1}{r^{d-1}}\left(\mathcal{W}^A_{(d-1)}+\texttt{W}^A_{(d-1)}\log r \right)+\dots\nonumber\\
\mathcal{U}&=r \mathcal{U}_{(-1)}+\mathcal{U}_{(0)}+\sum_{{p=0}}^{a+p<d-3}\frac{\mathcal{U}_{(a+p)}}{r^{a+p}}+\frac{1}{r^{d-3}}\left(\mathcal{U}_{(d-3)}+\texttt{{U}}_{(d-3)}\log r\right)+\dots	\nonumber
\end{align}
The details of the expansions can be found in Appendix \ref{app.rf}. Here we make some comments and list the expressions relevant in the next parts. The expansion is shown up to the order of the free functions $\mathcal{W}^A_{(d-1)}$ and $\mathcal{U}_{(d-3)}$ which are universal (i.e. their functional form does not depend on the spacetime dimension) and are given by
\begin{align}\label{eq.mathcalW}
\mathcal{W}^A_{(d-1)}&=-\frac{2}{d-1} e^{2\b_{(0)}} h_{(0)}^{AC}N_C,\\
\mathcal{U}_{(d-3)}&=-\frac{\kappa^2}{(d-2)\Omega_{d-2}}m,\label{eq.mathcalU}
\end{align}
where we have chosen a normalization for $\mathcal{U}_{(d-3)}$ with $\kappa^2=16\pi$ and $\Omega_{d-2}=2\pi^{\frac{d-1}{2}}/\G(\frac{d-1}{2})$. The functions $m$ and $N^A$ are free and correspond to the mass and angular momentum aspects in asymptotically Minkowskian spacetimes with radiative asymptotics\footnote{Notice that the angular momentum in higher dimensional spacetimes with the more general Einstein boundary conditions of \cite{Hollands2013cva} has not been studied.}. 

The leading terms of the expansion are universal in the sense that they take the same form for both values of $a$
\begin{equation}
\b_{(2a)}=-\frac{a}{16(d-2)}h_{(a)AB}h_{(a)}^{AB},\qquad W_{(1)}^A=2 e^{2\b_{(0)}}h_{(0)}^{AB}\partial_{B}\b_{(0)},
\end{equation}
\begin{equation}\label{eq.Wa1}
W^A_{(a+1)}=\frac{e^{2\b_{(0)}}}{a+1}\left(\frac{a}{a+2-d}h_{(0)}^{AC}\overset{\scriptscriptstyle (0)}{D}_{B}h_{(a)C}^B-e^{-2\b_{(0)}}h_{(a)}^{AB}W_{(1)B}\right)
\end{equation}
\begin{equation}\label{eq.U1}
\mathcal{U}_{(-1)}=\frac{2}{d-2}l
\end{equation}
\begin{align}\label{eq.U0}
\mathcal{U}_{(0)}&=\frac{e^{2\b_{(0)}}}{(d-3)(d-2)}\left(\overset{\scriptscriptstyle (0)}{R}+2(d-3)e^{-2\b_{(0)}}\overset{\scriptscriptstyle (0)}{D}_A W^A_{(1)}\right),\\
&\quad \overset{\scriptscriptstyle (0)}{D}_A W^A_{(1)}=2e^{2\b_{(0)}}\left(\overset{\scriptscriptstyle (0)}{D^2}\b_{(0)}+2\partial^A\b_{(0)}\partial_A\b_{(0)}\right).\nonumber
\end{align}
Except $d=4$ ($a=1$) we also get\footnote{This is given with $\overset{\scriptscriptstyle (a)}{\G^{A}}_{AC}=0$ following from the determinant condition.}
\begin{align}\label{eq.Ua}
\mathcal{U}_{(a)}&=\frac{1}{(d-2)(d-3-a)}\left[e^{2\b_{(0)}}(\delta R)_{(a)}+(2d-5-a)\overset{\scriptscriptstyle (0)}{D}_A W^A_{(a+1)}+\tilde{\mathcal{F}}_{(a+2)}[W_{(1)}]\right]\nonumber\\
\tilde{\mathcal{F}}_{(a+2)}&=h_{(a)}^{AB}\overset{\scriptscriptstyle (0)}{D}_AW_{(1)B}-h_{(0)}^{AB}\overset{\scriptscriptstyle (a)}{\Gamma^{C}}_{AB}W_{(1)C}-2e^{2\b_{(0)}}(\tilde{n}^A_{(2)}\tilde{n}_{(2+a)A}+\tilde{n}_{(2)A}\tilde{n}_{(2+a)}^A)
\end{align}
where the term in square bracket is denoted as $\mathcal{F}_{(a+2)}$ with the notation of App. \ref{app.thirdeq} and $\tilde{\mathcal{F}}_{(a+2)}[W_{(1)}]$ is the part automatically vanishing when $W^A_{(1)}$ vanishes\footnote{We have used the identity $\overset{\scriptscriptstyle (0)}{D}_AW_{(1)B}\equiv 2e^{2\b_{(0)}}\left(\overset{\scriptscriptstyle (0)}{D}_A\partial_B\b_{(0)}+2\partial_A\b_{(0)}\partial_B\b_{(0)}\right)$.} because $\tilde{n}_{(2)}^A\sim W^A_{(1)}$ from App. \ref{app.secondeq} and where
\begin{align}
\overset{\scriptscriptstyle (a)}{R}_{AB}&=\frac{1}{2}\left(\overset{\scriptscriptstyle (0)}{D}_C\overset{\scriptscriptstyle (0)}{D}_Ah_{(a)B}^C+\overset{\scriptscriptstyle (0)}{D}_C\overset{\scriptscriptstyle (0)}{D}_Bh_{(a)A}^C-\overset{\scriptscriptstyle (0)}{D^2}h_{(a)AB}\right),\label{eq.PertRicci}\\
(\delta R)_{(a)}&=h_{(0)}^{AB}\overset{\scriptscriptstyle (a)}{R}_{AB}-h^{AB}_{(a)}\overset{\scriptscriptstyle (0)}{R}_{AB}=\overset{\scriptscriptstyle (0)}D_A \overset{\scriptscriptstyle (0)}D_Bh_{(a)}^{AB}-h^{AB}_{(a)}\overset{\scriptscriptstyle (0)}{R}_{AB},\label{eq.Raexpanded}
\end{align}
obtained upon using the determinant condition that imposes the vanishing of $h_{(a)C}^C$.

In $d=4$ ($a=1$) the term in square brackets of \eqref{eq.Ua} constitutes the coefficient of the logarithmic term $\texttt{U}_{(1)}$.  
 By substituting either $a=1$ or $a=\frac{d-2}{2}$ the terms of $\mathcal{F}_{(a+2)}$ that do not depend on $W^A_{(1)}$ are
 \begin{align}
a&=1&& \mathcal{F}_{(3)}\sim e^{2\b_{(0)}}\left[(\delta R)_{(1)}-\overset{\scriptscriptstyle (0)}{D}_A\overset{\scriptscriptstyle (0)}{D}_B h_{(1)}^{BA}\right] \\
a&=\frac{d-2}{2}&&\mathcal{F}_{(\frac{d+2}{2})}\sim e^{2\b_{(0)}}\left[(\delta R)_{(\frac{d-2}{2})}-\frac{8-3d}{d}\overset{\scriptscriptstyle (0)}{D}_A\overset{\scriptscriptstyle (0)}{D}_B h_{(\frac{d-2}{2})}^{BA}\right]
 \end{align}
With $W_{(1)}^A=0$, $\tilde{F}_{(a+2)}[W_{(1)}]=0$ and $\mathcal{F}_{(a+2)}$ vanishes only when $a=1$ provided that the Ricci tensor of $h_{(0)AB}$ is proportional to $h_{(0)AB}$ because of \eqref{eq.Raexpanded}. Hence with the standard conditions ($W_{(1)}^A=0$) but $a=1$ in any dimension we get
\begin{align}
d=4\; (W_{(1)}^A=0):&\quad \mathcal{F}_{(3)}\sim\texttt{U}_{(1)}\equiv 0\nonumber\\
d>4\; (W_{(1)}^A=0):&\quad \mathcal{F}_{(3)}\sim\mathcal{U}_{(1)}=0 \quad \text{iff}\quad h_{(0)} \text{ Einstein}
\end{align}
On the other hand, with radiative falloff behaviour in higher dimensions $a=\frac{d-2}{2}$, $\mathcal{F}_{(a+2)}\neq 0\sim \mathcal{U}_{(a)}\neq 0$ even with a Minkowskian boundary, as known.
\\
\\
Further subleading orders depend on $a$ and $p$, for example
\begin{equation}
\b_{(2a+p)}=-\frac{a(a+p)}{4(d-2)(2a+p)}h_{(a)AB}h^{AB}_{(a+p)}.
\end{equation}
\begin{equation}
\b_{(3a)}=\frac{a}{12(d-2)}h^A_{(a)B}h_{(a)}^{BC}h_{(a)CA}
\end{equation}
and the next order in $W^A$ splits into (we are expressing everything in terms of $\b_{(0)}$ rather than $W^A_{(1)}$ here)
\begin{equation}\label{eq.W11}
W^A_{(a+p_o+1)}=\frac{e^{2\b_{(0)}}}{a+p_o+1}\left(\frac{a+p_o}{(a+p_o)+2-d}h_{(0)}^{AC}\overset{\scriptscriptstyle (0)}{D}_Bh^B_{(a+p_o)C}-2h^{AC}_{(a+p_o)}\partial_C\b_{(0)}\right),
\end{equation}
\begin{eqnarray}\label{eq.W12}
W^A_{(2a+1)}&=&\frac{e^{2\b_{(0)}}}{2a+1}\left[\frac{2h_{(0)}^{AC}}{d-2-2a}\left((d-3+2a)\partial_C\b_{(2a)}+\frac{a}{2}h_{(a)}^{BD}D_Bh_{(a)DC}\right)\right.\nonumber\\
&&\qquad\qquad+\left. \frac{a}{d-2-a} h_{(0)}^{AC}D_{B}h^B_{(0)C}-4\b_{(2a)}h_{(0)}^{AC}\partial_C\b_{(0)}\right],
\end{eqnarray}
according to the spacetime dimension. The above 
terms are valid if the denominators do not vanish, otherwise they contribute to the logarithmic term in $W^A$.

They first appear at order $r^{1-d}$ in the expansion of $W^A$ and the coefficient of this leading logarithmic term  is given for any $a$ by (see Appendix \ref{app.secondeq})\footnote{Again, by the determinant condition $\overset{\scriptsize{(a+p)}}{\Gamma^{A}}_{AD}=0$ for any $p$. }
\begin{align}
\texttt{W}^A_{(d-1)}&=-\frac{2}{d-1}e^{2\b_{(0)}}h_{(0)}^{AB}\tilde{n}_{(d)_B}\\
\tilde{n}_{(d)B}&=-\left[\overset{\scriptsize (0)}{D}_C\tilde{K}^C_{(d-1)B}+2(d-2)\partial_B\b_{(d-2)}-\sum_{p+m=d-2-2a}\overset{\scriptsize{(a+p)}}{\G^D}_{BA}K^A_{(m)D}\right]\nonumber
\end{align}
With the non-radiative (NR) falloff ($d>4$) $a=1$ we can organise this (as any other coefficient of the metric expansion) as
\begin{equation}
\tilde{n}_{(d)A}=\tilde{n}^{(R)}_{(d)A}+\tilde{n}^{(NR)}_{(d)A}
\end{equation}
with $\tilde{n}^{(NR)}_{(d)A}$ vanishing when restricting to the radiative behaviour of $h$ (its expansion starting from $a=\frac{d-2}{2}$). In four dimensions, clearly there is no distinction between $\tilde{n}_{(4)A}$ and $\tilde{n}_{(4)A}^{(R)}$. The functional dependence of $\tilde{n}_{(d)A}^{(R)}$ on the orders of $h$ is as follows
\begin{equation}
\tilde{n}_{(d)A}^{(R)}=\tilde{n}_{(d)A}^{(R)}[h_{(d-2)},h_{(\frac{d-2}{2})}].
\end{equation}
In particular, in any number of dimension $d\ge4$ we get for $\tilde{n}_{(d)A}$ the form
\begin{equation}
\tilde{n}_{(d)A}=\frac{d-2}{2}\overset{\scriptstyle (0)}{D}_Ch_{(d-2)A}^{(tf)C}+\text{radiative}+\text{non-radiative}
\end{equation}
where $h^{(tf)}_{(d-2)AB}$ denotes the trace-free part of $h_{(d-2)AB}$, whose trace depends on the previous orders by the determinant condition. 

In four dimensions this simplifies and only the first term survives, so that one gets $\texttt{W}_{(3)}^A=-\frac{2}{3}e^{2\b_{(0)}} \overset{\scriptstyle (0)}{D}_Ch_{(2)}^{(tf)AC}$ as in \cite{Barnich2010}, the exponential factor being a trivial effect of allowing a generic $\b_{(0)}$.  As an example of the difference with higher dimensions, we can compare the explicit expressions of $\tilde{n}_{(d)A}^{(R)}$ in the cases $d=4,5,6$
\begin{equation}\label{eq.nR4}
\tilde{n}^{(R)}_{(4)A} \;(\equiv \tilde{n}_{(4)A})=\overset{\scriptstyle (0)}{D}_Ch_{(2)A}^{C}-\frac{1}{2}\overset{\scriptstyle {(0)}}{D}_B(h^{BC}_{(1)}h_{(1)CA})=\overset{\scriptstyle (0)}{D}_Ch_{(2)A}^{(tf)C}
\end{equation}
\begin{align}
\tilde{n}^{(R)}_{(5)A}&=\frac{3}{2}\left[\overset{\scriptstyle (0)}{D}_Ch_{(3)A}^{C}-\frac{1}{2}\overset{\scriptstyle {(0)}}{D}_B(h^{BC}_{(\frac{3}{2})}h_{(\frac{3}{2})CA})\right]\nonumber\\&=\frac{3}{2}\left[\overset{\scriptstyle (0)}{D}_Ch_{(3)A}^{(tf)C}+\frac{1}{6}\overset{\scriptstyle (0)}{D}_A(h^{CD}_{(\frac{3}{2})}h_{(\frac{3}{2})CD})-\frac{1}{2}\overset{\scriptstyle {(0)}}{D}_B(h^{BC}_{(\frac{3}{2})}h_{(\frac{3}{2})CA})\right],
\end{align}
\begin{align}
\tilde{n}_{(6)A}^{(R)}&=2\left[\overset{\scriptstyle (0)}{D}_Ch_{(4)A}^{C}-\frac{1}{2}\overset{\scriptstyle {(0)}}{D}_B(h^{BC}_{(2)}h_{(2)CA})\right]\nonumber\\
&=2\left[\overset{\scriptstyle (0)}{D}_Ch_{(4)A}^{(tf)C}+\frac{1}{8}\overset{\scriptstyle (0)}{D}_A(h^{CD}_{(2)}h_{(2)CD})-\frac{1}{2}\overset{\scriptstyle {(0)}}{D}_B(h^{BC}_{(2)}h_{(2)CA})\right]
\end{align}
The last equality in \eqref{eq.nR4} follows from using the property \eqref{eq.2by2property} of two-dimensional symmetric, traceless matrices applied to $h_{(1)}$. This property does not hold for higher dimensional symmetric and traceless matrices and can be applied neither to $h_{(\frac{3}{2})AB}$ nor to $h_{(2)AB}$ (with non radiative falloff the latter also fails to be traceless), hence the difference in the final form of $\tilde{n}_{(4)A}^{(R)}$ from the higher dimensional case\footnote{From the trace of $h_{(3)AB}$ and $h_{(2)AB}$ we have removed the contributions depending on $h_{(1)}$, as they are absorbed in $\tilde{n}^{(NR)}_{(d)A}$. As a side comment, referring to the notation in Appendix \ref{app.secondeq}, the only term that contributes to $n_{(d)A}^R$ after all the manipulations is seen to be $\tilde{n}^{\star}_{(2+a+p)A}$.}.

The vanishing of the logarithmic coefficient $\texttt{W}_{(d-1)}$ in $d=4$ constrains $h_{(d-2)AB}^{(tf)}$ to be covariantly constant with respect to the covariant derivative of $h_{(0)AB}$. As observed in \cite{Sachs1962a}, and stressed in \cite{Barnich2010} in relation to BT-superrotations, this condition induces a physical singularity at the north pole of the celestial sphere unless the stronger condition $h_{(2)AB}^{(Tf)}=0$ is imposed\footnote{Bondi's original paper takes a diagonal $h_{AB}$ to start with and the condition we are discussing holds automatically by setting the order $r^{-2}$ to zero. But notice that with a non-diagonal $h_{AB}$ the constraint is only on the trace-free part of $h_{(2)AB}$ because the trace is made up by $h_{(1)AB}$ and remains free.}. In higher dimensions the constraint imposed by the vanishing of this logarithmic term is more involved. The covariant derivative of $h_{(d-2)AB}^{(tf)}$ is sourced by the radiative (as well as non-radiative in the more general case) orders. The presence or absence of this logarithmic term does not affect the Bondi mass because it is subleading, but would affect the angular momentum order\footnote{For example, the condition $h_{(d-2)AB}^{(tf)}=0$ in $d>4$, as in Bondi-Sachs original works, is rather unnatural and does not suffice in removing this logarithmic term. It is unnatural because it does not preserve the freedom of the radiative order of $h_{AB}$ if also the vanishing of the logarithmic term is imposed. It would force a condition similar to \eqref{eq.2by2property} to hold for higher dimensional matrices. It is not hard to find generic examples which fail to satisfy the condition.}.  
\\
\\
Turning to the logarithmic terms $\mathtt{U}_{(d-3)}$ in $d>4$, according to the notation in Appendix \ref{app.rf} we need the coefficient $\mathcal{F}_{(d-1)}$, which is of the form
\begin{equation}
\mathcal{F}_{(d-1)}= e^{2\b_{(0)}}(\delta R)_{(d-3)}+(d-2)\overset{\scriptscriptstyle (0)}{D}_AW^A_{(d-2)}+\tilde{F}_{(d-1)}[W_{(1)}]+\text{above radiative}
\end{equation}
and consistently reduce to $\mathcal{F}_{(3)}$ in $d=4$ where there are no terms above the radiative order.
\\
\\
Unless the logarithmic terms in $\mathcal{U}$ and $W^A$ are set to zero by constraining the metric expansion of $h_{AB}$, they induce respectively two logarithmic terms at order $r^{-d/2}$ and $r^{-(d+2)/2}$ in $h_{AB}$ from the fourth main equation (see next section) and such logarithmic terms propagate down in the asymptotic expansion of all the metric functions, so that the minimally polyhomogeneous expansion is obtained. 

\subsection{Fourth equation: $L^A_B$}\label{sec.L}
In order to discuss the asymptotic expansion of the general solution of the fourth main equation
\begin{equation}
L^D_B= e^{-\Theta^{DA}_{CB}}\left(\frac{1}{2}N^C_A+\bar{L}^C_A\right),
\end{equation}
it is useful to collect the expansions of the intermediate quantities $\Theta^{DA}_{CB}$
\begin{eqnarray}
\Theta^{DA}_{CB}=\int {\mathcal{O}}^{DA}_{CB}dr=-\int(\delta^D_C\tilde{K}^A_B-\tilde{K}^D_C\delta^A_B)dr=:\frac{1}{r^a}\sum_p \frac{\Theta^{DA}_{(a+p)CB}}{r^p},
\end{eqnarray}
and\footnote{We have cancelled a term using the leading solution \eqref{sol.eq_gen} $l_{(0)B}^A\propto \delta^A_B$, which we are going to derive next. } $J^D_{B}$
\begin{equation}\label{eq.Jsecondtimewritten}
J^D_B=r^{\frac{d-2}{2}}\left[\frac{1}{2}\left(\mathcal{H}^D_B-lk^D_B\right)+\cancel{\left(\delta^D_C\tilde{K}^A_B-\tilde{K}^D_C \delta^A_B\right)l_{(0)A}^C}-\frac{1}{r}\frac{(d-2)}{2}\delta^D_C\delta^A_Bl^C_{(0)A}\right]
\end{equation}
\begin{equation}\label{eq.expH}
\mathcal{H}^D_B=\frac{\mathcal{H}^D_{(1)B}}{r}+\frac{\mathcal{H}^D_{(2)B}}{r^2}+\sum_{p=0}\frac{\mathcal{H}^D_{(a+1+p)B}}{r^{a+1+p}}+\frac{\log r}{r^{d-1}}\texttt{H}_{(d-1)}[\texttt{U}_{(d-3)}]+\frac{\log r}{r^d}\texttt{H}_{(d)}[\texttt{W}_{(d-1)}]+\dots
\end{equation}
\begin{equation}
lk^D_B=\frac{l\delta^D_B}{r}+\frac{1}{r^{a+1}}\sum_{p=0}\frac{lK^D_{(p)B}}{r^{p}}.
\end{equation}
Notice that no logarithms are generated in $\Theta$ and that the first order at which the logarithmic term appears in $\mathcal{H}^D_B$ comes from the first logarithmic term in $\mathcal{U}$ (which can be automatically vanishing without serious restrictions) and the second logarithmic term is induced from the first logarithmic term in $W^A$. The expansion of $J^D_B$ is organised as
\begin{align}\label{eq-expansionJ}
J^D_B=&r^{\frac{d-4}{2}}J^D_{(-\frac{d-4}{2})B}+r^{\frac{d-6}{2}}\frac{\mathcal{H}^D_{(2)B}}{2}+\sum_{p=0}r^{\frac{d-4-2(a+p)}{2}}J^D_{(-\frac{d-4-2(a+p)}{2})B}\\
&+\frac{\log r}{r^{\frac{d}{2}}}\frac{\texttt{H}_{(d-1)}[\texttt{U}_{(d-3)}]}{2}+\frac{\log r}{r^{\frac{d+2}{2}}}\texttt{H}_{(d)}[\texttt{W}_{(d-1)}]+\dots\nonumber
\end{align}
where 
\begin{align}\label{eq.Jleading}
&J^D_{(-\frac{d-4}{2})B}=\frac{1}{2}\left(\mathcal{H}^D_{(1)B}-l\delta^D_B-{(d-2)}l^D_{(0)B}\right),\qquad \mathcal{H}^D_{(1)B}=(d-2)\delta^D_B\mathcal{U}_{(-1)}\\
\label{eq.Jsubleading}
&\mathcal{H}^D_{(2)B}=(d-3)\delta^D_B\mathcal{U}_{(0)}-e^{2\b_{(0)}}\overset{\scriptscriptstyle (0)}{R^D}_B+\frac{4-d}{2}\left(\overset{\scriptscriptstyle (0)}{D_B}W_{(1)}^D+\overset{\scriptscriptstyle (0)}{D^D}W_{(1)B}\right)-\overset{\scriptscriptstyle (0)}{D_A}W_{(1)}^A\delta^D_B\\
&\overset{\scriptscriptstyle (0)}{D^D}W_{(1)B}=2e^{2\b_{(0)}}\left(\overset{\scriptscriptstyle (0)}{D^D}\partial_B\b_{(0)}+2\partial^D\b_{(0)}\partial_B\b_{(0)}\right)\label{eq.Bb0}\\
&J^D_{(-\frac{d-4-2(a+p)}{2})B}=\frac{1}{2}\left(\mathcal{H}_{(a+1+p)B}^D-l{K}^D_{(p)B}\right)+\cancel{\left(l^D_{(0)A}K^A_{(p)B}-K^D_{(p)C}l^C_{(0)B}\right)}\label{eq.Jsubsub}
\end{align}
Substituting \eqref{eq.U0}  in \eqref{eq.Jsubleading} and lowering indices and using that $W^A_{(1)}$ is a gradient vector, we get
\begin{equation}\label{eq.Jsubleading2}
\mathcal{H}_{(2)AB}=-e^{2\b_{(0)}}\mathcal{R}_{AB}+(4-d)\mathcal{B}_{AB}[W_{(1)}]
\end{equation}
where
\begin{equation}
\mathcal{R}_{AB}=\overset{\scriptscriptstyle (0)}{R}_{AB}-\frac{h_{(0)AB}}{d-2}\overset{\scriptscriptstyle (0)}{R}
\end{equation}
\begin{equation}
\mathcal{B}_{AB}[W_{(1)}]=\frac{1}{2}\left(\overset{\scriptscriptstyle (0)}{D_A}W_{(1)B}+\overset{\scriptscriptstyle (0)}{D_B}W_{(1)A}\right)-\frac{h_{(0)AB}}{d-2}\overset{\scriptscriptstyle (0)}{D_C}W_{(1)}^C
\end{equation}
are both traceless. $\mathcal{B}_{AB}[W_{(1)}]$ can equivalently be written as
\begin{equation}
\mathcal{B}_{AB}[\b_{(0)}]=2e^{2\b_{(0)}}\left[\overset{\scriptscriptstyle (0)}{D_A}\partial_B \b_{(0)}+2\partial_A\b_{(0)}\partial_B\b_{(0)}-\frac{h_{(0)AB}}{d-2}\left(\overset{\scriptscriptstyle (0)}{D^2}\b_{(0)}+2\partial_C\b_{(0)}\partial^C\b_{(0)}\right)\right]
\end{equation}
The first term of \eqref{eq.Jsubsub} is
\begin{equation}\label{eq.Jsubsubleading}
J^D_{(-\frac{d-4-2a}{2})B}=\frac{1}{2}\left(\mathcal{H}_{(a+1)B}^D-l{K}^D_{(0)B}\right)+\cancel{\left(l^D_{(0)A}K^A_{(0)B}-K^D_{(0)C}l^C_{(0)B}\right)}
\end{equation}
\begin{equation}
\mathcal{H}^D_{(a+1)B}=(d-2-a){K}^D_{(0)B}\mathcal{U}_{(-1)}.
\end{equation}
So that
\begin{equation}\label{eq.Jsubsubleading_sol}
J^D_{(-\frac{d-4-2a}{2})B}=\frac{2(d-2-a)-(d-2)}{2(d-2)}l{K}^D_{(0)B}=  \begin{cases}
0 & \text{if } a=\frac{d-2}{2}\\
\frac{d-4}{2(d-2)}l{K}^D_{(0)B} & \text{if } a=1,\, d>4
\end{cases}.
\end{equation}
It can be convenient to expand explicitly the integral $\int J^D_B=:I^D_B $, which enters the definition of $\bar{L}^D_B$ as $\bar{L}^D_B=I+\dots$, where the dots represents terms depending on the various powers of $\Theta$. 
\begin{equation}
a=\frac{d-2}{2}:\quad I^D_B=\int r^{\frac{d-4}{2}}J^D_{(-\frac{d-4}{2})B}+r^{\frac{d-6}{2}}\frac{\mathcal{H}^D_{(2)B}}{2}+\sum_{p=0}\frac{J^D_{(1+p)B}}{r^{1+p}}+\frac{\log r}{r^{\frac{d}{2}}}\frac{\texttt{H}^D_{(d-1)B}}{2}+\dots
\end{equation}
\begin{equation}
a=1:\quad I^D_B=\int  r^{\frac{d-4}{2}}J^D_{(-\frac{d-4}{2})B}+r^{\frac{d-6}{2}}\frac{\mathcal{H}^D_{(2)B}}{2}+\sum_{p=0}r^{\frac{d-6}{2}-p}J^D_{(-\frac{d-6}{2}+p)B}+\frac{\log r}{r^{\frac{d}{2}}}\frac{\texttt{H}^D_{(d-1)B}}{2}+\dots
\end{equation}
So that 
\begin{align}
d>4:\quad I^D_B=r^\frac{d-2}{2}I^D_{(-\frac{d-2}{2})B}&+r^\frac{d-4}{2}\textcolor{red}{I^D_{(-\frac{d-4}{2})B}}+\textcolor{blue}{\texttt{I}^D_{[d>4]B}}\log{r}\\
&+ \begin{cases}
-\sum_{p>0}\frac{1}{r^p}\frac{J^D_B}{p} & \text{if } a=\frac{d-2}{2}\\
\sum_{p>0,\neq \frac{d-4}{2}}r^{\frac{d-4}{2}-p}\frac{2J^D_B}{d-4-2p} & \text{if } a=1
\end{cases} \nonumber\\
&-\frac{2}{(d-2)^2r^{\frac{d-2}{2}}}(2+(d-2)\log r)\texttt{H}^D_B+\dots \nonumber
\end{align}
\begin{equation}\label{eq.d=4}
d=4:\quad I^D_B=rI^D_{(-1)B}+\textcolor{byzantine}{\texttt{I}^D_{[d=4]B}}\log r-\sum_{p>0}\frac{1}{r^p}\frac{J^D_B}{p}-\frac{(1+\log r)}{r^{\frac{d-2}{2}}}\texttt{H}^D_B+\dots.
\end{equation}
For brevity we have only included in $J$ the logarithmic term corresponding to $\texttt{U}_{(d-3)}$; the logarithmic term corresponding to $\texttt{W}^A_{(d-1)}$ appears at order $r^{-\frac{d}{2}}$ upon integration. In any case we are not going to need these terms in the next part as we limit our considerations to the terms up to $r^0$. 
In the above we have
\begin{align}
d\ge 4:\qquad&  I^D_{(-\frac{d-2}{2})B}&&:=\frac{2}{d-2}J^D_{(-\frac{d-4}{2})B},\label{eq.coefficient1}\\
&  \textcolor{red}{I^D_{(-\frac{d-4}{2})B}}&&:=\begin{cases}
d>4:\quad\frac{1}{d-4}\mathcal{H}^D_{(2)B} &\text{if } a=\frac{d-2}{2}\\
d\ge4:\quad\frac{2}{d-4}\underbrace{\left(\frac{\mathcal{H}^D_{(2)B}}{2}+J^D_{(-\frac{d-6}{2})B}\right)}_{J^{D(tot)}_{(-\frac{d-6}{2})B}}=\frac{2}{d-4}\textcolor{byzantine}{\texttt{I}^D_{[d=4]B}}  &\text{if } a=1 \end{cases}\label{eq.coefficient2} \\
d>4:\qquad & \textcolor{blue}{\texttt{I}^D_{[d>4]B}}&&:=\begin{cases}
J^D_{(-\frac{d-4-2(a+p)}{2})B}|_{p=0} &\text{if } a=\frac{d-2}{2}\\
J^D_{(-\frac{d-4-2(a+p)}{2})B}|_{p=\frac{d-4}{2}}& \text{if } a=1 \end{cases}\label{eq.coefficient3}
\end{align}
Clearly $\textcolor{blue}{\texttt{I}^D_{[d>4]B}}$ and $\textcolor{byzantine}{\texttt{I}^D_{[d=4]B}}$ are just the coefficient $J_{(1)B}^D$ in \eqref{eq-expansionJ} for the given value of $a$ and the given dimension $d$. The colour assigned to the logarithmic coefficients is to signal that $\textcolor{byzantine}{\texttt{I}^D_{[d=4]B}}$ is the sum of the red and blue coefficients of the expansion with $a=(d-2)/2$ in $d=4$ (stripping off the numerical factor stemming from the integral).

The general solution of the fourth main equation
\begin{equation}
L^D_B= e^{-\Theta^{DA}_{CB}}\left(\frac{1}{2}N^C_A+\bar{L}^C_A\right)
\end{equation}
can thus be expanded as
\begin{equation}\label{eq.Lsol1}
L^D_B=r^\frac{d-2}{2}L^D_{(-\frac{d-2}{2})B}+\sum_{p=0}^{(a+p)\neq(d-2)/2}r^{\frac{d-2-2(a+p)}{2}}L^D_{(-\frac{d-2-2(a+p)}{2})B}+\texttt{L}^D_{[d\ge4]B}\log{r}+L_{(0)B}^D+\dots,
\end{equation}
For any $a$
\begin{align}
&L^D_{(-\frac{d-2}{2})B}=I^D_{(-\frac{d-2}{2})B}\\
&L_{(0)B}^D=\frac{N^D_B}{2}+\sum_{k>0}^{\frac{d-2}{2}}([e^{-\Theta}]_{(k)} [\bar{L}]_{(-k)})^D_B\label{eq.L0}
\end{align}
notice the form of $L_{(0)}$: $\bar{L}$ is always a sum of powers greater or less than $0$ and logs, so $L_{(0)}$ is given by the free function $N$ and the appropriate combinations of orders of $e^{-\Theta}$, the sums stops when $k$ gives the most leading term in $\bar{L}$, which is obtained for $k=\frac{d-2}{2}$. We anticipate from the leading solution \eqref{eq.gen} that $\bar{L}_{(-\frac{d-2}{2})}$ is always zero on shell. The next-to-leading order is
\begin{align}
d>4:\qquad
& L^D_{(-\frac{d-4}{2})B}&&:=\begin{cases}
\textcolor{red}{I^D_{(-\frac{d-4}{2})B}} &\text{if } a=\frac{d-2}{2} \\
\textcolor{red}{I^D_{(-\frac{d-4}{2})B}}+c(\Theta_{(1)}L_{(-\frac{d-2}{2})})^D_B  &\text{if } a=1 \label{eq.Lnext>4}\end{cases} \\
d=4:\qquad & \texttt{L}^D_{[d=4]B}&&=\textcolor{byzantine}{\texttt{I}^D_{(d=4)B}}+(J_{(0)}\Theta_{(1)})^D_B \label{eq.Llog4}
\end{align}
where $c$ is a numerical factor. The logarithmic coefficient in $d>4$ is instead
\begin{equation}\label{eq.Llog>4}
\texttt{L}^D_{[d>4]B}=\textcolor{blue}{\texttt{I}^D_{[d>4]B}}+\sum_{k>0,l\neq 1}([e^\Theta]_{(k)} J_{(l)})|^D_{(k+l=1)B},
\end{equation}
where $k>0$ and $l\neq 1$ because the term corresponding to $k=0$ and $l=1$ has already been included in $I$. In $d=5,6$ the logarithmic term appears immediately after the coefficient  $L^D_{(-\frac{d-4}{2})B}$, while in $d=4$ the $r^0\log r$ term is the next-to-leading term. The next term after $r^0\log r$ is, in any dimension, $r^0$ where the free function $N^D_B$ appears: the radiative order. The above expressions are thus all we need in $d=4,5,6$ to reach the radiative order. In dimensions higher than six, the sum in \eqref{eq.Lsol1} produces further terms between $L^D_{(-\frac{d-4}{2})B}$ and $r^0\log{r}$. 


\subsection{Fourth equation: $\partial_u h_{AB}$}\label{sec.partialu}
In order to translate the results for $L_{B}^D$ in terms of $l_{D}^B$ and $\partial_u h_{AB}$, the expression of $L^D_B$ obtained from integration \eqref{eq.Lsol1} - here noted as ${}^{\text{solution}}L^D_B$ - is to be equated to the defining expression of $L^D_B$ (\eqref{eq.deftildeL} and \eqref{eq.deftildeL2}), here noted as ${}^{\text{definition}}L^D_B$
\begin{equation}\label{eq.scheme}
{}^{\text{solution}}L^D_B={}^{\text{definition}}L^D_B.
\end{equation}
More explicitly ${}^{\text{definition}}L^D_B$ is given by
\begin{align}\label{eq.Ldef_expanded}
{}^{\text{definition}}L^D_B=r^{\frac{d-2}{2}}\tilde{L}^D_B&={r^{\frac{d-2}{2}}}\left(l^D_B-l^D_{(0)B}\right)\nonumber\\&=r^{\frac{d-2}{2}-a}l_{(a)B}^D+\sum_p r^{\frac{d-2}{2}-a-p}l_{(a+p)B}^D\nonumber\\&
=\sum_{p=0}r^{\frac{d-2-2a-2p}{2}}\frac{1}{2}\left(
h^{DE}_{(0)}\partial_u h_{(a+p)EB}-h^{DE}_{(a+p)}\partial_u h_{(0)EB}\right)\nonumber\\
&+\text{non-linear terms},
\end{align}
In Section \ref{Sec.NR} the notation
\begin{equation}
\bar{l}_{(a+p)B}^D:=\frac{1}{2}\left(
h^{DE}_{(0)}\partial_u h_{(a+p)EB}-h^{DE}_{(a+p)}\partial_u h_{(0)EB}\right).
\end{equation}
will be used. The non-linearities at each order $(a+p)$ (for appropriate $a$ and $p$) come from the full inverse of $h_{AB}$, while $\bar{l}_{(a+p)B}^D$ is only defined with respect to $h_{(a+p)AB}$ with raised indices. 

From the leading order of \eqref{eq.scheme} we get, in any $d\ge 4$ and for any $a$
\begin{equation}\label{eq.gen}
{L}^D_{(-\frac{d-2}{2})B}=0\Rightarrow {I}^D_{(-\frac{d-2}{2})B}=0.
\end{equation}
With \eqref{eq.Jleading} and \eqref{eq.U1} it gives
\begin{equation}\label{sol.eq_gen}
	l^D_{(0)B}=\frac{l\delta^D_B}{d-2}\Leftrightarrow\partial_u h_{(0)AB}=\frac{2l}{d-2}h_{(0)AB}\,,
\end{equation}
namely
\begin{equation}\label{eq.h0u}
h_{(0)AB}(u,x)=e^{2\varphi(u,x)}\hat{h}_{(0)AB}(x)\,,\quad \partial_u\varphi=\frac{l}{d-2}=\frac{\partial_uq}{(d-2)2q}.
\end{equation}
We have proved \eqref{eq.uh0}, consistently with \cite{Barnich2010,Podolsky:2006du}.

As a consequence of \eqref{sol.eq_gen} we have
\begin{equation}
\left(l^D_{(0)A}K^A_{(p)B}-K^D_{(p)C}l^C_{(0)B}\right)=0
\end{equation}
so that the barred terms in the expressions of the previous section (i.e. \eqref{eq.Jsubsub}) are justified.

With the given $l$, $\mathcal{U}_{(1)}$ can also be expressed as
\begin{equation}
\mathcal{U}_{(-1)}=2\partial_u\varphi\,.
\end{equation} 
We now discuss the subleading solutions considering separately the radiative case $a=\frac{d-2}{2}$ and the non-radiative case $a=1$ in $d>4$. In  $d=4$ there is no distinction between the two cases and we conveniently include this case in the radiative section.
\subsubsection{Radiative falloff $a=\frac{d-2}{2}$}\label{sec.rad}
With radiative falloff conditions, the expansion of $L^D_B$ up to order $r^0$ in $d>4$ automatically collapses to the sum of the leading term, the next-to-leading, the $r^0\log r$ term and the term of order $r^0$. In $d=4$ we have the leading term, the $r^0\log r$ term and $r^0$.

Upon using the leading solution \eqref{eq.gen}, we get in $d>4$ the following equations from \eqref{eq.scheme}
\begin{equation}\label{eq.d>4_red}
L^D_{(-\frac{d-4}{2})B}=0\Rightarrow\textcolor{red}{I^D_{(-\frac{d-4}{2})B}}=0,
\end{equation}
\begin{equation}\label{eq.J1rad}
\texttt{L}^D_{[d>4]B}=0\Rightarrow\textcolor{blue}{\texttt{I}^D_{[d>4]B}}=0,
\end{equation}
\begin{equation}\label{eq.d>4_N}
N^D_B=h^{DE}_{(0)}\partial_u {h}_{(\frac{d-2}{2})EB}-{h}^{DE}_{(\frac{d-2}{2})}\partial_u h_{(0)EB},
\end{equation}
while, in $d=4$
\begin{equation}\label{eq.d=4_purple}
\texttt{L}^D_{[d=4]B}=0\Rightarrow\textcolor{byzantine}{\texttt{I}_{[d=4]B}^D}=0,
\end{equation}
\begin{equation}\label{eq.d=4_N}
N^D_B=h^{DE}_{(0)}\partial_u h_{(1)EB}-h^{DE}_{(1)}\partial_u h_{(0)EB}.
\end{equation}
The coefficients of the logarithmic terms in ${}^{\text{solution}}L^D_B$ are equated to zero because, by our original assumption,  ${}^{\text{definition}}L^D_B$ contains only powers of $r$. Notice, however, that \eqref{eq.coefficient3} with \eqref{eq.Jsubsubleading_sol} implies that \eqref{eq.J1rad} is trivially satisfied
\begin{equation}\label{eq.J1rad_trivial}
\textcolor{blue}{\texttt{I}^D_{[d>4]B}}\equiv0.
\end{equation}
This, also implies, referring to our colour convention \eqref{eq.coefficient2}, that equation \eqref{eq.d=4_purple} reduces \eqref{eq.d>4_red} to
\begin{equation}
\textcolor{byzantine}{\texttt{I}_{[d=4]B}^D}\sim \mathcal{H}^D_{(2)B}=0
\end{equation}
The same condition ${H}^D_{(2)B}=0$ is seen to be imposed by \eqref{eq.d>4_red} because of \eqref{eq.coefficient2}. 

At order $r^0$, \eqref{eq.d>4_N} and \eqref{eq.d=4_N} have the same structure with $d\ge 4$ and, using \eqref{sol.eq_gen}, we get
\begin{equation}\label{eq.radnews}
	N_{AB}=\partial_u h_{(\frac{d-2}{2})AB}-\frac{2l}{d-2}h_{(\frac{d-2}{2})AB},
\end{equation}
which generalises the definition of the '\emph{news tensor}' for $u$-dependent  $h_{(0)AB}$ to any $d$ (cfr. \cite{Barnich2010,Tanabe2011}). We use the quotation marks because the notion of a news tensor (as the object that carries information on gravity waves and mass loss) with a time dependent boundary metric has never been formalised.
\\
\\
To recap, with radiative falloffs - imposing that no logarithmic terms are generated by the fourth main equation - the constraints are solved by \eqref{sol.eq_gen} and $\mathcal{H}^D_{(2)B}=0$. The news tensor take the usual linear form in $h_{(\frac{d-2}{2})AB}$.

In four dimensions, as already discussed, $\mathcal{H}^D_{(2)B}\equiv 0$ trivially, and this implies that the $r^0\log r$ term is not generated by the integration. In higher dimensions, the $r^0\log r$ is again trivially not generated, but the condition $\mathcal{H}^D_{(2)B}\equiv 0$ is to be imposed on $h_{(0)AB}$ and $\b_{(0)}$ to ensure that no overleading powers with respect to the radiative order are generated. The discussion of the implications of $\mathcal{H}^D_{(2)B}=0$ in $d>4$ was presented in sections \ref{sec.h1discussion} and \ref{Sec.LeadingExp}.

\subsubsection{Non-radiative falloff $a=1$ in $d>4$}\label{Sec.NR}
The leading solution \eqref{eq.gen} still holds. With this, the next-to-leading order of \eqref{eq.scheme} produces \eqref{eq.core}. We have
\begin{equation}\label{eq.firstsubnr}
L^D_{(-\frac{d-4}{2})B}=\frac{1}{2}\left(h^{DE}_{(0)}\partial_u h_{(1)EB}-h^{DE}_{(1)}\partial_u h_{(0)EB}\right)
\end{equation} 
where  $L^D_{(-\frac{d-4}{2})B}=\textcolor{red}{I^D_{(-\frac{d-4}{2})B}}$ because the term $(\Theta_{(1)}L_{(-\frac{d-2}{2})})^D_B$ in the second line of \eqref{eq.Lnext>4} vanishes by the leading solution and at this order, $l^D_{(1)B}=\bar{l}^D_{(1)B}$. Using  \eqref{eq.coefficient2} and \eqref{eq.Jsubsubleading_sol}, \eqref{eq.firstsubnr} is 
\begin{equation}
\frac{2}{d-4}\left(\frac{\mathcal{H}^D_{(2)B}}{2}+\frac{d-4}{2(d-2)}l{K}^D_{(0)B}\right)=\frac{1}{2}\left(h^{DE}_{(0)}\partial_u h_{(1)EB}-h^{DE}_{(1)}\partial_u h_{(0)EB}\right),
\end{equation}
Using  \eqref{eq.gen}, we then get equation \eqref{eq.core}:
\begin{equation}\label{eq.partial_u}
\partial_u h_{(1)AB}-\frac{l}{d-2}h_{(1)AB}=\frac{2}{d-4}\mathcal{H}_{(2)AB}.
\end{equation}
It can be equivalently written in terms of $K^D_{(0)B}$ as
\begin{equation}
\partial_u K_{(0)AB}-\frac{l}{d-2}K_{(0)AB}=-\frac{1}{d-4}\mathcal{H}_{(2)B}^D.
\end{equation}
The formal solution of this equation is
\begin{equation}\label{eq.duh1_2}
h_{(1)AB}(u,x)=e^{\varphi(u,x)}\hat{h}_{(1)AB}(x)+e^{\varphi(u,x)}\int e^{-\varphi(u',x)}\hat{\mathcal{H}}_{(2)AB}du'\,,
\end{equation}
where we have absorbed the factor $2/(d-4)$ into $\hat{\mathcal{H}}_{(2)AB}$ and we have used \eqref{eq.h0u}.
\\
\\
With this, we have completed the analysis of the first two equations of the cascade
\begin{align}
L^D_{(-\frac{d-2}{2})B}&=0\label{eq.1}\\
L^D_{(-\frac{d-4}{2})B}&=l_{(1)B}^D\label{eq.2}\\
\vdots \nonumber
\end{align}
stemming from \eqref{eq.scheme} ${}^{\text{solution}}L^D_B={}^{\text{definition}}L^D_B$. The subleading orders proceeds in steps of one unit in even dimensions and half unit in odd dimensions, so that other few are
\begin{align}
L^D_{(-\frac{d-5}{2})B}&=l_{(\frac{3}{2})B}^D\label{eq.3}\\
L^D_{(-\frac{d-6}{2})B}&=l_{(2)B}^D\label{eq.4}\\
L^D_{(-\frac{d-7}{2})B}&=l_{(\frac{5}{2})B}^D\label{eq.5}\\
\vdots\nonumber\\
L^D_{(0)B}&=l^D_{(\frac{d-2}{2})B}\label{eq.0}\\
\texttt{L}^D_{[d>4]B}&=0\label{eq.log}
\end{align}
where the odd figures $5,7,\dots$ only appear if the dimension is odd. It is understood that the equations appear iteratively up to when $L_{(-\frac{d-n}{2})}=L_{(0)}$, which is the radiative order. The equation for the logarithmic coefficient also appear at this order. Thus in $d=5$ and $d=6$, \eqref{eq.1} and \eqref{eq.2} are the only equations above the radiative order, as well as \eqref{eq.log}. In $d=7$, for example, we need to discuss also \eqref{eq.3} and \eqref{eq.4} before the radiative and the log order. As said, here we do not consider equations which are more subleading than radiative in the asymptotic expansion.

We now turn to the discussion of the radiative \eqref{eq.0} and the logarithmic order \eqref{eq.log}. For $d=5$ and $6$ the analysis is quick and exemplifies the case of higher odd and even dimensions respectively. 

\paragraph{$\boldsymbol{d=5}$ and odd.}
The radiative order in five dimensions is $r^{-\frac{3}{2}}$ and hence $l_{(\frac{3}{2})}$ in \eqref{eq.0} is simply $\bar{l}_{(\frac{3}{2})}$ with no non-linear contribution. The tensor $N_{AB}$ satisfies the same expression as the radiative news tensor discussed above
\begin{equation}
d=5:\qquad N_{AB}=\partial_u h_{(\frac{3}{2})AB}-\frac{2l}{3}h_{(\frac{3}{2})AB}.
\end{equation}
This is the same equation as in the radiative case \eqref{eq.radnews}.
\\
\\
The equation for the vanishing of the logarithmic coefficient \eqref{eq.log} is automatically satisfied. Recall that $\texttt{L}_{[d>4]B}^D$ is given by \eqref{eq.Llog>4}. Referring to that equation, in five dimensions only $\textcolor{blue}{\texttt{I}^D_{[d=5]B}}$ contributes. This is given by \eqref{eq.coefficient3} and \eqref{eq.Jsubsub}, which in five dimensions reads
\begin{equation}
J^D_{(1)B}=\frac{1}{2}\left(\mathcal{H}^D_{(\frac{5}{2})B}-l\tilde{K}_{(\frac{5}{2})}\right)=\frac{1}{2}\left(\mathcal{H}^D_{(\frac{5}{2})B}+\frac{3l}{4}h^D_{(\frac{3}{4})B}\right)\equiv 0
\end{equation}
because  $\mathcal{H}^D_{(\frac{5}{2})B}=l\tilde{K}^D_{(\frac{5}{2})B}$.
\\
\\
The behaviour of the radiative order and the vanishing of this logarithmic term can be shown to be valid in any odd dimensions provided that only integer orders appear before the radiative order, as written in \eqref{eq.exph}:
\begin{equation}\label{eq.oddexp}
h_{AB}=h_{(0)AB}+\sum_{p\in \mathbb{N}}^{<\frac{d-2}{2}}\frac{h_{(1)AB}}{r}+\frac{h_{(\frac{d-2}{2})AB}}{r^{\frac{d-2}{2}}}+\dots
\end{equation}
The proof in any dimension proceeds by induction, but the explicit $d=7$ case should suffice here to see the iterative structure of the terms. Consider $L_{(0)B}^D$ from \eqref{eq.L0} and $\texttt{L}^D_{[d>4]B}$ from \eqref{eq.Llog>4} and expand with both integers and half-integers powers starting from $r^{-1}$  (we avoid indices for brevity)
\begin{align}
L_{(0)}&=\frac{N}{2}-\Theta_{(1)}\bar{L}_{(-1)}-\Theta_{(\frac{3}{2})}\bar{L}_{(-\frac{3}{2})}-\Theta_{(2)}\bar{L}_{(-2)}-\Theta_{(\frac{5}{2})}\cancel{\bar{L}_{-(\frac{5}{2})}}\\
\texttt{L}^D_{[d=7]B}&=J_{(1)}+J_{(0)}\Theta_{(1)}+J_{(-\frac{1}{2})}\Theta_{(\frac{3}{2})}+J_{(-1)}\Theta_{(2)}+\cancel{J_{(\frac{3}{2})}}\Theta_{(\frac{5}{2})}
\end{align}
The barred terms are those which are automatically zero on-shell by the leading solutions. All the $\Theta$s that multiply other terms which are not vanishing for the previous reason and have half-integer labels cancel when $p\in \mathbb{N}$ in \eqref{eq.oddexp}. Furthermore from the formulas of the previous sections, $\bar{L}_{(-2)}$ and $J_{(-1)}$ do not exist in $d=7$. We are thus left with 
\begin{align}
L_{(0)}&=\frac{N}{2}-\Theta_{(1)}\bar{L}_{(-1)}-\Theta_{(\frac{3}{2})}\bar{L}_{(-\frac{3}{2})}\label{eq.L0odddim}\\
\texttt{L}^D_{[d=7]B}&=J_{(1)}+J_{(0)}\Theta_{(1)}+J_{(-\frac{1}{2})}\Theta_{(\frac{3}{2})}
\end{align}
The term $\Theta_{(\frac{3}{2})}$ is zero if $h_{(\frac{3}{2})AB}=0$. Notice also that $\bar{L}_{(-1)}\propto I_{(-1)}\propto J_{(0)}$. $J_{(1)}$ is obtained for $p=\frac{3}{2}$, while $J_{(0)}$ is obtained when $p=\frac{1}{2}$. Both these terms can be seen to be zero if $h_{(\frac{3}{2})AB}=0$, thus completing the proof. The same approach can be used to show that these terms are not trivial in even dimensions greater than four, as the six dimensional example discussed next shows.
\\
\\
We consider the expansion \eqref{eq.oddexp} \eqref{eq.exph} more natural than the one with both integers and half integers before the radiative order because it is implied by Einstein equation. As we have seen, Einstein field equations always implies a $r^{-1}$ falloff and that the free function appears at order $r^{-\frac{d-2}{2}}$. Unless half-integer powers are included by hand in the initial $h_{AB}$, no half integer powers before the radiative order are induced by the integration.

\paragraph{$\boldsymbol{d=6}$ and even.}
In $d=6$ the left hand side of \eqref{eq.0} is given by \eqref{eq.L0} with $k=1$ and the equation reads as
\begin{equation}
\frac{N^D_{B}}{2}-[\Theta_{(1)}\bar{L}_{(-1)}]^{D}_B=\bar{l}^D_{(2)B}-\frac{1}{2}h_{(1)}^{DE}\left(\partial_u-\frac{l}{2}\right)h_{(1)EB}
\end{equation}
Notice that the second term in each side of the equations is not independent from \eqref{eq.2}, which has already been analysed. In particular
\begin{equation}
-\frac{1}{2}h_{(1)}^{DE}\left(\partial_u-\frac{l}{2}\right)h_{(1)EB}=-\frac{1}{2}h_{(1)}^{DE}\mathcal{H}_{(2)EB}+\frac{l}{8}h_{(1)}^{DE}h_{(1)EB}=K^{DE}_{(0)}\mathcal{H}_{(2)EB}+\frac{l}{2}K^{DE}_{(0)}K_{(0)EB}.
\end{equation}
On the other hand, since $\bar{L}^D_{(-1)B}=\textcolor{red}{I^D_{(-\frac{d-4}{2})B}}|_{d=6}=\frac{2}{d-4}|_{d=6}J^{D(tot)}_{(0)B}$ we have
\begin{equation}\label{eq.TJ0d6}
[\Theta_{(1)}\bar{L}_{(-1)}]^{D}_B=\frac{1}{2}\left(\mathcal{H}^D_{(2)A}K^A_{(0)B}-\mathcal{H}^C_{(2)B}K^D_{(0)C}\right).
\end{equation} 
Hence we have
\begin{equation}\label{eq.N6}
\frac{N^D_{B}}{2}=\bar{l}_{(2)B}^D+\frac{1}{2}\left(K_{(0)}^{DE}\mathcal{H}_{(2)EB}+\mathcal{H}^D_{(2)A}K^A_{(0)B}+lK^{DE}_{(0)}K_{(0)EB}\right)
\end{equation}
The trace of the equation is\footnote{Independently of the computation, this is consistent because $N^A_B$ is obtained as the integration function of the fourth main equation (4.6), which is the trace-free part of Einstein equations on the transverse directions. Writing it as (4.19), its trace is $\partial_r L = J$ where $J=0$ at each order using the solutions of the previous three main equations. Thus the trace of $L$ can only be in a $r$-independent term, which is in contrast with the definition of $L^A_B$ (4.17) and (4.18). Furthermore $L=0$ can be explicitly checked using its definition and as a consequence of the Bondi-gauge determinant condition that fixes the traces of $h_{(n)AB}$ $n\ge 2$.}
\begin{equation}\label{eq.traceofN}
N=0
\end{equation}
We can write (1.1) in a manifestly trace-free form as
\begin{align}\label{eq.N6better}
N_{AB}&=\partial_u h_{(2)AB}^{(Tf)}-\frac{l}{2}h_{(2)AB}^{(Tf)}\nonumber\\
&+2K^E_{(0)(A}\mathcal{H}_{(2)B)E}+lK^E_{(0)(A}K_{(0)B)E}-\frac{h_{(0)AB}}{2}\left(\mathcal{H}_{(2)C}^DK_{(0)D}^C+\frac{l}{2}K_{(0)C}^DK_{(0)D}^C\right),
\end{align}
where $h^{(Tf)}_{(2)AB}$ denotes the trace-free part of $h_{(2)AB}$ and to match with \eqref{eq.N6better0} recall that $K^A_{(0)B}=-\frac{1}{2}h_{(1)B}^A$. The first line in (1.3) represents the definition of news tensor within a radiative expansion of $h_{AB}$ and time-dependent boundary metric. The second line captures the corrections induced by the generalised boundary conditions and their large gauge transformations coupled to the overleading terms of the asymptotic expansion.

Calling the ``news tensor'' in this non radiative situation as $N_{AB}^{(NR)}$, and the news tensor in the radiative case as $N_{AB}^{(R)}$, we get the following structure
\begin{equation}
N_{AB}^{(NR)}=N_{AB}^{(R)}+\text{non-radiative terms depending on leading orders}
\end{equation}
such non-radiative terms may lead to non-linearities in higher dimensions, but this is not evident from the $d=6$ example. The identification of $N_{AB}^{(NR)}$ with a news tensor is obscure because of the coupling of the  overleading terms with the boundary conditions/large gauge transformations\footnote{Suppose that we insist on imposing the vanishing of the non-linear terms so that $N_{AB}^{(NR)}$ reduces to a radiative news tensor despite the non-radiative expansion. Take $d=6$ as an example, we must impose that the term in parenthesis in \eqref{eq.N6} vanishes. If $\b_{(0)}=0$ and $h_{(0)AB}$ is Einstein, then only $[lK^{DE}_{(0)}K_{(0)EB}]^{(Tf)}$ remains to be equated to zero. In general, however, we see that with $\b_{(0)}\neq 0$ or a non-Einstein $h_{(0)AB}$, even if time independent, the constraint is more involved.}. The relevant degrees of freedom (in the hard sector) should correspond to gauge-independent versions of this object.
\\
\\
Notice from \eqref{eq.L0odddim} that this discussion (and the following) applies also in odd dimensions greater than five, as long as the metric $h_{AB}$ is expanded in integers and half-integer powers starting from $r^{-1}$, and in particular $h_{(\frac{3}{2})AB}\neq 0$: it is this term that couples with $\mathcal{H}_{(2)AB}$ in such dimensions. Otherwise, the situation is somewhat more similar to the four-dimensional case, as shown below \eqref{eq.L0odddim}, and the only thing that is left to do to define a news tensor is to build a gauge-invariant object from  $N_{AB}$ and an appropriate definition of higher dimensional Geroch tensor. 
\\
\\
Recall that in four dimensions a gauge-invariant definition of the news tensor is reached via the Geroch tensor \cite{Geroch1977} (and in particular its trace-free part \cite{Campiglia2020qvc}) that is added to the Bondi frame news tensor. Note that $\Phi_{AB}$ in \eqref{eq.TfGeroch} is formally the same as the trace-free part of the Geroch tensor\footnote{It corresponds to $T_{AB}$ in \cite{Campiglia2020qvc} and $N_{AB}^{\text{vac}}$ in \cite{Compere2018}, taking into account the different conventions on the conformal factor.}, although its manifestation in the higher dimensional context via equation \eqref{eq.core0} is different from the four-dimensional case where \eqref{eq.core0} does not apply: namely, in higher dimensions $\Phi_{AB}$ appears in a dynamical equation above the radiative order while in four dimensions it is to be added by hand to build a gauge-invariant radiative news. The formal similarity is exact when \eqref{eq.core0} is integrated assuming the conditions in \ref{clm.nonE} with the further restriction that $h_{(0)AB}$ is conformal to a sphere (this case is compatible with the comments made on limits to $i^0$ at the end of section \ref{sec.limit}). In such a case $h_{(1)AB}$ is linear in $u$ and its part linear in $u$ is pure gauge. Given the definition of $\mathcal{H}_{(2)AB}$ (see \eqref{eq.core} and \eqref{eq.defH}), is evident that $\Phi_{AB}$ enters $N_{AB}^{(NR)}$ in even dimensions. Notice, however, that one probably would like to call ``higher dimensional Geroch tensor'' the object that is to be added to $N_{AB}^{(NR)}$ to define a gauge-invariant object, not $\Phi_{AB}$ itself.
\\
\\
Equation \eqref{eq.log} in $d=6$ is
\begin{equation}\label{eq.nontriviallog0}
J_{(1)B}^D+[\Theta_{(1)}J^{(tot)}_{(0)}]_{B}^D=0,
\end{equation}
where $[\Theta_{(1)}J^{(tot)}_{(0)}]_{B}^D$ is exactly given by \eqref{eq.TJ0d6} because the numerical factor differentiating $J_{(0)}$ and $L_{(-1)}$ is $1$ in $d=6$ and $J_{(1)B}^D$ is given by \eqref{eq.coefficient3} and \eqref{eq.Jsubsub}
\begin{equation}
J_{(1)B}^D=\frac{1}{2}\left(\mathcal{H}_{(3)B}^D-l\tilde{K}^D_{(3)B}\right).
\end{equation}
Differently from $d=5$, this is not generically zero. For the sake of clarity let us consider the case $\b_{(0)}=0$. This does not affect the main conclusion because we can always split a part that vanishes if $\b_{(0)}=0$ from the part which does not:
\begin{equation}
\mathcal{H}_{(3)B}^D=e^{2\b_{(0)}}\bar{\mathcal{H}}_{(3)B}^D\left[\overset{{\scriptscriptstyle(0)}}{R},h_{(1)}\right]+\mathring{\mathcal{B}}_{(3)B}^D[\b_{(0)}]
\end{equation}
where $\bar{\mathcal{H}}_{(3)B}^D$ does not depend on $\b_{(0)}$ but on the Ricci curvature and scalar of $h_{(0)}$ and on covariant combinations of $h_{(1)AB}$ and  $\mathring{\mathcal{B}}_{(3)B}^D[\b_{(0)}]=0$ whenever $\partial_A\b_{(0)}$ or $\b_{(0)}$ are zero (we have placed a circle on $\mathcal{B}$ to overstress the obvious fact that it is different from the analogous term appearing at the previous order). We have
\begin{align}
\bar{\mathcal{H}}_{(3)B}^D=&-\overset{{\scriptscriptstyle(1)}}{R^D}_B+h_{(1)}^{DE}\overset{{\scriptscriptstyle(0)}}{R}_{EB}+2\left(\tilde{K}^D_{(3)B}\mathcal{U}_{(-1)}+\tilde{K}^D_{(2)B}\mathcal{U}_{(0)}+\delta^D_{B}\mathcal{U}_{(1)}\right)\nonumber\\
&-\overset{{\scriptscriptstyle(0)}}{D}_BW^D_{(2)}-\overset{{\scriptscriptstyle(0)}}{D^D}W_{(2)B}-\overset{{\scriptscriptstyle(0)}}{D}_CW^C_{(2)}\delta^D_B
\end{align}
Which, using $W^A_{(2)}$ from \eqref{eq.Wa1}, $\mathcal{U}_{(-1)}$ from \eqref{eq.U1}, $\mathcal{U}_{(0)}$ from \eqref{eq.U0} and $\mathcal{U}_{(1)}=-\frac{1}{8}h_{(1)}^{EF}\overset{{\scriptscriptstyle(0)}}{R}_{EF}$ from \eqref{eq.Ua} and \eqref{eq.Raexpanded}, as well as $\overset{{\scriptscriptstyle(1)}}{R}_{AB}$ from \eqref{eq.PertRicci}, reads 
\begin{align}\label{eq.nontrivallog}
\bar{\mathcal{H}}_{(3)B}^D-l\tilde{K}_{(3)B}^D=&-\frac{1}{2}\left(\overset{\scriptscriptstyle (0)}{D}_C\overset{\scriptscriptstyle (0)}{D^D}h_{(1)B}^C+\overset{\scriptscriptstyle (0)}{D}_C\overset{\scriptscriptstyle (0)}{D}_Bh_{(1)}^{DC}-\overset{\scriptscriptstyle (0)}{D^2}h_{(1)B}^D\right)+h_{(1)}^{DE}\overset{{\scriptscriptstyle(0)}}{R}_{EB}\nonumber\\
&-\frac{1}{12}h_{(1)B}^D\overset{{\scriptscriptstyle(0)}}{R}-\frac{1}{4}h_{(1)}^{EF}\overset{{\scriptscriptstyle(0)}}{R}_{EF}\delta^D_B\\
&+\frac{1}{6}\overset{{\scriptscriptstyle(0)}}{D}_B\overset{{\scriptscriptstyle(0)}}{D}_Ch_{(1)}^{CD}+\frac{1}{6}\overset{{\scriptscriptstyle(0)}}{D^D}\overset{{\scriptscriptstyle(0)}}{D}_Ch_{(1)B}^{C}+\frac{1}{6}\overset{{\scriptscriptstyle(0)}}{D}_C\overset{{\scriptscriptstyle(0)}}{D}_Ah_{(1)}^{CA}\delta^D_B\nonumber
\end{align}
By sorting covariant derivatives we can combine the first and the last line, but they are not going to cancel each other. This term is not automatically zero even if $h_{(0)AB}$ is Einstein.
\\
\\
This analysis suggests that with the non-radiative falloff $r^{-1}$ in $h$, a maximal polyhomogeneous expansion is to be considered in $d\ge6$ even. Alternatively, the imposition of \eqref{eq.nontriviallog0} produces a differential constraint on $h_{(1)AB}$, which upon linearization around the round-sphere metric reduces to the constraint (3.32) of \cite{Campoleoni2020}. It is interesting to explicitly note that in odd dimensions there are no constraints analogous to \eqref{eq.nontriviallog0}, unless the falloff conditions with half-integer powers before the radiaive order (in addition to the integer ones) are assumed, as proved before.

\section{Asymptotic solutions: leading-log seed}\label{sec.maxpol}
In the previous section we solved the fourth main equation imposing that the logarithmic term at radiative order in $\partial_u h_{AB}$ is zero. We saw that this is always possible with radiative falloff conditions in any dimensions and with $r^{-1}$ falloff in odd dimensions. 

Let us assume 
\begin{equation}
h_{AB}=h_{(0)AB}+\frac{h_{(1)AB}}{r}+\sum_{p\in \mathbb{N}}^{p<\frac{d-4}{2}} \frac{h_{(1+p)AB}}{r^{1+p}}+\frac{h_{(\frac{d-2}{2})AB}}{r^{\frac{d-2}{2}}}+\frac{\log r}{r^{\frac{d-2}{2}}}\left(\texttt{h}_{(\frac{d-2}{2})AB}+\log r \texttt{h}^{\{2\}}_{(\frac{d-2}{2})AB} +\dots\right)+\dots.
\end{equation}
to analyse the consequences of a maximal polyhomogeneous expansion. We have
\begin{equation}
k^A_B=\frac{\delta^A_B}{r}+\frac{K^A_{(2)B}}{r^2}+\dots+\frac{1}{r^{\frac{d}{2}}}\left(K^A_{(\frac{d}{2})B}+\log r\texttt{K}_{(\frac{d}{2})}+\dots\right)+\dots
\end{equation}
where clearly $K^A_{(\frac{d}{2})B}$ is modified with respect to the previous cases by additional terms with $\texttt{h}_{(\frac{d-2}{2})AB}$ factors,
\begin{align}
\b&=\b_{(0)}+\sum_{p=2}^{<\frac{d}{2}}\frac{\beta_{(p)}}{r^p}+\frac{1}{r^{\frac{d}{2}}}\left(\b_{(\frac{d}{2})}+\log r\texttt{b}_{(\frac{d}{2})}+\dots\right)+\dots+\frac{1}{r^{d-2}}\left(\beta_{(d-2)}+\log r \texttt{b}_{(d-2)}+\dots\right)+\dots \label{eq.b}\\
W^A&=\sum_{p=1}^{<\frac{d}{2}}\frac{W_{(p)}^A}{r^p}+\frac{1}{r^\frac{d}{2}}\left(W^A_{(\frac{d}{2})}+\log r\texttt{W}^A_{(\frac{d}{2})}+\dots\right)+\dots+\frac{1}{r^{d-1}}\left(\mathcal{W}^A_{(d-1)}+\texttt{W}^A_{(d-1)}\log r+\dots \right)+\dots\label{eq.W}\\
\mathcal{U}&=\sum_{p=-1}^{p<\frac{d-2}{2}}\frac{\mathcal{U}_{(p)}}{r^p}+\frac{1}{r^{\frac{d-2}{2}}}\left(\mathcal{U}_{(\frac{d-2}{2})}+\log r\texttt{U}_{(\frac{d-2}{2})}+\dots\right)+\dots+\frac{1}{r^{d-3}}\left(\mathcal{U}_{(d-3)}+\texttt{{U}}_{(d-3)}\log r+\dots\right)+\dots\label{eq.U}
\end{align}
where $\texttt{h}$ contributes to $\b_{(\frac{d}{2})}$, $\b_{(d-2)}$, $W^A_{(\frac{d}{2})}$, $\mathcal{U}_{(\frac{d-2}{2})}$. With the above, the coefficients of the power-law expansion of $\mathcal{H}^D_B$ \eqref{eq.expH} are not modified up to the order $r^{-\frac{d+2}{2}}$. The coefficient of this order is modified by the terms of $\b$, $\mathcal{U}$ and $W^A$ that are affected by the leading logarithm, and the logarithmic terms themselves induce the new term $\texttt{H}^D_B[\texttt{h}]r^{-\frac{d+2}{2}}\log r$ with respect to the previous case. Hence $J^D_B$ \eqref{eq.Jsecondtimewritten} gets logarithmic contributions at order $r^{-1}$ and $r^{-2}$
\begin{equation}
\frac{\log{r}}{r}\texttt{J}^D_{(1)B}\quad \text{and} \quad \frac{\log{r}}{r^2}\texttt{J}^D_{(2)B}
\end{equation}
with
\begin{equation}
\texttt{J}^D_{(1)B}=-\frac{l}{2}\texttt{K}^D_{(\frac{d}{2})B} \quad \text{and} \quad \texttt{J}^D_{(2)B}=\frac{\texttt{H}^D_B[\texttt{h}]}{2},
\end{equation}
and its order $r^{-2}$ is modified by the modification of $\mathcal{H}_{(\frac{d+2}{2})}$.
On the other hand, $\Theta^{AB}_{CD}$ gets a logarithmic contribution at order $r^{-\frac{d-2}{2}}$, call it $\mathtt{\Theta}^{AB}_{(\frac{d-2}{2})CB}$. 

These results also holds with radiative falloff conditions, because the first logarithmic term in $h_{AB}$ is always induced at radiative order. With a radiative and maximally polyhomogeneous $h_{AB}$, the first power in the expansions of $\b$, $W^A$ and $\mathcal{U}$, apart from those that are universal, are respectively $r^{-(d-2)}$, $r^{-\frac{d}{2}}$ and $r^{-\frac{d-2}{2}}$. 
\\
\\
Because of the discussed strcutre, $\texttt{L}_{[d>4]B}^D$ \eqref{eq.Llog>4} is not modified by the presence of $\texttt{h}$. Indeed it could only get contributions from $-[\mathtt{\Theta}_{(\frac{d-2}{2})}J_{(-\frac{d-4}{2})}]^D_B$, which however vanishes upon the imposition of the leading order equation \eqref{eq.gen}. Equation \eqref{eq.log} now reads as
\begin{equation}\label{eq.newradlog}
\texttt{L}_{[d>4]B}^D=\texttt{l}^D_{B}
\end{equation}
to which we come back momentarily.

However, the first term $\texttt{J}^D_{(1)B}$ with $\mathtt{\Theta}^{AB}_{(\frac{d-2}{2})CB}$
contribute to $L^D_B$ with a new $\log^2 r$ term at order $r^0$ 
\begin{equation}\label{eq.newlog}
L^D_B \ni \frac{1}{2}\left(\texttt{J}^D_{(1)B}+[\mathtt{\Theta}_{(\frac{d-2}{2})}J_{(-\frac{d-4}{2})}]^D_B\right)\log^2 r=\frac{\texttt{J}^D_{(1)B}}{2}\log^2 r,
\end{equation}
the last equality following from the imposition of \eqref{eq.gen}, and induces further $\log r$ terms down in the asymptotic expansions as well as new additional terms in the coefficients of the pure power terms. The second term $\texttt{J}^D_{(2)B}$ instead contributes to $L^D_B$ from order $r^{-1}$ onwards. 

The term \eqref{eq.newlog} induces in $h_{AB}$ a further $r^{-\frac{d-2}{2}}\log^2r$ term and by the same reasoning, we obtain all higher powers of $\log r$ at the same order and at subleading orders. We end up with a maximal polyhomogeneous expansion of the form \eqref{eq.pol} with an infinte sum over the powers of logs at each order $r^{-j}\texttt{h}_{(j,i)AB}\log^i r$. The $u$-dependence of each $\texttt{h}_{(j,i)AB}$ ($i>1$) is determined by the $u$-dependence of $h_{(0)AB}$ and by  $\texttt{h}_{(j,i-1)AB}$.  The infinite sum of logarithmic terms is avoided if the metric $h_{(0)AB}$ is time independent because $\texttt{J}^D_{(1)B}=0$ as $l=0$. 
\\
\\
Coming back to \eqref{eq.newradlog}, $\texttt{l}^D_{B}$ is given by (not imposing $l=0$)
\begin{equation}
\texttt{l}^D_{B}=\frac{1}{2}\left(h^{DE}_{(0)}\partial_u \texttt{h}_{(\frac{d-2}{2})EB}-\texttt{h}^{DE}_{(\frac{d-2}{2})}\partial_u h_{(0)EB}\right),
\end{equation}
and  $\texttt{L}_{[d>4]B}^D$ by $\eqref{eq.Llog>4}$.

As we have discussed in subsection \ref{sec.rad}, $\texttt{L}_{[d>4]B}^D=0$ trivially with radiative falloffs, hence 
\begin{equation}\label{eq.leadinglog>4}
d\ge4:\qquad\texttt{h}_{(\frac{d-2}{2})AB}(u,x)=e^{2\varphi(u,x)}\hat{\texttt{h}}_{(\frac{d-2}{2})AB}(x).
\end{equation}
Also in $d=4$ we have from \eqref{eq.d=4_purple}
\begin{equation}\label{eq.leadinglog=4}
d=4:\qquad\frac{1}{2}\left(h^{DE}_{(0)}\partial_u \texttt{h}_{(1)EB}-\texttt{h}^{DE}_{(1)}\partial_u h_{(0)EB}\right)=0.
\end{equation}
With a time independent boundary metric $(l=0)$ $\texttt{h}_{(\frac{d-2}{2})AB}$ is $u$-independent, consistently with the analysis of \cite{Kroon2001}. However, notice that while \eqref{eq.leadinglog=4} is valid for any pair $(h_{(0)AB},\b_{(0)})$ in four dimensions, the analogous equation in higher dimensions \eqref{eq.leadinglog>4} holds automatically only with radiative falloff conditions and it is in general modified by non-radiative falloffs.

With non-radiative falloffs, $\texttt{L}_{[d>4]B}^D$ follows the behaviour discussed in the previous subsection. The asymptotic expansion that makes this term automatically zero is natural in odd dimensions and hence $\texttt{h}_{(\frac{d-2}{2})AB}$ satisfies the above equation also in this case. In even dimensions, there is a non-trivial time dependence of $\texttt{h}_{(\frac{d-2}{2})AB}$ regardless of the condition $l=0$.

\section{Asymptotic Killing fields}\label{sec.akvALM2}
Up to conformal rescalings of $g_{AB}$, 
Bondi-Sachs gauge is preserved by asymptotic diffeomorphisms satisfying
\begin{equation}\label{eq.1condition.high1}
\mathfrak{L}_\xi g_{rr}=0 \, , \quad \mathfrak{L}_\xi g_{rA}=0 \, , \quad g^{AB}\mathfrak{L}_\xi g_{AB}=2(d-2){\omega}(u,x)\,.
\end{equation}
The exact Killing equations are solved by the vector field
\begin{equation}
\xi=\xi^u\partial_u+\xi^r\partial_r+\xi^A\partial_A
\end{equation}
\begin{equation}\label{eq.Killing.general}
\begin{cases}
\xi^u=f(u,x^A) \, ,\\
\xi^r= 	-\dfrac{r}{d-2}\left[{}^ {(d-2)}{D}_A\xi^A-W^C\partial_Cf+fl-(d-2){\omega}\right] ,\\
\xi^A=Y^A(u,x^B)-\partial_Bf\int_{r}^{\infty}dRe^{2\beta}g^{AB}\,,
\end{cases}
\end{equation} 
with $f$, $Y^A$ and $\omega$ arbitrary. 
They act on the remaining metric components as 
\begin{align}
\mathfrak{L}_\xi g_{AB}&=r^2(\delta_\xi g_{AB})_{(2)}+r(\delta_\xi g_{AB})_{(1)}+r^{2-a}(\delta_\xi g_{AB})_{(2-a)}+r^{1-a}(\delta_\xi g_{AB})_{(1-a)}+\dots\label{eq.Kab}\\
\mathfrak{L}_\xi g_{uu}&=r(\delta_\xi g_{uu})_{(1)}+r^{1-a}\sum_{p=0}r^{-p}(\delta_\xi g_{uu})_{(1-a-p)}\label{eq.Kuu}\\
\mathfrak{L}_\xi g_{ur}&=(\delta_\xi g_{ur})_{(0)}+ r^{-a-1}\sum_{p=0} r^{-p}(\delta_\xi g_{ur})_{(-a-1-p)}\label{eq.Kur}\\
\mathfrak{L}_\xi g_{uA}&=r^2(\delta_\xi g_{uA})_{(2)}+r(\delta_\xi g_{uA})_{(1)}+r^{2-a}\sum_{p>0}r^{-p}(\delta_\xi g_{uA})_{(2-a-p)}\label{eq.KuA}
\end{align}
All except \eqref{eq.KuA} preserve the leading order of the metric expansion. The leading order parts are the transformation rules of the leading order data, but we have already fixed $W^A_{(0)}=0$ thus making $(\delta_\xi g_{uA})_{(2)}$ inconsistent with the configuration space unless it is zero. This results in
\begin{equation}\label{eq.YAu}
\partial_uY^A=0\,,
\end{equation}
while the first, the third and the second give respectively 
\begin{align}
\delta_\xi h_{(0)AB}&=\mathfrak{L}_Yh_{(0)AB}-\frac{2}{d-2}\left(D_CY^C-(d-2)\omega\right)h_{(0)AB}\label{eq.h0killingcond}\\
\delta_\xi\b_{(0)}&=(f\partial_u+\mathfrak{L}_Y)\b_{(0)}+\frac{1}{2}(\partial_uf+\xi^r_{(-1)})\label{eq.b0killingcond},\\
\delta_\xi \mathcal{U}_{(-1)}&=(f\partial_u+\mathfrak{L}_Y)\mathcal{U}_{(-1)}-\partial_uf\mathcal{U}_{(-1)}-2\partial_u\xi^r_{(-1)} \label{eq.U-1killing}
\end{align}
where in \eqref{eq.h0killingcond} we have explicitly used that 
\begin{equation}
\xi^r_{(-1)}=-\frac{1}{d-2}\left(D_AY^A+fl-(d-2)\omega\right)
\end{equation}
and the definition \eqref{eq.uh0} of $l$ to eliminate the terms involving $\partial_u h_{(0)AB}$. Here $D$ is the covariant derivative compatible with $h_{(0)}$, we remove the superscript $(0)$  used in the previous sections over $D$ because no confusion arise.

Equations \eqref{eq.h0killingcond}, \eqref{eq.b0killingcond}, \eqref{eq.U-1killing} can be used to locally fix the gauge to the standard conditions employed in the definitions of asymptotic flatness and the asymptotic symmetries usually considered in the four dimensional literature can be adapted to higher dimensions as ($a$ again parametrises the leading falloff behaviour $a=1$ or $a=\frac{d-2}{2}$)
\begin{itemize}
	\item[CL)] \textbf{Campiglia-Laddha}: $\mathscr{I}=\mathbb{R}\times S^{d-2}$, $h_{(0)AB}$ $u$-independent and free except for its fixed determinant, $\b_{(0)}=0$ fixed
	\begin{equation}\label{eq.CLk}
	\mathfrak{L}_{\xi}g_{AB}=O(r^{2})\,,\quad
	\mathfrak{L}_{\xi}g_{uu}=O(r^0)\,,\quad
	\mathfrak{L}_{\xi}g_{ur}=O(r^{-2a})\,,\quad
	\mathfrak{L}_{\xi}g_{uA}=O(r^{1-a}).\quad 
	\end{equation}
	\item[BS)] \textbf{Bondi-Sachs}:  $\mathscr{I}=\mathbb{R}\times S^{d-2}$, $h_{(0)AB}=\gamma_{AB}$ round sphere metric, $u$-independent and $\b_{(0)}=0$ fixed
	\begin{equation}
	\mathfrak{L}_{\xi}g_{AB}=O(r^{2-a})\,,\quad
	\mathfrak{L}_{\xi}g_{uu}=O(r^{-a})\,,\quad
	\mathfrak{L}_{\xi}g_{ur}=O(r^{-2a})\,,\quad \mathfrak{L}_{\xi}g_{uA}=O(r^{1-a}).\quad
	\end{equation}
\end{itemize}
As stressed previously, any other cross section of $\mathscr{I}$ can be taken to define the configuration space. Of course, there may be no asymptotic symmetries at all both when there are no restrictions on the leading order data and when the restrictions are too mild, such as when the topology changes with time evolution (see \cite{Foster1987} on this in four dimensional spacetimes).

A set of conditions that one may consider in light of our previous considerations on $\b_{(0)}$ is
\begin{equation}\label{eq.mgbc}
\mathscr{I}=\mathbb{R}\times \mathbb{B}^{d-2},\,\,\, h_{(0)AB}(u,x)\,\,\text{free within a conformal class,}\,\,\,\b_{(0)}(u,x)\,\, \text{fixed}.
\end{equation}
This partially restricts the freedom in \eqref{eq.h0killingcond}, \eqref{eq.b0killingcond}, but is more general than the CL or BS conditions. The falloffs \eqref{eq.CLk} remains the same except for
\begin{equation}\label{eq.guuk}
\mathfrak{L}_{\xi}g_{uu}=O(r),
\end{equation}
which produces the transformation law \eqref{eq.U-1killing} of $\mathcal{U}_{(-1)}$ (or $\partial_u \varphi$).

The above \eqref{eq.mgbc} can be taken as the conditions defining, at least locally, the spacetime with $\mathbb{R}\times \mathbb{B}^{d-2}$ asymptotics and a fixed $\b_{(0)}$ and reflect the discussion around \eqref{eq.Ash}.

To satisfy \eqref{eq.mgbc} we impose $\delta_\xi\b_{(0)}=0$ 
so that, from \eqref{eq.b0killingcond}
\begin{align}\label{eq.generalf}
f(u,x)&=e^{\varphi-2\b_{(0)}}\alpha(x)+e^{\varphi-2\b_{(0)}}\int^u e^{-(\varphi-2\b_{(0)})}\left(\frac{F}{d-2}-\omega-2(d-2)\mathfrak{L}_Y\b_{(0)}\right)\dif u'\nonumber\\
&F(u,x):=D_AY^A(x)=(d-2)Y^A\partial_A\varphi +\underbrace{\left(\frac{1}{2}\partial_A \log|\hat{q}|+\partial_A\right)Y^A}_{\hat{D}_AY^A}
\end{align}
The scalar $\alpha$ is an arbitrary function on $\mathbb{B}^{d-2}$ and we have defined $F$ as the leading covariant divergence of $Y^A$, which splits in a part depending on $u$ and a covariant divergence with respect to the $u$-independent factor of $h_{(0)AB}$, which we denoted with a hat and hence $\hat{q}$ is its determinant.  With \eqref{eq.YAu}, \eqref{eq.h0killingcond} and \eqref{eq.generalf} all the leading order conditions in \eqref{eq.CLk}, \eqref{eq.guuk}. To proceed in the analysis of asymptotic symmetries we must satisfy all the other subleading equations.

We do not discuss the asymptotic symmetries of this case in detail. This would imply a though analysis depending on the topology of $\mathbb{B}^{d-2}$ and the metric, similarly to the cases considered in \cite{Foster1987}. We limit ourselves to show the rather obvious fact that the presence of $\b_{(0)}$ generically implies constraints on $f$ and $Y^A$, so that the identification of supertranslations/superrotations is not in general possible. This statement is not surprising and is similar to the case in which a time-dependent conformal factor $\varphi$ is attached to the time-independent $\hat{h}$ in the background structure of null infinity: $f$ dos not contain the supertranslation part, because $\alpha$ is multiplied by a time-dependent factor. 

With \eqref{eq.YAu}, \eqref{eq.h0killingcond} and \eqref{eq.generalf} all the leading order conditions are solved. With $a=1$, $\delta_\xi g_{ur}=O(r^{-2})$ is now automatically satisfied because the $O(r^{-1})$ component is trivially zero. On the other hand, $(\delta_\xi g_{uA})_{(1)}=0$ reads
\begin{align}\label{eq.killingremaining}
&\frac{1}{d-2}\partial_A\left[F+fl-(d-2)\omega\right]e^{2\b_{(0)}}-\partial_u\left(\partial_Cf h_{(0)}^{CB}e^{2\b_{(0)}}\right)h_{(0)BA}-\frac{2l}{d-2}e^{2\b_{(0)}}\partial_A f\nonumber\\
&+W_{(1)A}\left[(\partial_u+1)f+\frac{1}{d-2}\left(F+lf-(d-2)\omega\right)\right]-Y^B\partial_BW_{(1)A}-\partial_A(Y^B)W_{(1)B}=0
\end{align}
This equation is automatically satisfied when $\b_{(0)}=0$ because $exp(2\b_{(0)})=1$ and the first line can be seen to vanish after simple manipulations, while the second line vanishes automatically because $W_{(1)}=0$. In the general case of a non vanishing $\b_{(0)}$ we get a constraint relating $f$ and $\b_{(0)}$. The subcase where $W_{(1)}=0$ but $\b_{(0)}\neq 0$ (i.e. when $\partial_A\b_{(0)}=0$) imposes a constraint of the form $\partial_A f\partial_u\b_{(0)}=0$, which is then trivially satisfied by a constant $\b_{(0)}$, with no need of constraining $f$. This case thus trivially reduces to the case $\b_{(0)}=0$.

When $d$ is even this completes the analysis of $\mathfrak{L}_\xi g_{uA}=O(r^0)$, but when $d$ odd there is a potential further condition at order $r^{1/2}$ (compare with \eqref{eq.KuA}). It is easy to check that this may only come from $\partial_u \xi^B g_{AB}$ in the explicit expression of $\mathfrak{L}_\xi g_{uA}$ and that it automatically vanishes because of \eqref{eq.YAu}.
\\
\\
At subleading orders we find the transformation laws of the subleading terms of the metric expansion. To complete the goal set in the introduction of the chapter,  the most relevant to address is $\delta_\xi h_{(1)AB}$. Notice that with $a=1$, both
\begin{equation}
(\delta_\xi g_{AB})_{(1)}=e^{2\b_{(0)}}\left(\frac{2}{d-2}D^2f-2D_AD_Bf\right)+\frac{4}{d-2}h_{(0)AB}W^E_{(1)}\partial_Ef-4W_{(1)(A}\partial_{B)}f
\end{equation}
and 
\begin{equation}
(\delta_\xi g_{AB})_{(2-a)}=f\partial_uh_{(a)AB}-\frac{(2-a)}{d-2}\left(D_CY^C+fl-(d-2)\omega\right)h_{(a)AB}+\mathfrak{L}_Yh_{(a)AB}
\end{equation}
contribute to the end result. We get
\begin{align}
\delta_\xi h_{(1)AB}&=2e^{2\b_{(0)}}\left(\frac{1}{d-2}h_{(0)AB}D^2-D_AD_B\right)f+f\underbrace{\partial_uh_{(1)AB}}_{\substack{\eqref{eq.partial_u}\, d>4\\\eqref{eq.radnews}\, d=4}}-\frac{F}{d-2}h_{(1)AB}+\mathfrak{L}_{Y}h_{(1)AB} \nonumber \\
&-\left(\frac{f l}{d-2}+\omega\right)h_{(1)AB}+2\left[\frac{\partial^C fW_{(1)C}}{d-2}h_{(0)AB}-2\partial_{(B}f W_{(1)A)}\right]
\end{align}
The term in square brackets vanishes for the standard boundary condition on $\b_{(0)}$.

To conclude we consider the CL/BS subcases with $a=1$ and $a=\frac{d-2}{2}$ and  $\omega=0$.
\paragraph{\textbf{CL \&} $\boldsymbol{a=1}$, $\mathbf{d>4}$.} 
This is the case in \cite{Colferai2020} for even $d$. The above expressions reduce to
\begin{equation}
f(u,x)=\alpha(x)+\frac{u}{d-2}F(x),\quad F(x):=\hat{D}_AY^A(x)
\end{equation}
and
\begin{equation}
\delta_\xi h_{(1)AB}=2\left(\frac{1}{d-2}h_{(0)AB}\hat{D}^2-\hat{D}_A\hat{D}_B\right)f-f\cancel{\partial_uh_{(1)AB}}-\frac{F}{d-2}h_{(1)AB}+\mathfrak{L}_{Y}h_{(1)AB}
\end{equation}
which, restricted to $(\alpha=0, Y)$ gives\footnote{The last term in the next equation does not appear in \cite{Colferai2020}. However, our expressions are fully consistent with \cite{Compere2018} when restricted to $d=4$.}
\begin{equation}\label{eq.incons}
\delta_Y h_{(1)AB}=\frac{2u}{(d-2)^2}\left(h_{(0)AB}\hat{D}^2-(d-2)\hat{D}_A\hat{D}_B\right)F+\left(\mathfrak{L}_{Y}-\frac{F}{d-2}\right)h_{(1)AB}
\end{equation}
which contains the $u$-dependent piece that is the reason of the issue noticed in \cite{Colferai2020} with KLPS conditions \eqref{eq.KLPS}. 

As we have restricted to $\b_{(0)}=0$ and time independent $h_{(0)}$, \eqref{eq.incons} will be consistent on the configuration space if $h_{(0)}$ is not Einstein. Since $\alpha$ is time independent, we can say that in this case the asymptotic transformations of the configuration space comprise supertranslations and diffeomorphisms of the cross sections of $\mathscr{I}$ (we use supertranslations with the proviso that their physical interpretation as generalisations of translations does depend on the form of the boundary metric). The generalisation to time-dependent $h_{(0)}$ and generic topology of the cross section is consistent, as seen above. 

\paragraph{BS \& $\boldsymbol{a=\frac{d-2}{2}}$.}
This corresponds to the TKS analysis \cite{Tanabe2011}. From \eqref{eq.generalf} we again recover \cite{Tanabe2011,Barnich2010}
\begin{equation}
f(u,x)=\alpha(x)+\frac{u}{d-2}F(x),\quad F(x):=\hat{D}_AY^A(x)
\end{equation}
but now, from the vanishing of \eqref{eq.h0killingcond}, we get
\begin{equation}
\delta_Yh_{(0)AB}=\frac{2}{d-2}Fh_{(0)AB},
\end{equation}
so that $Y$ is a conformal Killing vector. 

With this value of $a$, $(\delta_\xi g_{AB}^{(1)})$ is of the same order of $(\delta_\xi g_{AB}^{(2-a)})$ only in $d=4$ and hence
\begin{equation}
\delta_\xi h_{(1)AB}=2\left(\frac{1}{d-2}h_{(0)AB}\hat{D}^2-\hat{D}_A\hat{D}_B\right)f-f\underbrace{\partial_uh_{(1)AB}}_{N_{AB}}-\frac{F}{d-2}h_{(1)AB}+\mathfrak{L}_{Y}h_{(1)AB}
\end{equation}
only in $d=4$. The function $\alpha$ remains arbitrary. We get the $\mathfrak{bms}$ algebra (either global or local).

When $d>4$, we have a constraint on $f$ coming from  $(\delta_\xi g_{AB}^{(1)})=0$ 
\begin{equation}\label{eq.constraintf}
\frac{2}{d-2}h_{(0)AB}\hat{D}^2f-2\hat{D}_A\hat{D}_Bf=0,
\end{equation}
from which we get 
\begin{equation}\label{eq.constraintalpha}
\frac{2}{d-2}h_{(0)AB}\hat{D}^2\alpha-2\hat{D}_A\hat{D}_B\alpha=0,
\end{equation}
as $F$ can be proved to satisfy
\begin{equation}\label{eq.constraintF}
\hat{D}_A\hat{D}_BF-\frac{1}{d-2}\hat{D}^2F h_{(0)AB}=0
\end{equation}
from the properties of the Riemann tensor of a maximally symmetric space \cite{Tanabe2011}. 
$F$ is given by the $l=1$ modes of the scalar harmonics on the hypersphere and $\alpha$ by $l=0,1$ modes. Given this, the subleading conditions 
\begin{equation}\label{eq.subconstraintTKS}
(\delta_\xi g_{ur})_{(-a-1-p)}=0\quad \forall\,p<a-1,\quad (\delta_\xi g_{uA})_{(2-a-p)}\, \quad \forall\, p<1
\end{equation}
are automatically satisfied as in \cite{Tanabe2011}.  Let us take for example $(\delta_\xi g_{ur})_{(-a-1-p)}=0$.  For each $p<a-1$ it is of the form
\begin{equation}\label{eq.autsat}
h_{(a+p)AB}\hat{D}^A\hat{D}^Bf=0,
\end{equation}
which holds using \eqref{eq.constraintf} and the fact that $h_{(a+p)AB}$ is traceless for any $p<a$.

The resulting algebra of asymptotic symmetries is Poincaré.  The action of the asymptotic Killing on the radiative data follow from  $(\delta_\xi g_{AB}^{(2-a)})$ as
\begin{equation}
\delta_\xi h_{(\frac{d-2}{2})AB}=-f\underbrace{\partial_uh_{(\frac{d-2}{2})AB}}_{N_{AB}}-\frac{F}{d-2}h_{(\frac{d-2}{2})AB}+\mathfrak{L}_{Y}h_{(\frac{d-2}{2})AB}
\end{equation}
For $\mathbb{B}^{d-2}\neq S^{d-2}$  in $d=4$ the asymptotic symmetries have been studied in \cite{Foster1987}.

\paragraph{BS in $\boldsymbol{d>4}$ \& $\boldsymbol{a=1}$.} In even spacetime dimensions this is the case considered by KLPS \cite{Kapec2015}. The analysis is the same as the above except that \eqref{eq.constraintf} does not apply. Thus $\alpha$ is free and $h_{(1)AB}$ transforms formally as in $d=4$
\begin{equation}
\delta_\xi h_{(1)AB}=2\left(\frac{1}{d-2}h_{(0)AB}D^2-D_AD_B\right)f-f\cancel{\partial_uh_{(1)AB}}-\frac{F}{d-2}h_{(1)AB}+\mathfrak{L}_{Y}h_{(1)AB}
\end{equation}
Notice that since $Y$ generates Lorentz transformations in this case, the remark after \eqref{eq.incons} does not apply because the $u$-dependent piece of \eqref{eq.incons} automatically vanishes. 
The action of the Killing field on the radiative data is read by pushing the expansion of \eqref{eq.Kab} up to $p=\frac{d-4}{2}$. So for example in $d=5$ it is found at $p=1/2$, as it should as $a(=1)+p(=1/2)=3/2$.
\paragraph{CL \& $\boldsymbol{a=\frac{d-2}{2}}$.} To conclude, we consider the possibility of having superrotations without supertranslations in $d>4$. This may seem plausible since we have repeatedly stated that superrotations only depend on the boundary conditions, while supertranslations depend on the falloff conditions. So can
\begin{equation}\label{eq.possible}
Diff(S^{d-2})\ltimes T
\end{equation}
be an asymptotic symmetry group?

Given the CL conditions \eqref{eq.CLk}, \eqref{eq.constraintf} does not apply. However, the conditions \eqref{eq.subconstraintTKS} must be considered. While they were automatically satisfied by the solutions of \eqref{eq.constraintf}, they are not so now. For example consider again \eqref{eq.autsat}. Since $\alpha$ exponentiates to translations by assumption (and hence satisfies \eqref{eq.constraintalpha}), \eqref{eq.autsat} 
\begin{equation}
h_{(a+p)AB}\hat{D}^A\hat{D}^B F =0
\end{equation}
is not automatically satisfied but constrains $Y^A$. A solution of this equation is given by \eqref{eq.constraintF}, which collapses to the standard Bondi-Sachs case. A trivial solution is $F=0$, that is $Y$ is divergence-free. We have not investigated other solutions. 

Thus, \eqref{eq.possible} cannot be an asymptotic symmetry group: we are either forced to the Poincaré group or to some other group where the diffeomorphisms generated by $Y^A$ are restricted. It would have been puzzling otherwise, because we would have been able, in principle, to recover the subleading soft theorem from the $Diff(S^{d-2})$ part of the asymptotic symmetries but not the leading one, as there is not enough symmetry in the Abelian factor of the group. 
 
\section{Conclusions and outlook}\label{sec.concl}
In this paper we have explored the asymptotics of Ricci flat spacetimes at null infinity in any dimension $d\ge4$ with general boundary conditions to answer two questions that arose from recent literature regarding the extension of supertranslations and Campiglia-Laddha superrotations to dimensions higher than four, which are considered relevant in view of the known relationship of these symmetry structures with soft scattering theorems in four-dimensional spacetimes.

As we have reviewed in subsection \ref{sec.KLPS}, while supertranslations can be defined at the linear level in higher dimensions, superrotations cannot be defined on the same configuration space. In section \ref{sec.res} we have discussed the boundary conditions that allow to bypass the issues experienced by previous literature (sections \ref{sec.bc} and \ref{sec.h1discussion}) 
and we have recognised the case in claim \ref{clm.nonE} as corresponding to the case discussed only in five spacetime dimension in \cite{Capone2019} and at a linear level in \cite{Campoleoni2020}. 

In some sense, the most conservative possibility allowing for superrotation-like and supertranslation-like transformations is the one of claim \ref{clm.nonE}. This is however bound to issues - at least of technical nature - in the discussion of asymptotic charges. Another possibility is the time-dependence of an Einstein boundary metric (claim \ref{clm.timeh}), but this also leads to severe problems in the interpretation of charges and the global definitions of (stable) infinity with limits to $i^0$, as briefly seen in subsection \ref{sec.limit}.

A comment is in order here. We have not discussed, as in previous analysis \cite{Kapec2015,Hollands2016,Campoleoni2020}, the gauge character of $h_{(1)AB}$ and the other overleading terms before the radiative order. As we can see from \cite{Kapec2015,Hollands2016,Campoleoni2020}, the proof that $h_{(1)AB}$ is pure gauge strictly depends on the assumption that the boundary metric $h_{(0)AB}$ is the time-independent round sphere metric, on the properties of the Laplacian associated to this metric and on the linearised versions of the constraints \eqref{eq.core}, \eqref{eq.nontriviallog0} around the boundary round sphere. Both of them are needed in the proof: without \eqref{eq.nontriviallog0} a part of $h_{(1)AB}$ is left undetermined. The main text does not discuss further the constraint \eqref{eq.nontriviallog0} because the scope is to address the construction of the most general solution space and to point out the differences between even and odd dimensions. In this respect, we reiterate that in odd dimensions no analogous of \eqref{eq.nontriviallog0} exists.

Despite the points raised here, which are to be critically revised in order to push  the scattering picture \cite{Strominger2014} to higher dimensions, if we limit ourselves to local considerations of Ricci-flat asymptotics, we can hope to relate the configuration spaces here discussed with AdS/CFT in the spirit of de Boer and Solodukhin \cite{deBoer2000} or a flat limit approach in the spirit of \cite{Barnich2012,Bagchi2013,Fareghbal2013,Fareghbal2018,AMK,Compere2020}. We have envisaged in subsection \ref{sec.leadinglog} an interpretation of the leading logarithmic term appearing in the expansion of $h_{AB}$ in terms of anomalies of the dual field theory, which we plan to analyse further \cite{Capone2021}. This could provide a geometric approach to the topic of anomalies in BMS field theories \cite{Bagchi2021}. We however stress that the analogy drawn in subsection \ref{sec.leadinglog} is not conclusive because a proper phase space analysis and (holographic) renormalization procedure is needed. In all generality this touches upon the problem of well-posedness of variational principles with null asymptotics \cite{Parattu2015,Lehner2016,Chandrasekaran2020}

In this respect, a simple analysis \cite{Capone2021} of the presymplectic form shows that the divergences generalise the structure already discussed in \cite{Compere2018}, hence we expect similar considerations could be applied to higher dimensions under appropriate restrictions of the boundary conditions.  While this implies that a variant of the method of \cite{Compere2020} can be used to renormalise the charges, the issue on the globality of null infinity remains open as well as more formal analysis along the lines of \cite{Chrusciel2010}. 
\\
\\
We conclude with a further speculative comment. In three spacetime dimensions, the analysis of \cite{Ashtekar1996,Barnich2012} (see also discussion in subsection \ref{sec.bc}) shows that a proper definition of asymptotic flatness includes point particles in the phase space. In four spacetime dimensions, the extension of the BMS group to include BT-superrotations led to the proposal that the standard asymptotically flat phase space should be extended to include cosmic strings\footnote{At least a single straight string, the situation with more general boost-rotation symmetric spacetimes or network of them is less clear.} \cite{Strominger2017}, which are usually called ``asymptotically locally flat spacetimes'' in four dimensions (because of the incompleteness of $\mathscr{I}$). The interplay between BT and CL superrotations has been studied geometrically in \cite{Adjei2019} and from the celestial conformal field theory perspective in \cite{Donnay2020}. In five spacetime dimensions, the proposal was made to include cosmic $(d-3)$-branes in the configuration space, to produce effects to be interpreted as related to superrotations. The boundary conditions here used are more general than those allowing for such branes, but may potentially allow for transitions among configurations with and without branes. 

Cosmic branes, cosmic strings and point particles are codimension-2 objects in $d>4$, $d=4$ and $d=3$ dimensions respectively. We could suggest that configuration spaces with consistent actions of superrotations (whatever this name means according to the dimension) include such codimension-2 objects. 

\section*{Acknowledgements}
This work is funded by the EPSRC Doctoral Prize Award EP/T517859/1. I would like to thank A. Poole and M. Taylor, for useful discussions and comments on the manuscript, as well as S. Hollands and A. Ashtekar for some related mail correspondence. I also wish to thank K. Skenderis and L. Too for related discussions. I am grateful to J. Simon and D. Turton for their comments on an early draft discussed during my PhD examination.

\appendix
\section{Derivation of Einstein's equations}\label{app.Details}
The Ricci tensor
\begin{equation}
R_{\mu\nu}=\partial_\rho \Gamma^\rho_{\mu\nu}+\Gamma^\rho_{\rho\sigma}\Gamma^{\sigma}_{\mu\nu}-\partial_\mu \Gamma^\rho_{\rho\nu}-\Gamma^\rho_{\mu\sigma}\Gamma^\sigma_{\rho\nu}
\end{equation}
is conveniently computed using
\begin{equation}
\Gamma^\rho_{\rho\mu}=\frac{1}{\sqrt{|{}^{\scriptscriptstyle(d)}g|}}\partial_\mu\sqrt{|{}^{\scriptscriptstyle(d)}g|}\,,\quad |{}^{\scriptscriptstyle(d)}g|=e^{4\b}|{}^{\scriptscriptstyle(d-2)}g|=e^{4\b}r^{2(d-2)}\underbrace{|h_{(0)}|}_{q}
\end{equation}
such that 
\begin{equation}
\Gamma^\rho_{\rho\mu}=\partial_\mu\left(2\b+\frac{1}{2}\log|{}^{\scriptscriptstyle(d-2)}g|\right)=:\partial_\mu \mathcal{O}_2
\end{equation}
and
\begin{equation}
R_{\mu\nu}=\underbrace{\left[\partial_\rho+\partial_\rho\left(2\b+\frac{1}{2}\log|{}^{\scriptscriptstyle(d-2)}g|\right)\right]}_{\mathcal{O}_\rho}\Gamma_{\mu\nu}^\rho-\partial_\mu\partial_\nu\left(2\b+\frac{1}{2}\log|{}^{\scriptscriptstyle(d-2)}g|\right)-\Gamma^\rho_{\mu\sigma}\Gamma^\sigma_{\rho\nu}.
\end{equation}
Some useful relationships involving $\text{det}{g}_{AB}=|{}^{\scriptscriptstyle(d-2)}g|$ are found using
\begin{equation}
g^{AB}\partial_\mu g_{AB}=\frac{\partial_\mu (\text{det} g_{AB})}{\text{det} g_{AB}}
\end{equation}
so that\footnote{I.e.: $
	\underbrace{\partial_r|g_{AB}|}_{2(d-2)r^{2(d-2)-1}q} = \quad\underbrace{|g_{AB}|}_{r^{2(d-2)}q}g^{AB}\partial_r g_{AB}$, $g^{AB}\partial_r g_{AB}=\frac{h^{AB}}{r^2}\partial_r\left(r^2h_{AB}\right)=\frac{2(d-2)}{r}+h^{AB}\partial_r h_{AB}$}
\begin{eqnarray}
&g^{AB}\partial_r g_{AB}=\frac{2(d-2)}{r} \quad &\text{and} \quad h^{AB}\partial_rh_{AB}=0\nonumber\\
&g^{AB}\partial_u g_{AB}=\frac{\partial_u q}{q}\quad &\text{and} \quad h^{AB}\partial_u h_{AB}=\dfrac{\partial_u q}{q} \\
&g^{AB}\partial_{C} g_{AB}=\dfrac{\partial_C q}{q}\quad &\text{and} \quad h^{AB}\partial_C h_{AB}=\dfrac{\partial_C q}{q}\nonumber
\end{eqnarray}
To make contact with \cite{Barnich2010} we define
\begin{equation}
l_{AB}=\frac{1}{2}\partial_u g_{AB}\,,\quad k_{AB}=\frac{1}{2}\partial_rg_{AB}\,,\quad n_A=\frac{1}{2}e^{-2\b}g_{AB}\partial_rW^B
\end{equation}
whose indices are raised and lowered with $g_{AB}$ and such that
\begin{equation}
l^{AB}=-\frac{1}{2}\partial_ug^{AB}\,,\quad k^{AB}=-\frac{1}{2}\partial_rg^{AB}
\end{equation}
The Christoffel symbols are given in the following as
\begin{equation}
\qquad\begin{bmatrix}
\Gamma^\mu_{uu} & \Gamma^\mu_{ur} & \Gamma^\mu_{uC}\\
\Gamma^\mu_{ru} & \Gamma^\mu_{rr} & \Gamma^\mu_{rC}\\
\Gamma^\mu_{Au} & \Gamma^\mu_{Ar} & \Gamma^\mu_{BC}
\end{bmatrix}\,, \quad \mu = u,r, A
\end{equation}
\begin{eqnarray}
&\Gamma^u_{uu}=2\partial_u \b -\frac{1}{2}(2\partial_r\b+\partial_r)\mathcal{U}+2n_AW^A+e^{-2\b}k_{CD}W^CW^D\,,\nonumber\\
&\quad \Gamma^u_{ur}=0\,,\quad \Gamma^u_{uA}=\partial_A\b-n_A-e^{-2\b}k_{AB}W^B\,,\nonumber\\
&\Gamma^u_{rr}=0\,,\quad \Gamma^u_{rA}=0\,,\quad \Gamma^u_{AB}=e^{-2\b}k_{AB}\,,\nonumber\\
&\nonumber\\
&\Gamma^r_{uu}=\frac{1}{2}(\partial_u-2\partial_u\b)\mathcal{U}+\frac{1}{2}\mathcal{U}(\partial_r+2\partial_r\b)\mathcal{U}-\frac{1}{2}W^A(\partial_A+2\partial_A\b)\mathcal{U}-2\,\mathcal{U}\,n_AW^A\nonumber\\&+e^{-2\b}\left(\mathcal{U}k_{AB}W^AW^B+l_{AB}W^AW^B+W^AW^B{}^{\scriptscriptstyle(d-2)}D_AW_B\right)\nonumber\\
&\Gamma^r_{ur}= \frac{1}{2}\left(2\partial_r\b+\partial_r\right)\mathcal{U}-\left(\partial_A\b+n_A\right)W^A\,,\quad  \nonumber\\
&
\Gamma^r_{uA}=\frac{1}{2}(\partial_A+2n_A)\mathcal{U}-\frac{1}{2}e^{-2\b}W^B\left({}^{\scriptscriptstyle(d-2)}D_AW_B+{}^{\scriptscriptstyle(d-2)}D_BW_A+2l_{AB}-2k_{AB}\mathcal{U}\right) \nonumber \\
&\Gamma^r_{rr}=2\partial_r\b\,,\quad \Gamma^r_{rA}=\frac{1}{2}e^{-2\b}g_{AC} \partial_r W^C+\partial_A\b\,,\nonumber\\
& \Gamma^r_{AB}=\frac{e^{-2\b}}{2}\left({}^{\scriptscriptstyle(d-2)}D_BW_A+{}^{\scriptscriptstyle(d-2)}D_AW_B+2l_{AB}-2\mathcal{U}k_{AB}\right)\nonumber\\
\end{eqnarray}
\begin{eqnarray}
&\Gamma^A_{uu}=2W^A\partial_u\b-\frac{1}{2}W^A(\partial_r+2\partial_r\b)\mathcal{U}+2n_BW^BW^A-\partial_uW^A\nonumber-2l^A_CW^C\\
&+e^{-2\b}k_{BC}W^AW^BW^C+\frac{1}{2}e^{2\b}(2\partial^A\b+\partial^A)\mathcal{U}+\frac{1}{2}{}^{\scriptscriptstyle(d-2)}D^A(W^2)\nonumber\\
&\Gamma^{A}_{ur}=-k^A_CW^C+e^{2\b}\partial^A\b-\frac{1}{2}\delta^A_C\partial_rW^C=-k^A_CW^C+e^{2\b}(\partial^A\b-n^A)\,,\nonumber\\
& \Gamma^A_{uB}= W^A(\partial_B\b-n_{B})-e^{-2\b}k_{BC} W^AW^C+ l^A_B+\frac{1}{2}{}^{\scriptscriptstyle(d-2)}D^AW_B-\frac{1}{2}{}^{\scriptscriptstyle(d-2)}D_BW^A\nonumber\\
&\Gamma^A_{rr}=0\,,\quad \Gamma^A_{rB}=k^A_B\,,\quad  \Gamma^A_{BC}=e^{-2\b}W^Ak_{BC}+{}^{\scriptscriptstyle(d-2)}\Gamma^A_{BC}\,,\nonumber
\end{eqnarray}

The relevant Ricci tensor components giving rise to the main equations are
\begin{equation}\label{eq.RrrApp}
R_{rr}=\mathcal{O}_r\Gamma^r_{rr}-\partial_r^2\mathcal{O}_2-\Gamma^r_{rr}\Gamma^r_{rr}-\Gamma^r_{rA}\Gamma^A_{rr}+\Gamma^B_{rA}\Gamma^A_{Br}
\end{equation}
\begin{equation}\label{eq.RrAApp}
R_{rA}=\mathcal{O}_r\Gamma^r_{rA}+\mathcal{O}_B\Gamma^B_{rA}-\partial_r\partial_A\mathcal{O}_2-\Gamma^r_{rr}\Gamma^r_{rA}-\Gamma^r_{rB}\Gamma^B_{rA}+\Gamma^C_{rB}\Gamma^B_{CA}+\Gamma^C_{ru}\Gamma^u_{CA}
\end{equation}
\begin{eqnarray}\label{eq.RabApp}
R_{AB}&=& \mathcal{O}_u \Gamma^u_{AB}+\mathcal{O}_r \Gamma^r_{AB}+\mathcal{O}_C\Gamma^C_{AB}-\partial_A\partial_B\mathcal{O}_2\\&-&\Gamma^u_{Au}\Gamma^u_{uB}-\Gamma^u_{AC}\Gamma^C_{uB}-\Gamma^r_{AC}\Gamma^C_{rB}-\Gamma^r_{Ar}\Gamma^r_{rB}-\Gamma^C_{Au}\Gamma^u_{CB}-\Gamma^C_{Ar}\Gamma^r_{CB}-
\Gamma^C_{AD}\Gamma^D_{CB}\nonumber 
\end{eqnarray}
The components \eqref{eq.RrrApp},\eqref{eq.RrAApp} are easily evaulated. The symbol $K^A_{B}$ can be further defined as $K^A_B=\frac{r^2}{2}h^{AC}\partial_rh_{CB}$ so that $k^A_B=\frac{1}{r}\delta^A_B+\frac{1}{r^2}K^A_B$ and $R_{rr}=\frac{1}{2(d-2)r^3}K^A_BK^B_A$ can be directly compared with (4.33) of \cite{Barnich2010}.

In \ref{eq.RabApp}, the third, fourth and latter terms can be arranged to
\begin{eqnarray}
\mathcal{O}_C\Gamma^C_{AB}-\partial_A\partial_B\mathcal{O}_2-
\Gamma^C_{AD}\Gamma^D_{CB}&=&{}^{\scriptscriptstyle(d-2)}R_{AB}+\partial_C\left(e^{-2\b}W^Ck_{AB}\right)\\&+&2e^{-2\b}\partial_C\b W^Ck_{AB}+{}^{\scriptscriptstyle(d-2)}\Gamma^D_{DC}e^{-2\b}W^Ck_{AB}\nonumber\\&+&2{}^{\scriptscriptstyle(d-2)}\Gamma^C_{AB}\partial_C\b-2\partial_A\partial_B\b-e^{-4\b}W^CW^Dk_{AD}k_{CB}\nonumber\\
&-&e^{-2\b}W^Ck_{AD}{}^{\scriptscriptstyle(d-2)}\Gamma^D_{CB}-e^{-2\b}W^Ck_{DB}{}^{\scriptscriptstyle(d-2)}\Gamma^D_{AC}.\nonumber
\end{eqnarray}
so that
\begin{eqnarray}
R_{AB}&=&\left(\partial_r+2\partial_r\b+\frac{d-2}{r}\right)\Gamma^r_{AB}+\left(\partial_u+2\partial_u\b+l\right)\Gamma^u_{AB}-2^{(d-2)}D_A\partial_B\b+{}^{\scriptscriptstyle(d-2)}R_{AB}\nonumber\\
&+&{}^{\scriptscriptstyle(d-2)}D_C\left(e^{-2\b}W^Ck_{AB}\right)+2e^{-2\b}\partial_C\b W^Ck_{AB}-e^{-4\b}W^CW^Ek_{AE}k_{CB}\nonumber\\
&-& \Gamma^u_{Au}\Gamma^u_{uB}-\Gamma^u_{AC}\Gamma^C_{uB}-\Gamma^r_{AC}\Gamma^C_{rB}-\Gamma^r_{Ar}\Gamma^r_{rB}-\Gamma^C_{Au}\Gamma^u_{CB}-\Gamma^C_{Ar}\Gamma^r_{CB}
\end{eqnarray}
and by contraction 
\begin{eqnarray}
g^{DA}R_{AB}&=&{}^{\scriptscriptstyle(d-2)}R^D_B-2\left({}^{\scriptscriptstyle(d-2)}D^D\partial_B\b+\partial^D\b\partial_B\b+n^Dn_B\right)\nonumber\\
&+& e^{-2\b}\left(\partial_r +\frac{d-2}{r}\right)\left(\frac{1}{2}{}^{\scriptscriptstyle(d-2)}D^DW_B+\frac{1}{2}{}^{\scriptscriptstyle(d-2)}D_BW^D+l^D_B-k^D_B\mathcal{U}\right)\\
&+&e^{-2\b}\left[\left(\partial_u+l\right)k^D_B+{}^{\scriptscriptstyle(d-2)}D_C(W^Ck^D_B)+k^D_A{}^{\scriptscriptstyle(d-2)}D_BW^A-k^A_B{}^{\scriptscriptstyle(d-2)}D_AW^B\right]\nonumber
\end{eqnarray}
leading to \eqref{eq.traceless}.
\section{Recursive formulae for power-law seed}\label{app.rf}
In this Appendix we write the asymptotic expansions of the metric functions starting from a  non-polyhomogeneous $h_{AB}$. We warn the reader that in producing the closed expressions the objects with more than one index have been manipulated as functions: this highlights the orders but care is needed when reading off the coefficients of such orders. 
\\
\\
Given $h_{AB}$ as \eqref{eq.exp} 
\begin{equation}\label{eq.expApp}
h_{AB}(u,r,x)=h_{(0)AB}(u,x)+\sum_{p}\frac{h_{(a+p)AB}(u,x)}{r^{a+p}}.
\end{equation} 
its inverse is recursively given by
\begin{eqnarray}
h^{-1}&=&h_{(0)}^{-1}+\sum_{n=1}^{\lfloor a+p_{o}\rfloor}(-1)^n\left(\sum_{p}^{p_o}\frac{h_{(0)}^{-1}h_{(a+p)}}{r^{a+p}}\right)^nh_{(0)}^{-1}\nonumber\\
&=& h_{(0)}^{-1}+\sum_{n}^{\lfloor a+p_{o}\rfloor}\frac{(-1)^n}{r^{an}}\sum_{\oplus j_{t}=n}\binom{n}{j_{[0,p_o]}}\prod_{t=0}^{p_o}\left(\frac{h_{(0)}^{-1}h_{(a+t)}}{r^t}\right)^{j_t}h_{(0)}^{-1}
\end{eqnarray}
where $n$ is integer and $\lfloor a+p_{o}\rfloor$ is the floor of the maximal power $a+p_{o}$ we keep in \eqref{eq.expApp}.
Eventually discard from $h^{-1}$ terms of order higher than $a+p_o$. In the second line we have used the multinomial theorem and defined
\begin{equation}
\binom{n}{j_{[0,p_o]}}:=\binom{n}{j_0,\dots,j_t,\dots,j_{p_o}}:=\frac{n!}{\prod_{t=0}^{p_o} (j_{t}!)}
\end{equation}
where $j_{[0,p_o]}$ is the collection of non negative integer indices $j_p$ associated to the object $r^{-p}h_{(0)}^{-1}h_{(a+p)}$ for each $p$. The sum $\sum_{\oplus j_p=n}$ is taken on any combination of $j_p$ such that their total sum $\oplus j_{p}$ is equal to $n$. The tensor indices must match in the contraction and each inverse is taken with respect to $h_{(0)}$, so that
$h_{(0)}h_{(0)}^{-1}=\delta$, $h_{(0)}^{-1}h_{(a+p)}h_{(0)}^{-1}=h_{(a+p)}^{-1}$.

These conventions translates the practice. To find the explicit expressions of the metric functions we have to fix $a$ and expand up to the relevant order, so up to a $p=p_o$. For example in $d=4$ we have $a=1$ and it is appropriate to take the maximal $p$ to be $p_o=2$ (but for many purposes $p_o=1$ suffices in $d=4$). 

The same conventions apply to the expansion of any other object, so in the following we simplify notation where no confusion arise.  

$\tilde{K}^A_B=\frac{1}{2}h^{AC}\partial_rh_{CB}$ now follows as (in manipulating these we pretend that these objects behaves as numbers)
\begin{eqnarray}
\tilde{K}^A_B&=&-\sum_p\frac{a+p}{2r^{a+p+1}}h^A_{(a+p)B}-\frac{1}{2}\sum_n (-1)^n\left(\sum_{p}\frac{h_{(0)}^{-1}h_{(a+p)}}{r^{a+p}}\right)^nh_{(0)}^{-1}\sum_q\frac{a+q}{r^{a+q+1}}h_{(a+q)}\nonumber\\
&=&\frac{1}{r^{a+1}}\sum_{p}\frac{\tilde{K}_{(a+1+p)F}^E}{r^p}\left[\delta^A_E\delta^F_B+\sum_n\frac{(-1)^n}{r^{an}}\left(\sum_q\frac{h_{(a+q)E}^A}{r^q}\right)^n\delta^F_B\right]\nonumber\\
&=&\frac{1}{r^{a+1}}\sum_{p}^{p_o}\frac{\tilde{K}_{(a+1+p)F}^E}{r^p}\left[\delta^A_E\delta^F_B+\sum_{n=1}^{\sim a+p_o}\frac{(-1)^n}{r^{an}}\sum_{\oplus j_t=n}\binom{n}{j_{[0,q_o]}}\prod_{t=0}^{p_o}\left(\frac{h_{(a+q)E}^A}{r^t}\right)^{j_t}\delta^F_B\right]\nonumber\\
&=&\frac{1}{r^{a+1}}\sum_{p}^{p_o}\frac{\tilde{K}_{(a+1+p)F}^E}{r^p}\left[\delta^A_E\delta^F_B+\sum_{n=1}^{\sim a+p_o}\frac{(-1)^n}{r^{an}}\sum_{\oplus j_t=n}\binom{n}{j_{[0,q_o]}}r^{-\sum t j_t}\prod_{t=0}^{p_o}(h_{(a+q)E}^A)^{j_t}\delta^F_B\right]\nonumber\\
&&
\end{eqnarray}
at convenience we reshuffle indices and write
\begin{equation}
\tilde{K}^A_B = \frac{1}{r^{a+1}}\sum_{p=0}\frac{K^A_{(p)B}}{r^p}
\end{equation}
Christoffel symbols are expanded as
\begin{align}
{}^{\scriptscriptstyle(d-2)}\Gamma^A_{BC}&=\overset{\scriptscriptstyle (0)}{\G}{}^A_{BC}+\frac{1}{2}\sum_p\frac{1}{r^{a+p}}h_{(0)}^{AD}\left(\partial_Bh_{(a+p)DC}+\partial_Ch_{(a+p)BD}-\partial_Dh_{(a+p)BC}\right)\\\
&+\frac{1}{2}\sum_{n=1}(-1)^n\left(\sum_{p}\frac{h_{(0)}^{AE}h_{(a+p)EF}}{r^{a+p}}\right)^nh_{(0)}^{FD}\left(\partial_Bh_{(0)DC}+\partial_Ch_{(0)BD}-\partial_Dh_{(0)BC}\right)\nonumber\\
+\frac{1}{2}\sum_{n=1}&(-1)^n\left(\sum_{p}\frac{h_{(0)}^{AE}h_{(a+p)EF}}{r^{a+p}}\right)^nh_{(0)}^{FD}\sum_{q=1}\frac{1}{r^{a+q}}\left(\partial_Bh_{(a+q)DC}+\partial_Ch_{(a+q)BD}-\partial_Dh_{(a+q)BC}\right)\nonumber\nonumber
\end{align}
and appropriately rewrites each order in terms of the quantites at the previous orders, so for example
\begin{equation}
{}^{(a)}\Gamma^A_{BC}=\frac{1}{2}\left(\overset{\scriptscriptstyle (0)}{D}_Bh^A_{(a)C}+\overset{\scriptscriptstyle (0)}{D}_Ch^A_{(a)B}-\overset{\scriptscriptstyle (0)}{D}{}^Ah_{(a)BC}\right).
\end{equation}
Curvature tensors and scalar are expanded as
\begin{equation}
{}^{\scriptscriptstyle(d-2)}R=r^{-2}\overset{\scriptscriptstyle (0)}{R}+r^{-(2+a)}\sum_{p}r^{-p}\overset{\mathrm{(a+p)}}{R}.
\end{equation}

\subsection{First equation: $\b$}
The integrand of \eqref{eq.intb} is
\begin{equation}
r\tilde{K}^A_B\tilde{K}^B_A=\frac{1}{r^{2a+1}}\sum_{p,q} K^A_{(p)B}K_{(q)A}^B 
\end{equation}
and $\b$ is its integral 
As $a$ is fixed, all the terms can be reorganised as
\begin{equation}
\b=\b_{(0)}+\sum_{k>0}\frac{\b_{(2a+k)}}{r^{2a+k}}
\end{equation}
where $k$, as $p$ moves forward by half integer steps if $d$ is odd and by integer steps if $d$ is even.
\subsection{Second equation: $W^A$}\label{app.secondeq}
Referring to \eqref{eq.tilden} we write
\begin{align}
\tilde{n}_A=\frac{\tilde{n}_{(2)A}}{r^2}&+\sum_{p}^{2+a+p\neq d}\frac{\tilde{n}^{\star}_{(2+a+p)A}}{r^{2+a+p}}\nonumber\\
&+\sum_{p}^{2+2a+p\neq d}\frac{\tilde{n}^{\circ}_{(2+2a+p)A}}{r^{2+2a+p}}+\sum_{p,m}^{2+2a+(p+m)\neq d}\frac{\tilde{n}^{\triangleleft}_{(2+2a+(m+p))}}{r^{2+2a+(m+p)A}}+\frac{\tilde{\texttt{n}}_A}{r^d}\log r+\frac{N_A}{r^d}
\end{align}
where the coefficient $\texttt{n}$ of the logarithmic term is given by $\tilde{n}^{\star}$, $\tilde{n}^{\circ}$ $\tilde{n}^{\triangleleft}$ (stripped off the denominators in their expressions) whenever the power in the sum where they appear is equal to $d$
\begin{align}
\tilde{n}^{\star}:& \quad p=\frac{d-2}{2} \quad \text{if} \quad a=\frac{d-2}{2};\quad& p=d-3\quad \text{if} \quad a=1\nonumber\\
\tilde{n}^{\circ}:& \quad p=0 \qquad\quad \text{if} \quad a=\frac{d-2}{2};\quad& p=d-4\quad \text{if} \quad a=1\\
\tilde{n}^{\triangleleft}:& \quad p+m=0\quad \text{if} \quad a=\frac{d-2}{2};\quad& p+m=d-4\quad \text{if} \quad a=1\nonumber
\end{align}
because the denominators in the following expressions vanish
\begin{align}
&\tilde{n}_{(2)A}=-\partial_A\b_{(0)}\\
&\tilde{n}^{\star}_{(2+a+p)A}=-\frac{1}{d-2-(a+p)}	\overset{\scriptsize{(0)}}{D}_{B}\tilde{K}^B_{(a+1+p)A}\\
&\tilde{n}^{\circ}_{(2+2a+p)A}=-\frac{d-2+2a+p}{d-2-(2a+p)}\partial_A \b_{(2a+p)}\\
&\tilde{n}^{\triangleleft}_{(2+2a+(m+p))A}=\frac{1}{d-2-2a-(m+p)}\overset{\scriptsize{(a+p)}}{\G^D}_{BA}\tilde{K}^B_{(a+1+m)D}
\end{align}
In the latter we use the determinant constraint to set $\overset{\scriptsize{(a+p)}}{\G^B}_{BD}=0$ for any $p$.
These expressions are obtained by expanding the integral in \eqref{eq.tilden} using 
 $\mathcal{G}_A $ given in \eqref{eq.RrA} and renaming the various terms with $\tilde{n}^{\star}$, $\tilde{n}^{\circ}$ $\tilde{n}^{\triangleleft}$ according to the number of powers of $a$ and free indices $p,m$. Using the definition of $W^A$
 \begin{equation}\label{eq.WdefApp}
 W^A=2\int dr e^{2\b}h^{AB}\tilde{n}_B
 \end{equation}
the expansion
\begin{equation}
W^A=W^A_{(0)}+\frac{W^A_{(1)}}{r}+\sum_{p=0}^{d-2-a}\frac{W^A_{(a+1+p)}}{r^{a+1+p}}+\frac{1}{r^{d-1}}\left(\mathcal{W}^A_{(d-1)}+\texttt{W}^A_{(d-1+p)}\log r \right)+\dots
\end{equation}
follows. In particular, the first logarithmic coefficient $\texttt{W}_{(d-1)}$ is obtained from $\texttt{n}_{(d)}\log{r}/r^d$
\begin{equation}
\texttt{W}^A_{(d-1)}=-\frac{2}{d-1}e^{2\b_{(0)}}h_{(0)}^{AB}\tilde{n}_{(d)B}
\end{equation}
while 
\begin{equation}
W_{(d-1)}=-\frac{2}{(d-1)^2}e^{2\b_{(0)}}h_{(0)}^{AB}\tilde{n}_{(d)B}+\dots
\end{equation}
where $\dots$ are the terms coming from the integration \eqref{eq.WdefApp} when the integrand is of order $r^{-d}$. This also gives $\mathcal{W}_{(d-1)}^A$ which contains the free function $N^A$ \eqref{eq.mathcalW}.
\subsection{Third equation: $\mathcal{U}$}\label{app.thirdeq}
Now move to the equation for $\mathcal{U}$ \eqref{eq.trace}. Using the above results, $\mathcal{F}$ is expanded
\begin{equation}
\mathcal{F}=\frac{\mathcal{F}_{(1)}}{r}+\frac{\mathcal{F}_{(2)}}{r^2}+\sum_p\left(\frac{\mathcal{F}_{(a+p+2)}}{r^{a+p+2}}+\frac{1}{r^{d+p}}\left(\mathcal{F}_{(d+p)}+\log r{\texttt{F}}_{(d+p)}[\texttt{W}]\right)\right)+\dots,
\end{equation}
where we specify that the coefficient of the logarithmic term here depends on that of $W^A$. The integrand in \eqref{eq.intU} is thus
\begin{equation}\label{eq.integrandUexpanded}
r^{d-2} \mathcal{F}=r^{d-3}\mathcal{F}_{(1)}+r^{d-4}\mathcal{F}_{(2)}+\sum_p r^{d-4-a-p}\mathcal{F}_{(a+p+2)}+r^{-2-p}\left(\mathcal{F}_{(d+p)}+\log r\texttt{F}_{(d+p)}[\texttt{W}]\right)+\dots\,.
\end{equation}
Notice that $d-4-a-p>-2-p$ for both values of $a$ we are considering, while
\begin{equation}
d-4-a-p=-1\Leftrightarrow p=d-a-3\Rightarrow p=\begin{cases}
d-\frac{d-2}{2}-3=\frac{d-4}{2}\\
d-1-3=d-4
\end{cases}
\end{equation}
So the solution \eqref{eq.intU} of \eqref{eq.trace} includes a logarithmic term independent of $\texttt{W}$ and appearing before the one induced by $\texttt{W}$. The expansion is thus organised as
\begin{equation}\label{eq.UsolApp}
\mathcal{U}=r \mathcal{U}_{(-1)}+\mathcal{U}_{(0)}+\sum_{p=0}^{a+p<d-3}\frac{\mathcal{U}_{(a+p)}}{r^{a+p}}+\frac{1}{r^{d-3}}\left(\mathcal{U}_{(d-3)}+\texttt{U}_{(d-3)}\log r\right)+\dots
\end{equation}
with
\begin{align}
&\mathcal{U}_{(-1)}=\frac{1}{(d-2)^2}\mathcal{F}_{(1)}\\
&\mathcal{U}_{(0)}=\frac{1}{(d-2)(d-3)}\mathcal{F}_{(2)}\\
&\mathcal{U}_{(a+p)}=\frac{1}{(d-2)(d-3-(a+p))}\mathcal{F}_{(a+p+2)}\\
&\texttt{U}_{(d-3)}=\frac{1}{d-2}\mathcal{F}_{(a+p+2)}|_{p=d-3-a}
\end{align}

\section{Other ansatzes}
To compare the result \eqref{eq.exph} with other ansatzes used in literature we collect the most used here.
\paragraph{Radiative falloff ansatz.}
The radiative falloff ansatz used in \cite{Tanabe2010,Tanabe2011} is
\begin{equation}
h_{AB}=h_{(0)AB}+\sum_{p}\frac{h_{(\frac{d-2}{2}+p)AB}(u,x)}{r^{\frac{d-2}{2}+p}},
\end{equation}
where $p\in \mathbb{N}_0$ if $d$ is even and $p \in\mathbb{N}_0/2$ if $d$ is odd.  In some places, some arguments have been made to  further impose the vanishing of some of the coefficients of $h_{AB}$ \cite{Tanabe2010}.

In literature we usually also find other kinds of considerations that restrict the orders at which both integer and half-integer powers coexist in odd $d$. We may argue that the expansion only contains half-integer powers of $r$ up to a certain point at which integer powers starts to contribute \cite{Tanabe2011,Godazgar2012,Wald2019}. Indeed, the mixture of both integer and half-integer powers can be attributed to non-linear effects, which are supposed to be negligible asymptotically.

For example, Wald and Satishchandran \cite{Wald2019} (not working in Bondi gauge) considered the following ansatz for odd $d$
\begin{equation}\label{eq.ansatzWald}
g_{\mu\nu}=\eta_{\mu\nu}+G_{\mu\nu}\,,\quad G_{\mu\nu}=\sum_{n=\frac{d-2}{2}}r^{-n}G^{(n)}_{\mu\nu}+\sum_{m=d-3}r^{-m}\tilde{G}^{(m)}_{\mu\nu}\,,
\end{equation}
where $n$ is half-integer and $m$ is integer and both sums proceed with unity steps. Thus, here the integer powers enter starting from the Coulombic order. It is important to stress - following \cite{Godazgar2012}, that it is not known when exactly the nonlinearities mixing integer and half-integer expansions kicks in. It is possible that they appear before the Coulombic order. 

\paragraph{Polyhomogeneous expansion.}
In four spacetime dimensions, the most general ansatz that has been used is
\begin{eqnarray}\label{eq.pol}
h_{AB}(u,r,x)&=&h_{(0)AB}(u,x)+\sum_{p}\frac{H_{(a+p)AB}(u,r,x)}{r^{a+p}}\,,\nonumber\\ H_{(a+p)AB}(u,r,x)&=&h_{(a+p)AB}(u,x)+\sum_j\log r^j \texttt{h}_{(a+p,j)AB}(u,x)
\end{eqnarray}
with $a=1$ $j\in \mathbb{N}_0$ (see for example \cite{Kroon2001,Kroon:1998tu}). We could repeat the analysis in higher dimensions with this ansatz and $a=\frac{d-2}{2}$ or $a=1$. In the main text we have shown where the first logarithmic term appears.

\section{Null infinity, superrotations and extended Carroll structures}\label{app.space}
\paragraph*{Definition.} An asymptote of a $d$-dimensional spacetime $(\bar{M},\bar{g})$ is a triplet $({M},{g},\Omega)$ plus a diffeomorphism $\psi:\bar{M}\rightarrow{M}\backslash\mathscr{I}$, where ${M}$ is a manifold with boundary $\mathscr{I}$, $\bar{g}$ is a smooth metric on ${M}$, $\psi$ identifies $M$ with the interior ${M}\backslash\mathscr{I}$ of ${M}$, and $\Omega:{M}\rightarrow\mathbb{R}$ is a smooth function which is strictly positive in the interior ${M}\backslash \mathscr{I}$ and such that
\begin{itemize}
	\item[i)] ${g}_{\mu\nu}=\Omega^2\bar{g}_{\mu\nu}$ on $\bar{M}$,
	\item[ii)] $\Omega=0$, $n:=d\Omega\neq0$ at $\mathscr{I}$.
\end{itemize}
This is a slight adaptation of Geroch's definition of asymptote of a spacetime \cite{Geroch1977}, where we take $d$ generic rather than $d=4$. The smoothness assumption is to be contrasted with the polyhomogeneity of the expansions. The last condition implies that ${g}$ is finite at infinity, $\Omega$ can be used as a coordinate on ${M}$ and defines the one form $n$ which is associated with the normal $n^\mu={g}^{\mu\nu}\bar{D}_\nu \Omega$ to the boundary. Manifestly, the given definition does not totally determine topology of $\mathscr{I}$. When $\Lambda=0$ the boundary topology is automatically restricted.

\paragraph*{Conformal freedom.} Given a spacetime $(\bar{M},\bar{g})$ and an asymptote $(M,g,\Omega)$ and any smooth positive scalar function $\omega$ on $M$, $(M,\omega^2 g,\omega\Omega)$ is an equivalent asymptote. Under the conformal transformation
\begin{equation}
\Omega\rightarrow \Omega':=\omega \Omega,\quad g_{\m\n}\rightarrow  g_{\m\n}'= \omega^2g_{\m\n}
\end{equation}
the normal $n^\m$ transforms as
\begin{equation}
n^\mu\rightarrow n'^\mu=\omega^{-1}n^\m+\omega^{-2}\Omega d^\m \omega.
\end{equation}
The function $\Omega$ is called ``defining function'' in mathematics literature \cite{FG1985} and the gauge/gravity duality literature \cite{Witten1998,SkenderisLec}. 
\paragraph*{Causal structure of the boundary.} Einstein's equations have to be imposed to infer the causal nature of $\mathscr{I}$. The boundary $\mathscr{I}$ is timelike if the spacetime solves Einstein's equations with $\Lambda<0$, null if $\Lambda=0$ and spacelike if $\Lambda>0$, because
\begin{equation}
|d\Omega|^2=-\frac{2\Lambda}{(d-1)(d-2)}=\mp\frac{1}{l^2},\quad |d\Omega|^2=g^{\m\n}\partial_\m\Omega\partial_\n\Omega
\end{equation}
where we have used the relation between $\Lambda$ and the characteristic length scale $l$ and the sign is $-$ is for $\Lambda>0$ and $+$ for $\Lambda<0$. 

The given relationship is true in vacuum or as long as the stress-energy tensor falloffs sufficiently fast at infinity. Some of the metrics discussed in this paper will not satisfy these conditions, as also cosmic strings - which play a role in the context of superrotations - do not \cite{Bicak1989,Strominger2017}; they are called ``asymptotically locally flat''.

\paragraph*{Universal structure.} The smoothness assumption is necessary to provide the necessary analytical tools to do tensor analysis on $\mathscr{I}$ as induced from $M$, but taking $\mathscr{I}$ abstractly as ``detached'' from $M$: i.e. we can safely define a pullback operation from $M$ to $\mathscr{I}$. We denote the pulled-back quantities with an over arrow pointing left, i.e. $\overset{\leftarrow}{{g}}_{\mu\nu},\; \overset{\leftarrow}{{n}}^{\mu}$.

The pulled-back fields which are shared by all spacetimes in the same class (i.e. asymptotically flat or asymptotically $(A)dS$) define the \emph{universal geometry}. Asymptotic symmetries preserve the universal geometry.

As discussed in the main text, the smoothness assumption overrestrict the space of solutions of both asymptotically flat and asymptotically AdS spacetimes and cannot be extended to odd $d>4$ asymptotically flat radiative spacetimes \cite{Hollands2003}. Cases with lower regularity $\mathscr{I}$ have been studied in literature.

\paragraph*{Topology of null infinity.} Since $\mathscr{I}$ is null, its normal $\overset{\leftarrow}{{n}}^{\mu}$ is both null and tangent and $\overset{\leftarrow}{{g}}_{\m\n}\overset{\leftarrow}{{n}}^{\n}=0$, meaning that $\overset{\leftarrow}{{g}}_{\m\n}$ is degenerate. The set $\mathbb{B}$ of all maximally extended integral curves of $\overset{\leftarrow}{{n}}^{\mu}$ can be given the structure of a manifold provided that for any given point $p$ along one such curve, the curve itself does not reenters sufficiently small neighborhood of $p$. This is accomplished by the mapping $\Pi:\mathscr{I}\rightarrow\mathbb{B}$ sending each $p\in\mathscr{I}$ to the integral curve to which it lies.
The manifold $\mathbb{B}$ is the base space of $\mathscr{I}$ and by abuse of terminology $\mathbb{B}$ is a cross section of $\mathscr{I}$. The topology of $\mathscr{I}$ is
\begin{equation}
\mathscr{I}\sim \mathbb{R}\times\mathbb{B}
\end{equation}
The usual definition of asymptotic flatness in $d=4$ and by extension $d>4$ takes $\mathbb{B}=S^{d-2}$ and the null generators of $\mathscr{I}$ to be complete. This is the \emph{asymptotically Minkowski} case.
\paragraph*{Bondi condition.} 
Due to the conformal freedom and Einstein's equations, $\overset{\leftarrow}{{g}} $ and $\overset{\leftarrow}{{n}}$ are related by $\mathfrak{L}_{\overset{\leftarrow}{{n}}}\overset{\leftarrow}{{g}}_{\m\n}=b\overset{\leftarrow}{{g}}_{\m\n}$for a positive function $b$. It is always possible to find $b$ locally such that $\mathfrak{L}_{\overset{\leftarrow}{{n}}}\overset{\leftarrow}{{g}}_{\m\n}=0$. This defines $\Omega$ as
\begin{equation}
D_\m n_\n=0\Leftrightarrow D_\m D_\n \Omega=0
\end{equation}
on $\mathscr{I}$ and defines the so-called Bondi frame. In this frame there is a residual conformal freedom given by
\begin{equation}                
\mathfrak{L}_n \overset{\leftarrow}{\omega} =0,\quad \overset{\leftarrow}{\omega}>0
\end{equation}
This is sufficient to show that in $d=4$ all asymptotically flat spacetimes have locally the same conformally flat boundary metric \cite{Geroch1977} and when $\mathbb{B}=S^2$ a natural choice is the standard round sphere metric. 

The phase space of asymptotically flat spacetimes is usually defined by such condtions. This immediately lead to BMS without CL-superrotation. Apart from the spherical case, the group of conformal motions of the other possible simply connected $\mathbb{B}^2$ has been studied in \cite{Foster1987}, but the analysis of asymptotic symmetries and charges in such cases lack. An explicit example of a spacetime with a non simply connected null boundary was found \cite{Schmidt1996} as an $A$-metric with toroidal $\mathbb{B}^2$.

\paragraph*{Abstract $\mathscr{I}$ as a Carroll structure.} A Carroll manifold is defined in \cite{Duval2014b} as a triple $(C,q,\c)$, where $C$ is a smooth $(d-1)$-dimensional manifold endowed with a twice-symmetric covariant positive tensor field $q$ whose kernel is generated by the nowhere vanishing, complete vector field $\x$. 

A generic Carroll structure is given by $C^{d-1}=\mathbb{B}^{d-2}\times \mathbb{R}$, $\c=\partial_s$ where $s$ is the $d$th coordinate known as Carrollian time. The standard Carroll manifold is defined by $\mathbb{B}^{d-2}=\mathbb{R}^{d-2}$ and $q_{\m\n}=\delta_{\m\n}$ (notice we use the same indices as before for brevity).

The isometry group of the Carroll manifold is the infinite dimensional group of transformations $x'^A=x^A$, $s'=s+\alpha(x)$. A conformal Carroll transformation of level $N$ is defined as the group of transformations preserving the tensor
\begin{equation}\label{eq.G}
\Gamma_{(N)}=q\otimes \c^{\otimes N}=q_{\m\n}\c^{\r_1}\dots \c^{\r_{N}},
\end{equation}
The conformal Carroll group transforms $q$ and $\c$ as
\begin{equation}\label{eq.Ctensor}
q_{\m\n}\rightarrow a^2q_{\m\n}\,,\quad \c^\m \rightarrow a^{-2/N}\c^\m\,.
\end{equation}
When $N=2$, \eqref{eq.G} with the identification of $C\equiv \mathscr{I}$ and $q\equiv \overset{\leftarrow}{{g}} $, $\c \equiv \overset{\leftarrow}{{n}}$ ($s\equiv u$), $a\equiv \overset{\leftarrow}{\omega}$ is the universal geometry of null infinity as defined by Geroch \cite{Geroch1977}. 

Conformal Carroll transformations of level two are (standard) BMS transformations \cite{Duval2014}. The interesting insight provided by the Carrollian language is that BMS arise as (a conformal extension) of a In\"onu-Wigner contraction of the Poincaré group.

The point to be stressed, however, is that null infinity in the conformal sense is only well defined for $d=4$ (or even). The identifications we made here between the abstract fields on the Carroll manifold (i.e. abstract $\mathscr{I}$) and the pull-backs of bulk fields are only allowed when the pull-back operation can be given a meaning.


\paragraph*{Extended Carroll structures and CL-superrotations.} 
The Carrollian picture can be easily extended to explicitly include CL-superrotations. We define the \emph{extended conformal Carroll group} of level $P$ as the group of transformations preserving the tensor
\begin{equation}\label{eq.extCtensor}
\tilde{\Gamma}_{(P)}=\epsilon\otimes\c^{\otimes P}=\epsilon_{\m_1\dots\m_{d-1}}\c^{\r_{1}}\dots\c^{\r_{P}}
\end{equation}
where $\epsilon$ is the volume element on $C$. The infinitesimal transformation acts on $\epsilon$ and $\c$ as
\begin{equation}\label{eq.cl}
\mathfrak{L}_\x \epsilon_{\m_1\dots\m_{d-1}}=\lambda \epsilon_{\m_1\dots\m_{d-1}},\quad \mathfrak{L}_\x \chi^{\m}=k \chi^{\m},\quad k=-\frac{\l}{P}
\end{equation}
If the Carroll manifold is the null boundary of a spacetime $C\equiv \mathscr{I}$ and we take $P=d-1$ with $\chi$ identified with the normal to $\mathscr{I}$, and $\epsilon$ taken as the pullback of the $(d-1)$-form induced by the spacetime volume element $\epsilon_{\m_1,\dots \m_{d}}=d\epsilon_{[\m_1\dots\m_{d-1}}n_{\m_{d}]}$, \eqref{eq.cl} constitute BMS extended with CL-superrotations. Indeed the above identifications correspond to choosing the normal and the induced volume form to $\mathscr{I}$ as universal structure of asymptotically flat spacetimes \cite{Flanagan2019}.

\bibliography{article_sr}

\providecommand{\href}[2]{#2}\begingroup\raggedright\begin{thebibliography}{100}

\bibitem{Sen2017}
A.~Laddha and A.~Sen, \emph{{Sub-subleading Soft Graviton Theorem in Generic
  Theories of Quantum Gravity}},
  \href{http://dx.doi.org/10.1007/JHEP10(2017)065}{\emph{JHEP} {\bf 10} (2017)
  065}, [\href{https://arxiv.org/abs/1706.00759}{{\tt 1706.00759}}].

\bibitem{Sen2017b}
S.~Chakrabarti, S.~P. Kashyap, B.~Sahoo, A.~Sen and M.~Verma, \emph{{Subleading
  Soft Theorem for Multiple Soft Gravitons}},
  \href{http://dx.doi.org/10.1007/JHEP12(2017)150}{\emph{JHEP} {\bf 12} (2017)
  150}, [\href{https://arxiv.org/abs/1707.06803}{{\tt 1707.06803}}].

\bibitem{Weinberg1965}
S.~Weinberg, \emph{Infrared photons and gravitons},
  \href{http://dx.doi.org/10.1103/PhysRev.140.B516}{\emph{Phys. Rev.} {\bf 140}
  (1965) B516--B524}.

\bibitem{Cachazo2014}
F.~{Cachazo} and A.~{Strominger}, \emph{{Evidence for a New Soft Graviton
  Theorem}}, {\emph{ArXiv e-prints} (Apr., 2014) },
  [\href{https://arxiv.org/abs/1404.4091}{{\tt 1404.4091}}].

\bibitem{Strominger2014}
A.~Strominger, \emph{On {BMS} invariance of gravitational scattering},
  \href{http://dx.doi.org/10.1007/JHEP07(2014)152}{\emph{JHEP} {\bf
  \textbf{2014}} (2014) 152}, [\href{https://arxiv.org/abs/1312.2229}{{\tt
  1312.2229}}].

\bibitem{He2015}
T.~He, V.~Lysov, P.~Mitra and A.~Strominger, \emph{{BMS} supertranslations and
  weinberg's soft graviton theorem},
  \href{http://dx.doi.org/10.1007/JHEP05(2015)151}{\emph{JHEP} {\bf \textbf{5}}
  (2015) 151}, [\href{https://arxiv.org/abs/1401.7026}{{\tt 1401.7026}}].

\bibitem{Kapec2014}
D.~Kapec, V.~Lysov, S.~Pasterski and A.~Strominger, \emph{{Semiclassical
  Virasoro symmetry of the quantum gravity $ \mathcal{S}$-matrix}},
  \href{http://dx.doi.org/10.1007/JHEP08(2014)058}{\emph{JHEP} {\bf
  \textbf{08}} (2014) }, [\href{https://arxiv.org/abs/1406.3312}{{\tt
  1406.3312}}].

\bibitem{Campiglia2014}
M.~Campiglia and A.~Laddha, \emph{Asymptotic symmetries and subleading soft
  graviton theorem},
  \href{http://dx.doi.org/10.1103/PhysRevD.90.124028}{\emph{Phys. Rev. D} {\bf
  90} (Dec, 2014) 124028}, [\href{https://arxiv.org/abs/1408.2228v3}{{\tt
  1408.2228v3}}].

\bibitem{Bondi1962}
H.~Bondi, M.~van~der Burg and A.~Metzner, \emph{Gravitational waves in general
  relativity. {VII}. waves from axi-symmetric isolated systems},
  \href{http://dx.doi.org/10.1098/rspa.1962.0161}{\emph{Proc. R. Soc. Lond. A}
  {\bf \textbf{269}} (1962) 21--52}.

\bibitem{Sachs1962}
R.~Sachs, \emph{Asymptotic symmetries in gravitational theory},
  \href{http://dx.doi.org/10.1103/PhysRev.128.2851}{\emph{Phys. Rev.} {\bf
  \textbf{128}} (1962) 2851--2864}.

\bibitem{Geroch1977}
R.~Geroch, \emph{Asymptotic structure of space-time},  in \emph{Asymptotic
  Structure of Space-Time} (F.~P. Esposito and L.~Witten, eds.), pp.~1--105.
\newblock Springer US, 1977.
\newblock \href{http://dx.doi.org/10.1007/978-1-4684-2343-3_1}{DOI}.

\bibitem{Barnich2010}
G.~Barnich and C.~Troessaert, \emph{Aspects of the {BMS/CFT} correspondence},
  \href{http://dx.doi.org/10.1007/JHEP05(2010)062}{\emph{JHEP} {\bf \textbf{5}}
  (2010) }, [\href{https://arxiv.org/abs/1001.1541}{{\tt 1001.1541}}].

\bibitem{Barnich2011a}
G.~Barnich and C.~Troessaert, \emph{{BMS} charge algebra},
  \href{http://dx.doi.org/10.1007/JHEP12(2011)105}{\emph{JHEP} {\bf
  \textbf{12}} (2011) 105}, [\href{https://arxiv.org/abs/1106.0213}{{\tt
  1106.0213}}].

\bibitem{Campiglia2015}
M.~Campiglia and A.~Laddha, \emph{{New symmetries for the Gravitational
  S-matrix}}, \href{http://dx.doi.org/10.1007/JHEP04(2015)076}{\emph{JHEP} {\bf
  04} (2015) 076}, [\href{https://arxiv.org/abs/1502.02318}{{\tt 1502.02318}}].

\bibitem{Compere2018}
G.~Comp\`ere, A.~Fiorucci and R.~Ruzziconi, \emph{{Superboost transitions,
  refraction memory and super-Lorentz charge algebra}},
  \href{http://dx.doi.org/10.1007/JHEP11(2018)200}{\emph{JHEP} {\bf 11} (2018)
  200}, [\href{https://arxiv.org/abs/1810.00377}{{\tt 1810.00377}}].

\bibitem{Freidel2021}
L.~Freidel, R.~Oliveri, D.~Pranzetti and S.~Speziale, \emph{{The Weyl BMS group
  and Einstein's equations}},  \href{https://arxiv.org/abs/2104.05793}{{\tt
  2104.05793}}.

\bibitem{Flanagan2019}
E.~E. Flanagan, K.~Prabhu and I.~Shehzad, \emph{{Extensions of the asymptotic
  symmetry algebra of general relativity}},
  \href{http://dx.doi.org/10.1007/JHEP01(2020)002}{\emph{JHEP} {\bf 01} (2020)
  002}, [\href{https://arxiv.org/abs/1910.04557}{{\tt 1910.04557}}].

\bibitem{Bicak1989}
J.~Bicak and B.~Schmidt, \emph{On the asymptotic structure of axisymmetric
  radiative spacetimes},
  \href{http://dx.doi.org/10.1088/0264-9381/6/11/010}{\emph{Class. Quantum
  Grav.} {\bf 6} (1989) 1547}.

\bibitem{Bicak1984}
J.~Bičák and B.~G. Schmidt, \emph{Isometries compatible with gravitational
  radiation}, \href{http://dx.doi.org/10.1063/1.526161}{\emph{J. Math. Phys.}
  {\bf 25} (1984) 600--606}.

\bibitem{Zel1974}
Y.~Zel'dovich and A.~Polnarev, \emph{{Radiation of gravitational waves by a
  cluster of superdense stars}}, {\emph{Sov.Ast.} {\bf \textbf{18}} (1974) 17}.

\bibitem{Chris_mem}
D.~Christodoulou, \emph{Nonlinear nature of gravitation and gravitational-wave
  experiments},
  \href{http://dx.doi.org/10.1103/PhysRevLett.67.1486}{\emph{Phys. Rev. Lett.}
  {\bf \textbf{67}} (1991) 1486--1489}.

\bibitem{Strominger2016}
A.~Strominger and A.~Zhiboedov, \emph{Gravitational memory, {BMS}
  supertranslations and soft theorems},
  \href{http://dx.doi.org/10.1007/JHEP01(2016)086}{\emph{JHEP} {\bf
  \textbf{01}} (2016) 86}, [\href{https://arxiv.org/abs/1411.5745}{{\tt
  1411.5745}}].

\bibitem{Strominger2017}
A.~Strominger and A.~Zhiboedov, \emph{Superrotations and black hole pair
  creation}, {\emph{Class. Quantum Grav.} {\bf \textbf{34}} (2017) 064002},
  [\href{https://arxiv.org/abs/1610.00639}{{\tt 1610.00639}}].

\bibitem{Cheung2016iub}
C.~Cheung, A.~de~la Fuente and R.~Sundrum, \emph{{4D scattering amplitudes and
  asymptotic symmetries from 2D CFT}},
  \href{http://dx.doi.org/10.1007/JHEP01(2017)112}{\emph{JHEP} {\bf 01} (2017)
  112}, [\href{https://arxiv.org/abs/1609.00732}{{\tt 1609.00732}}].

\bibitem{Pasterski2017}
S.~Pasterski and S.-H. Shao, \emph{{Conformal basis for flat space
  amplitudes}}, \href{http://dx.doi.org/10.1103/PhysRevD.96.065022}{\emph{Phys.
  Rev. D} {\bf 96} (2017) 065022},
  [\href{https://arxiv.org/abs/1705.01027}{{\tt 1705.01027}}].

\bibitem{Donnay2020}
L.~Donnay, S.~Pasterski and A.~Puhm, \emph{{Asymptotic Symmetries and Celestial
  CFT}}, \href{http://dx.doi.org/10.1007/JHEP09(2020)176}{\emph{JHEP} {\bf 09}
  (2020) 176}, [\href{https://arxiv.org/abs/2005.08990}{{\tt 2005.08990}}].

\bibitem{Bagchi2016}
A.~{Bagchi}, R.~{Basu}, A.~{Kakkar} and A.~{Mehra}, \emph{{Flat holography:
  aspects of the dual field theory}},
  \href{http://dx.doi.org/10.1007/JHEP12(2016)147}{\emph{JHEP} {\bf 12} (2016)
  147}, [\href{https://arxiv.org/abs/1609.06203}{{\tt 1609.06203}}].

\bibitem{Banerjee2020}
S.~Banerjee, S.~Ghosh and R.~Gonzo, \emph{{BMS symmetry of celestial OPE}},
  \href{http://dx.doi.org/10.1007/JHEP04(2020)130}{\emph{JHEP} {\bf 04} (2020)
  130}, [\href{https://arxiv.org/abs/2002.00975}{{\tt 2002.00975}}].

\bibitem{deBoer2003}
J.~de~Boer and S.~N. Solodukhin, \emph{{A Holographic reduction of Minkowski
  space-time}},
  \href{http://dx.doi.org/10.1016/S0550-3213(03)00494-2}{\emph{Nucl. Phys. B}
  {\bf 665} (2003) 545--593}, [\href{https://arxiv.org/abs/hep-th/0303006}{{\tt
  hep-th/0303006}}].

\bibitem{Penedones2017}
J.~{Penedones}, \emph{{TASI Lectures on AdS/CFT}},  in \emph{New Frontiers in
  Fields and Strings (TASI 2015} (J.~{Polchinski} and {et al.}, eds.),
  pp.~75--136, 2017.
\newblock \href{https://arxiv.org/abs/1608.04948}{{\tt 1608.04948}}.
\newblock \href{http://dx.doi.org/10.1142/9789813149441_0002}{DOI}.

\bibitem{Maldacena1997}
J.~M. Maldacena, \emph{{The Large N limit of superconformal field theories and
  supergravity}}, \href{http://dx.doi.org/10.1023/A:1026654312961}{\emph{Int.
  J. Theor. Phys.} {\bf \textbf{38}} (1999) 1113--1133},
  [\href{https://arxiv.org/abs/hep-th/9711200}{{\tt hep-th/9711200}}].

\bibitem{MAGOO}
O.~{Aharony}, S.~S. {Gubser}, J.~{Maldacena}, H.~{Ooguri} and Y.~{Oz},
  \emph{{Large N field theories, string theory and gravity}},
  \href{http://dx.doi.org/10.1016/S0370-1573(99)00083-6}{\emph{Phys. Rept.}
  {\bf 323} (2000) 183--386}, [\href{https://arxiv.org/abs/hep-th/9905111}{{\tt
  hep-th/9905111}}].

\bibitem{WittenStrings}
E.~Witten, \emph{Baryons and branes in anti de sitter space},  in \emph{Strings
  '98}, 1998.
\newblock
  \href{https://arxiv.org/abs/http://online.itp.ucsb.edu/online/strings98/witten/}{{\tt
  http://online.itp.ucsb.edu/online/strings98/witten/}}.

\bibitem{Witten:2001kn}
E.~Witten, \emph{{Quantum gravity in de Sitter space}},  in \emph{{Strings
  2001: International Conference}}, 6, 2001.
\newblock \href{https://arxiv.org/abs/hep-th/0106109}{{\tt hep-th/0106109}}.

\bibitem{SkenderisLec}
K.~Skenderis, \emph{{Lecture notes on holographic renormalization}},
  \href{http://dx.doi.org/10.1088/0264-9381/19/22/306}{\emph{Class. Quant.
  Grav.} {\bf 19} (2002) 5849--5876},
  [\href{https://arxiv.org/abs/hep-th/0209067}{{\tt hep-th/0209067}}].

\bibitem{Caldarelli2013aaa}
M.~M. Caldarelli, J.~Camps, B.~Gout\'eraux and K.~Skenderis,
  \emph{{AdS/Ricci-flat correspondence}},
  \href{http://dx.doi.org/10.1007/JHEP04(2014)071}{\emph{JHEP} {\bf 04} (2014)
  071}, [\href{https://arxiv.org/abs/1312.7874}{{\tt 1312.7874}}].

\bibitem{Bagchi2010}
A.~Bagchi, \emph{{Correspondence between Asymptotically Flat Spacetimes and
  Nonrelativistic Conformal Field Theories}},
  \href{http://dx.doi.org/10.1103/PhysRevLett.105.171601}{\emph{Phys. Rev.
  Lett.} {\bf 105} (2010) 171601}, [\href{https://arxiv.org/abs/1006.3354}{{\tt
  1006.3354}}].

\bibitem{Barnich2012}
G.~Barnich, A.~Gomberoff and H.~A. Gonzalez, \emph{{The Flat limit of three
  dimensional asymptotically anti-de Sitter spacetimes}},
  \href{http://dx.doi.org/10.1103/PhysRevD.86.024020}{\emph{Phys. Rev. D} {\bf
  86} (2012) 024020}, [\href{https://arxiv.org/abs/1204.3288}{{\tt
  1204.3288}}].

\bibitem{Bagchi2013}
A.~Bagchi, S.~Detournay, R.~Fareghbal and J.~Sim\'on, \emph{{Holography of 3D
  Flat Cosmological Horizons}},
  \href{http://dx.doi.org/10.1103/PhysRevLett.110.141302}{\emph{Phys. Rev.
  Lett.} {\bf 110} (2013) 141302}, [\href{https://arxiv.org/abs/1208.4372}{{\tt
  1208.4372}}].

\bibitem{Ciambelli2018b}
L.~Ciambelli, C.~Marteau, A.~C. Petkou, P.~M. Petropoulos and K.~Siampos,
  \emph{{Flat holography and Carrollian fluids}},
  \href{http://dx.doi.org/10.1007/JHEP07(2018)165}{\emph{JHEP} {\bf 07} (2018)
  165}, [\href{https://arxiv.org/abs/1802.06809}{{\tt 1802.06809}}].

\bibitem{Costa2012fm}
R.~N.~C. Costa, \emph{{Holographic Reconstruction and Renormalization in
  Asymptotically Ricci-flat Spacetimes}},
  \href{http://dx.doi.org/10.1007/JHEP11(2012)046}{\emph{JHEP} {\bf 11} (2012)
  046}, [\href{https://arxiv.org/abs/1206.3142}{{\tt 1206.3142}}].

\bibitem{Costa2013vza}
R.~N. Caldeira~Costa, \emph{{Aspects of the zero $\Lambda$ limit in the AdS/CFT
  correspondence}},
  \href{http://dx.doi.org/10.1103/PhysRevD.90.104018}{\emph{Phys. Rev. D} {\bf
  90} (2014) 104018}, [\href{https://arxiv.org/abs/1311.7339}{{\tt
  1311.7339}}].

\bibitem{Crnkovic1988}
C.~Crnkovic, \emph{Symplectic geometry of the convariant phase space},
  {\emph{Class. Quantum Grav.} {\bf \textbf{5}} (1988) 1557--1575}.

\bibitem{Christodoulou1993}
D.~Christodoulou and S.~Klainerman, \emph{The global nonlinear stability of the
  minkowski space}, {\emph{Princeton Math.} {\bf 41} (1993) }.

\bibitem{Chrusciel}
P.~T. Chruściel, M.~A.~H. MacCallum and D.~B. Singleton, \emph{Gravitational
  waves in general relativity {XIV}. bondi expansions and the
  {\textquoteleft}polyhomogeneity{\textquoteright} of scri},
  \href{http://dx.doi.org/10.1098/rsta.1995.0004}{\emph{Phil. Trans. R. Soc.
  Lond. A} {\bf \textbf{350}} (1995) 113--141},
  [\href{https://arxiv.org/abs/gr-qc/9305021}{{\tt gr-qc/9305021}}].

\bibitem{Friedrich2017cjg}
H.~Friedrich, \emph{{Peeling or not peeling\textemdash{}is that the
  question?}}, \href{http://dx.doi.org/10.1088/1361-6382/aaafdb}{\emph{Class.
  Quant. Grav.} {\bf 35} (2018) 083001},
  [\href{https://arxiv.org/abs/1709.07709}{{\tt 1709.07709}}].

\bibitem{Gibbons1976}
G.~W. Gibbons and S.~W. Hawking, \emph{{Action Integrals and Partition
  Functions in Quantum Gravity}},
  \href{http://dx.doi.org/10.1103/PhysRevD.15.2752}{\emph{Phys. Rev. D} {\bf
  15} (1977) 2752--2756}.

\bibitem{Henningson1998}
M.~Henningson and K.~Skenderis, \emph{{The Holographic Weyl anomaly}},
  \href{http://dx.doi.org/10.1088/1126-6708/1998/07/023}{\emph{JHEP} {\bf 07}
  (1998) 023}, [\href{https://arxiv.org/abs/hep-th/9806087}{{\tt
  hep-th/9806087}}].

\bibitem{Wald2000}
R.~Wald and A.~Zoupas, \emph{General definition of ``conserved quantities'' in
  general relativity and other theories of gravity},
  \href{http://dx.doi.org/10.1103/PhysRevD.61.084027}{\emph{Phys. Rev. D} {\bf
  \textbf{61}} (2000) 084027}, [\href{https://arxiv.org/abs/gr-qc/9911095}{{\tt
  gr-qc/9911095}}].

\bibitem{Godazgar2020}
M.~Godazgar and G.~Long, \emph{{BMS charges in polyhomogeneous spacetimes}},
  \href{http://dx.doi.org/10.1103/PhysRevD.102.064036}{\emph{Phys. Rev. D} {\bf
  102} (2020) 064036}, [\href{https://arxiv.org/abs/2007.15672}{{\tt
  2007.15672}}].

\bibitem{FlanaganBMS}
E.~Flanagan and D.~Nichols, \emph{{Conserved charges of the extended
  Bondi-Metzner-Sachs algebra}}, {\emph{Phys. Rev. D} {\bf \textbf{95}} (2015)
  }, [\href{https://arxiv.org/abs/1510.03386}{{\tt 1510.03386}}].

\bibitem{AshTalk}
A.~Ashtekar, ``The bms group, conservation laws, and soft gravitons.''
  \url{http://pirsa.org/16080055}, Talk at the Perimeter Institute (2016).

\bibitem{Prabhu2019}
K.~Prabhu, \emph{{Conservation of asymptotic charges from past to future null
  infinity: Supermomentum in general relativity}},
  \href{http://dx.doi.org/10.1007/JHEP03(2019)148}{\emph{JHEP} {\bf 03} (2019)
  148}, [\href{https://arxiv.org/abs/1902.08200}{{\tt 1902.08200}}].

\bibitem{Hollands2003}
S.~{Hollands} and A.~{Ishibashi}, \emph{{Asymptotic flatness at null infinity
  in higher dimensional gravity}},  in \emph{{Proceedings, 7th Hungarian
  Relativity Workshop (RW 2003)}}, pp.~51--61, 2004.
\newblock \href{https://arxiv.org/abs/hep-th/0311178}{{\tt hep-th/0311178}}.

\bibitem{Hollands2004}
S.~{Hollands} and R.~{Wald}, \emph{{Conformal null infinity does not exist for
  radiating solutions in odd spacetime dimensions}},
  \href{http://dx.doi.org/10.1088/0264-9381/21/22/008}{\emph{Class. Quantum
  Grav.} {\bf \textbf{21}} (2004) 5139--5145},
  [\href{https://arxiv.org/abs/gr-qc/0407014}{{\tt gr-qc/0407014}}].

\bibitem{Hollands2005}
S.~{Hollands} and A.~{Ishibashi}, \emph{{Asymptotic flatness and Bondi energy
  in higher dimensional gravity}},
  \href{http://dx.doi.org/10.1063/1.1829152}{\emph{J. Math. Phys.} {\bf
  \textbf{46}} (2005) }, [\href{https://arxiv.org/abs/gr-qc/0304054}{{\tt
  gr-qc/0304054}}].

\bibitem{Hollands2013cva}
S.~Hollands and A.~Thorne, \emph{{Bondi mass cannot become negative in higher
  dimensions}},
  \href{http://dx.doi.org/10.1007/s00220-014-2096-8}{\emph{Commun. Math. Phys.}
  {\bf 333} (2015) 1037--1059}, [\href{https://arxiv.org/abs/1307.1603}{{\tt
  1307.1603}}].

\bibitem{Tanabe2011}
K.~Tanabe, S.~Kinoshita and T.~Shiromizu, \emph{Asymptotic flatness at null
  infinity in arbitrary dimensions},
  \href{http://dx.doi.org/10.1103/PhysRevD.84.044055}{\emph{Phys. Rev. D} {\bf
  \textbf{84}} (2011) 044055}, [\href{https://arxiv.org/abs/1104.0303v2}{{\tt
  1104.0303v2}}].

\bibitem{Tanabe2012}
K.~Tanabe, S.~Kinoshita and T.~Shiromizu, \emph{{Angular momentum at null
  infinity in higher dimensions}},
  \href{http://dx.doi.org/10.1103/PhysRevD.85.124058}{\emph{Phys. Rev. D} {\bf
  \textbf{85}} (2012) }, [\href{https://arxiv.org/abs/1203.0452}{{\tt
  1203.0452}}].

\bibitem{Kapec2015}
D.~Kapec, V.~Lysov, S.~Pasterski and A.~Strominger, \emph{{Higher-dimensional
  supertranslations and Weinberg\textquoteright{}s soft graviton theorem}},
  \href{http://dx.doi.org/10.4310/AMSA.2017.v2.n1.a2}{\emph{Ann. Math. Sci.
  Appl.} {\bf 02} (2017) 69--94}, [\href{https://arxiv.org/abs/1502.07644}{{\tt
  1502.07644}}].

\bibitem{Aggarwal2019}
A.~Aggarwal, \emph{{Supertranslations in Higher Dimensions Revisited}},
  \href{http://dx.doi.org/10.1103/PhysRevD.99.026015}{\emph{Phys. Rev. D} {\bf
  99} (2019) 026015}, [\href{https://arxiv.org/abs/1811.00093}{{\tt
  1811.00093}}].

\bibitem{Avery2016}
S.~G. Avery and B.~U.~W. Schwab, \emph{Burg-metzner-sachs symmetry, string
  theory, and soft theorems},
  \href{http://dx.doi.org/10.1103/PhysRevD.93.026003}{\emph{Phys. Rev. D} {\bf
  93} (Jan, 2016) 026003}.

\bibitem{CaponeProcBMS}
F.~Capone, \emph{BMS Symmetries and Holography: An Introductory Overview},
  pp.~197--225.
\newblock Springer International Publishing, Birkh\"auser, Cham, 2019.
\newblock
  \href{https://link.springer.com/chapter/10.1007\%2F978-3-030-18061-4_6}{DOMOSCHOOL
  2018.}

\bibitem{Capone2019}
F.~Capone and M.~Taylor, \emph{{Cosmic branes and asymptotic structure}},
  \href{http://dx.doi.org/10.1007/JHEP10(2019)138}{\emph{JHEP} {\bf 10} (2019)
  138}, [\href{https://arxiv.org/abs/1904.04265}{{\tt 1904.04265}}].

\bibitem{Colferai2020}
D.~Colferai and S.~Lionetti, \emph{{Asymptotic symmetries and subleading soft
  graviton theorem in higher dimensions}},
  \href{https://arxiv.org/abs/2005.03439}{{\tt 2005.03439}}.

\bibitem{FG1985}
C.~Fefferman and C.~R. Graham, \emph{Conformal invariants},  in \emph{\'Elie
  Cartan et les math\'ematiques d'aujourd'hui - Lyon, 25-29 juin 1984},
  no.~S131 in Ast\'erisque.
\newblock Soci\'et\'e math\'ematique de France, 1985.

\bibitem{Capone2021}
F.~Capone and A.~Poole, \emph{Holographic anomalies and bondi-sachs
  asymptotics}, {\emph{xxxxxxx} (xxxx) }.

\bibitem{Sachs1962a}
R.~Sachs, \emph{Gravitational waves in general relativity. {VIII}. waves in
  asymptotically flat space-time},
  \href{http://dx.doi.org/10.1098/rspa.1962.0206}{\emph{Proc. R. Soc. Lond. A}
  {\bf \textbf{270}} (1962) 103--126}.

\bibitem{Tanabe2010}
K.~Tanabe, N.~Tanahashi and T.~Shiromizu, \emph{{On asymptotic structure at
  null infinity in five dimensions}},
  \href{http://dx.doi.org/10.1063/1.3429580}{\emph{J. Math. Phys.} {\bf 51}
  (2010) 062502}, [\href{https://arxiv.org/abs/0909.0426}{{\tt 0909.0426}}].

\bibitem{Campoleoni2020}
A.~{Campoleoni}, D.~{Francia} and C.~{Heissenberg}, \emph{{On asymptotic
  symmetries in higher dimensions for any spin}},
  \href{http://dx.doi.org/10.1007/JHEP12(2020)129}{\emph{JHEP} {\bf 2020}
  (2020) 129}, [\href{https://arxiv.org/abs/2011.04420}{{\tt 2011.04420}}].

\bibitem{NU1962}
E.~T. Newman and T.~W.~J. Unti, \emph{Behavior of asymptotically flat empty
  spaces}, \href{http://dx.doi.org/10.1063/1.1724303}{\emph{Journal of
  Mathematical Physics} {\bf 3} (1962) 891--901}.

\bibitem{Foster1987}
J.~{Foster}, \emph{{Asymptotic Symmetry and the Global Structure of Future Null
  Infinity}}, \href{http://dx.doi.org/10.1007/BF00669365}{\emph{Int. J. Th.
  Phys.} {\bf 26} (1987) 1107--1124}.

\bibitem{Kuiper1949}
N.~Kuiper, \emph{On conformally flat spaces in the large},
  \href{http://dx.doi.org/https://doi.org/10.2307/1969587}{\emph{Ann. Math.}
  {\bf 50} (1949) 916--924}.

\bibitem{Bohm1998}
C.~Böhm, \emph{Inhomogeneous einstein metrics on low-dimensional spheres and
  other low-dimensional spaces.},
  \href{http://dx.doi.org/https://doi.org/10.1007/s002220050261}{\emph{Invent.
  math.} {\bf 134} (1998) 145–176}.

\bibitem{Boyer2003}
C.~P. Boyer, K.~Galicki and J.~Kollar, \emph{{Einstein metrics on spheres}},
  \href{https://arxiv.org/abs/math/0309408}{{\tt math/0309408}}.

\bibitem{Campiglia2020qvc}
M.~Campiglia and J.~Peraza, \emph{{Generalized BMS charge algebra}},
  \href{http://dx.doi.org/10.1103/PhysRevD.101.104039}{\emph{Phys. Rev. D} {\bf
  101} (2020) 104039}, [\href{https://arxiv.org/abs/2002.06691}{{\tt
  2002.06691}}].

\bibitem{RT1960}
I.~Robinson and A.~Trautman, \emph{Spherical gravitational waves},
  \href{http://dx.doi.org/10.1103/PhysRevLett.4.431}{\emph{Phys. Rev. Lett.}
  {\bf 4} (Apr, 1960) 431--432}.

\bibitem{Boonstra1999}
H.~J. Boonstra, K.~Skenderis and P.~K. Townsend, \emph{{The domain wall / QFT
  correspondence}},
  \href{http://dx.doi.org/10.1088/1126-6708/1999/01/003}{\emph{JHEP} {\bf 01}
  (1999) 003}, [\href{https://arxiv.org/abs/hep-th/9807137}{{\tt
  hep-th/9807137}}].

\bibitem{Fareghbal2018}
R.~Fareghbal and I.~Mohammadi, \emph{{Flat-space holography and correlators of
  Robinson-Trautman stress tensor}},
  \href{http://dx.doi.org/10.1016/j.aop.2019.167960}{\emph{Annals Phys.} {\bf
  411} (2019) 167960}, [\href{https://arxiv.org/abs/1802.05445}{{\tt
  1802.05445}}].

\bibitem{Podolsky:2006du}
J.~Podolsky and M.~Ortaggio, \emph{{Robinson-Trautman spacetimes in higher
  dimensions}},
  \href{http://dx.doi.org/10.1088/0264-9381/23/20/002}{\emph{Class. Quant.
  Grav.} {\bf 23} (2006) 5785--5797},
  [\href{https://arxiv.org/abs/gr-qc/0605136}{{\tt gr-qc/0605136}}].

\bibitem{HT1987}
P.~Hogan and A.~Trautman, \emph{On gravitational radiation from bounded
  sources},  in \emph{Gravitation and Geometry} (A.~T. W.~Rindler, ed.),
  Bibliopolis, Napoli, 1987.

\bibitem{Hogan1985}
P.~Hogan, \emph{Asymptotic symmetries in general relativity}, {\emph{Lett Math
  Phys} {\bf 10} (1985) 283--288}.

\bibitem{Ashtekar1996}
A.~Ashtekar, J.~Bicak and B.~G. Schmidt, \emph{{Asymptotic structure of
  symmetry reduced general relativity}},
  \href{http://dx.doi.org/10.1103/PhysRevD.55.669}{\emph{Phys. Rev. D} {\bf 55}
  (1997) 669--686}, [\href{https://arxiv.org/abs/gr-qc/9608042}{{\tt
  gr-qc/9608042}}].

\bibitem{AMK}
A.~Poole, K.~Skenderis and M.~Taylor, \emph{{(A)dS$\mathbf{_4}$ in Bondi
  gauge}}, \href{http://dx.doi.org/10.1088/1361-6382/ab117c}{\emph{Class.
  Quant. Grav.} {\bf 36} (2019) 095005},
  [\href{https://arxiv.org/abs/1812.05369}{{\tt 1812.05369}}].

\bibitem{Compere2020}
G.~Comp\`ere, A.~Fiorucci and R.~Ruzziconi, \emph{{The $\Lambda$-BMS$_4$ charge
  algebra}}, \href{http://dx.doi.org/10.1007/JHEP10(2020)205}{\emph{JHEP} {\bf
  10} (2020) 205}, [\href{https://arxiv.org/abs/2004.10769}{{\tt 2004.10769}}].

\bibitem{Kroon2001}
J.~A. Valiente~Kroon, \emph{{Can one detect a nonsmooth null infinity?}},
  \href{http://dx.doi.org/10.1088/0264-9381/18/20/310}{\emph{Class. Quant.
  Grav.} {\bf 18} (2001) 4311--4316},
  [\href{https://arxiv.org/abs/gr-qc/0108049}{{\tt gr-qc/0108049}}].

\bibitem{Chrusciel2010}
P.~T. Chrusciel and R.~T. Wafo, \emph{{Solutions of quasi-linear wave equations
  polyhomogeneous at null infinity in high dimensions}},
  \href{http://dx.doi.org/10.1142/S0219891611002445}{\emph{J. Hyperbol. Diff.
  Equat.} {\bf 8} (2011) 269--346},
  [\href{https://arxiv.org/abs/1010.2387}{{\tt 1010.2387}}].

\bibitem{Ash2018}
A.~{Ashtekar}, M.~{Campiglia} and A.~{Laddha}, \emph{{Null infinity, the BMS
  group and infrared issues}},
  \href{http://dx.doi.org/10.1007/s10714-018-2464-3}{\emph{General Relativity
  and Gravitation} {\bf 50} (Nov., 2018) 140},
  [\href{https://arxiv.org/abs/1808.07093}{{\tt 1808.07093}}].

\bibitem{Ashtekar1978b}
A.~Ashtekar and R.~O. Hansen, \emph{A unified treatment of null and spatial
  infinity in general relativity. i. universal structure, asymptotic
  symmetries, and conserved quantities at spatial infinity},
  \href{http://dx.doi.org/10.1063/1.523863}{\emph{Journal of Mathematical
  Physics} {\bf 19} (1978) 1542--1566},
  [\href{https://arxiv.org/abs/https://doi.org/10.1063/1.523863}{{\tt
  https://doi.org/10.1063/1.523863}}].

\bibitem{Ashtekar1992}
A.~Ashtekar and J.~D. Romano, \emph{Spatial infinity as a boundary of
  spacetime},
  \href{http://dx.doi.org/10.1088/0264-9381/9/4/019}{\emph{Classical and
  Quantum Gravity} {\bf 9} (apr, 1992) 1069--1100}.

\bibitem{Beig1982}
R.~Beig and B.~Schmidt, \emph{Einstein's equations near spatial infinity},
  \href{http://dx.doi.org/https://doi.org/10.1007/BF01211056}{\emph{Commun.Math.
  Phys.} {\bf 87} (1982) 65–80}.

\bibitem{Troessaert2017}
C.~Troessaert, \emph{{The BMS4 algebra at spatial infinity}},
  \href{http://dx.doi.org/10.1088/1361-6382/aaae22}{\emph{Class. Quant. Grav.}
  {\bf 35} (2018) 074003}, [\href{https://arxiv.org/abs/1704.06223}{{\tt
  1704.06223}}].

\bibitem{Nguyen2021}
K.~Nguyen and J.~Salzer, \emph{{Celestial IR divergences and the effective
  action of supertranslation modes}},
  \href{https://arxiv.org/abs/2105.10526}{{\tt 2105.10526}}.

\bibitem{Kroon}
J.~A.~V. {Kroon}, \emph{{A Comment on the Outgoing Radiation Condition for the
  Gravitational Field and the Peeling Theorem}}, {\emph{Gen. Rel. Grav.} {\bf
  \textbf{31}} (Aug., 1999) 1219},
  [\href{https://arxiv.org/abs/gr-qc/9811034}{{\tt gr-qc/9811034}}].

\bibitem{deHaro2001}
S.~de~Haro, K.~Skenderis and S.~N. Solodukhin, \emph{Holographic reconstruction
  of spacetime and renormalization in the ads/cft correspondence},
  \href{http://dx.doi.org/10.1007/s002200100381}{\emph{Comm. Math. Phys.} {\bf
  217} (2001) 595--622}, [\href{https://arxiv.org/abs/hep-th/0002230}{{\tt
  hep-th/0002230}}].

\bibitem{Hollands2016}
S.~{Hollands}, A.~{Ishibashi} and R.~M. {Wald}, \emph{{BMS supertranslations
  and memory in four and higher dimensions}}, {\emph{Class. Quantum Grav.} {\bf
  \textbf{34}} (2017) }, [\href{https://arxiv.org/abs/1612.03290}{{\tt
  1612.03290}}].

\bibitem{deBoer2000}
J.~de~Boer, E.~P. Verlinde and H.~L. Verlinde, \emph{{On the holographic
  renormalization group}},
  \href{http://dx.doi.org/10.1088/1126-6708/2000/08/003}{\emph{JHEP} {\bf 08}
  (2000) 003}, [\href{https://arxiv.org/abs/hep-th/9912012}{{\tt
  hep-th/9912012}}].

\bibitem{Fareghbal2013}
R.~Fareghbal and A.~Naseh, \emph{{Flat-Space Energy-Momentum Tensor from
  BMS/GCA Correspondence}},
  \href{http://dx.doi.org/10.1007/JHEP03(2014)005}{\emph{JHEP} {\bf 03} (2014)
  005}, [\href{https://arxiv.org/abs/1312.2109}{{\tt 1312.2109}}].

\bibitem{Bagchi2021}
A.~Bagchi, S.~Dutta, K.~S. Kolekar and P.~Sharma, \emph{{BMS field theories and
  Weyl anomaly}}, \href{http://dx.doi.org/10.1007/JHEP07(2021)101}{\emph{JHEP}
  {\bf 07} (2021) 101}, [\href{https://arxiv.org/abs/2104.10405}{{\tt
  2104.10405}}].

\bibitem{Parattu2015}
K.~Parattu, S.~Chakraborty, B.~R. Majhi and T.~Padmanabhan, \emph{{A Boundary
  Term for the Gravitational Action with Null Boundaries}},
  \href{http://dx.doi.org/10.1007/s10714-016-2093-7}{\emph{Gen. Rel. Grav.}
  {\bf 48} (2016) 94}, [\href{https://arxiv.org/abs/1501.01053}{{\tt
  1501.01053}}].

\bibitem{Lehner2016}
L.~Lehner, R.~C. Myers, E.~Poisson and R.~D. Sorkin, \emph{{Gravitational
  action with null boundaries}},
  \href{http://dx.doi.org/10.1103/PhysRevD.94.084046}{\emph{Phys. Rev. D} {\bf
  94} (2016) 084046}, [\href{https://arxiv.org/abs/1609.00207}{{\tt
  1609.00207}}].

\bibitem{Chandrasekaran2020}
V.~Chandrasekaran and A.~J. Speranza, \emph{{Anomalies in gravitational charge
  algebras of null boundaries and black hole entropy}},
  \href{http://dx.doi.org/10.1007/JHEP01(2021)137}{\emph{JHEP} {\bf 01} (2021)
  137}, [\href{https://arxiv.org/abs/2009.10739}{{\tt 2009.10739}}].

\bibitem{Adjei2019}
E.~Adjei, W.~Donnelly, V.~Py and A.~J. Speranza, \emph{{Cosmic footballs from
  superrotations}},
  \href{http://dx.doi.org/10.1088/1361-6382/ab74f6}{\emph{Class. Quant. Grav.}
  {\bf 37} (2020) 075020}, [\href{https://arxiv.org/abs/1910.05435}{{\tt
  1910.05435}}].

\bibitem{Godazgar2012}
M.~Godazgar and H.~S. Reall, \emph{{Peeling of the Weyl tensor and
  gravitational radiation in higher dimensions}},
  \href{http://dx.doi.org/10.1103/PhysRevD.85.084021}{\emph{Phys. Rev. D} {\bf
  85} (2012) 084021}, [\href{https://arxiv.org/abs/1201.4373}{{\tt
  1201.4373}}].

\bibitem{Wald2019}
G.~Satishchandran and R.~M. Wald, \emph{{Asymptotic behavior of massless fields
  and the memory effect}},
  \href{http://dx.doi.org/10.1103/PhysRevD.99.084007}{\emph{Phys. Rev. D} {\bf
  99} (2019) 084007}, [\href{https://arxiv.org/abs/1901.05942}{{\tt
  1901.05942}}].

\bibitem{Kroon:1998tu}
J.~A.~V. Kroon, \emph{{Conserved quantities for polyhomogeneous space-times}},
  \href{http://dx.doi.org/10.1088/0264-9381/15/8/023}{\emph{Class. Quant.
  Grav.} {\bf 15} (1998) 2479--2491},
  [\href{https://arxiv.org/abs/gr-qc/9805094}{{\tt gr-qc/9805094}}].

\bibitem{Witten1998}
E.~Witten, \emph{{Anti-de Sitter space and holography}}, {\emph{Adv. Theor.
  Math. Phys.} {\bf \textbf{2}} (1998) 253--291},
  [\href{https://arxiv.org/abs/hep-th/9802150}{{\tt hep-th/9802150}}].

\bibitem{Schmidt1996}
B.~G. Schmidt, \emph{Vacuum spacetimes with toroidal null infinities},
  {\emph{Class. Quant. Grav.} {\bf 13} (1996) 2811}.

\bibitem{Duval2014b}
C.~Duval, G.~W. Gibbons and P.~A. Horvathy, \emph{{Conformal Carroll groups}},
  \href{http://dx.doi.org/10.1088/1751-8113/47/33/335204}{\emph{J. Phys. A}
  {\bf 47} (2014) 335204}, [\href{https://arxiv.org/abs/1403.4213}{{\tt
  1403.4213}}].

\bibitem{Duval2014}
C.~{Duval}, G.~W. {Gibbons} and P.~A. {Horvathy}, \emph{{Conformal Carroll
  groups and BMS symmetry}},
  \href{http://dx.doi.org/10.1088/0264-9381/31/9/092001}{\emph{Class. Quant.
  Grav.} {\bf 31} (2014) 092001}, [\href{https://arxiv.org/abs/1402.5894}{{\tt
  1402.5894}}].

\end{thebibliography}\endgroup

\end{document}